\documentclass[%
 aip,
 amsmath,amssymb,
preprint,%
]{revtex4-1}

\usepackage{graphicx}
\usepackage{dcolumn}

\usepackage[utf8]{inputenc}
\usepackage[T1]{fontenc}
\usepackage{mathptmx}
\usepackage{etoolbox}
\usepackage{xcolor}

\usepackage{caption}
\usepackage{subcaption}
\usepackage{graphicx}
\usepackage{multirow}
\usepackage{mhchem}

\makeatletter
\def\@email#1#2{%
 \endgroup
 \patchcmd{\titleblock@produce}
  {\frontmatter@RRAPformat}
  {\frontmatter@RRAPformat{\produce@RRAP{*#1\href{mailto:#2}{#2}}}\frontmatter@RRAPformat}
  {}{}
}%
\makeatother

\begin{document}

\preprint{AIP/123-QED}
\title{Analytical harmonic vibrational frequencies with VV10-containing density functionals:
Theory, efficient implementation, and benchmark assessments.}

\author{Jiashu Liang}
\affiliation{
	Kenneth S. Pitzer Center for Theoretical Chemistry,
	Department of Chemistry,
	University of California at Berkeley,
	Berkeley, CA 94720, USA
}
\author{Xintian Feng}
\affiliation{
	Q-Chem Inc.,
	Pleasanton, CA 94588, USA
}
\author{Xiao Liu}
\affiliation{
	Kenneth S. Pitzer Center for Theoretical Chemistry,
	Department of Chemistry,
	University of California at Berkeley,
	Berkeley, CA 94720, USA
}
\author{Martin Head-Gordon}
\affiliation{
	Kenneth S. Pitzer Center for Theoretical Chemistry,
	Department of Chemistry,
	University of California at Berkeley,
	Berkeley, CA 94720, USA
}
\affiliation{
	Chemical Sciences Division,
	Lawrence Berkeley National Laboratory,
	Berkeley, CA 94720, USA
}
\email{mhg@cchem.berkeley.edu}

\date{\today}

\begin{abstract}

VV10 is a powerful nonlocal density functional for long-range correlation that is used to include dispersion effects in many modern density functionals such as the meta-generalized gradient approximation (mGGA), B97M-V, the hybrid GGA, $\omega$B97X-V and the hybrid mGGA, $\omega$B97M-V. While energies and analytical gradients for VV10 are already widely available, this study reports the first derivation and efficient implementation of the analytical second derivatives of the VV10 energy. The additional compute cost of the VV10 contributions to analytical frequencies is shown to be small in all but the smallest basis sets for recommended grid sizes. This study also reports the assessment of VV10-containing functionals for predicting harmonic frequencies using the analytical second derivative code. The contribution of VV10 to simulating harmonic frequencies is shown to be small for small molecules but important for systems where weak interactions are important, such as water clusters. In the latter cases, B97M-V, $\omega$B97M-V, and $\omega$B97X-V perform very well. The convergence of frequencies with respect to grid size and atomic orbital basis set size is studied and recommendations reported. Finally, scaling factors to allow comparison of scaled harmonic frequencies with experimental fundamental frequencies and to predict zero-point vibrational energy are presented for some recently developed functionals (including r2SCAN, B97M-V, $\omega$B97X-V, M06-SX, and $\omega$B97M-V).

\end{abstract}

\maketitle

\clearpage

\section{Introduction} \label{sec:intro}

Density functional theory (DFT) is currently the most popular approach to predicting molecular properties, such as relative energies, equilibrium geometries, frequencies, excitation energies, permanent and induced moments, etc.\cite{mardirossian2017thirty, palafox2018dft, zapata2021meta, liang2022revisiting} However, standard semi-local exchange-correlation functionals cannot properly account for the long-range correlation effects such as dispersion (van der Waals) interactions.\cite{kristyan1994can, perez1995density, hobza1995density,najibi2018nonlocal} A range of methods have been developed to address this issue.\cite{klimevs2012perspective, stohr2019theory} One popular, effective, and computationally very inexpensive strategy to fix the problem is adding an empirical dispersion correction, such as the exchange-dipole model (XDM)\cite{becke2007exchange}, the Tkatchenko-Scheffler van der Waals method\cite{tkatchenko2009accurate} and the DFT-D family.\cite{grimme2006semiempirical, grimme2010consistent, grimme2011effect, grimme2016dispersion} However, strictly speaking, these methods are not density functionals, as they depend explicitly on nuclear positions. The most elaborate methods are orbital-dependent, like the random-phase approximation (RPA)\cite{nguyen2010first, zhu2010range, chen2017random} and double-hybrid density functionals,\cite{goerigk2014double,martin2020empirical} but they are very time-demanding. To strike an intermediate balance between rigor and computational tractability, researchers have devised a series of nonlocal correlation functionals that aim to capture only long-range dispersion effects. The first was the well-known vdW-DF functional,\cite{dion2004van} which was designed in 2004, followed by vdW-DF2 in 2010.\cite{lee2010higher} Vydrov and Van Voorhis also proposed the VV09 functional\cite{vydrov2009nonlocal} and its more successful successor, VV10,\cite{vydrov2010nonlocal} for the calculation of dispersion-containing interactions in molecules.

VV10 has since been widely used as an add-on to existing functionals.\cite{vydrov2010nonlocal,calbo2015nonlocal,grimme2016dispersion,stohr2019theory} Furthermore, VV10 has been incorporated as a component in self-consistently training new functionals, such as $\omega$B97X-V\cite{Mardirossian:2014} [range-separated hybrid (RSH) generalized-gradient approximation (GGA)], B97M-V\cite{Mardirossian:2015} [local meta-GGA (mGGA)], and $\omega$B97M-V\cite{Mardirossian:2016} (RSH mGGA). Their success has been demonstrated by improved accuracy in predicting energy differences across broad classes of molecules composed of main group elements, with particular improvements noted for intermolecular interactions\cite{Mardirossian:2016, mardirossian2017thirty, najibi2018nonlocal} In addition, for applications in condensed matter, VV10 was reformulated, leading to the closely related rVV10 functional,\cite{sabatini2013nonlocal} which offers significant implementation advantages in periodic codes, while yielding virtually identical results.

Analytical derivative theory\cite{pulay1987analytical,amos1989implementation} has long been essential for exploring potential energy surfaces using electronic structure methods. Indeed, the analytical gradient of VV10 is widely available, enabling geometry optimizations. Building on the theory of Hartree-Fock second derivatives\cite{pople1979derivative,frisch1990direct}, efficient algorithms and implementations of DFT analytical second derivatives are well established,\cite{johnson1994implementation,deglmann2002efficient,wolff2005analytical} with some development continuing.\cite{bykov2015efficient,delgado2016analytic,gu2021second} However, to the best of our knowledge, the VV10  analytical second derivative has not yet been implemented in any quantum chemistry software. There has been recent interest in analytic derivatives through automatic differentiation\cite{abbott2021arbitrary,kasim2022dqc} or symbolic algebra,\cite{salek2007self,mazur2016automatic,van2016comparison,lehtola2018recent} but VV10 Hessians have not been reported that way either, possibly due to its complexity (as shown in this paper and similarly for vdW-DF\cite{sabatini2016phonons, miwa2022linear}). In this context, the finite difference (FD) method has to be employed for frequency,\cite{liu2017accuracy} excitation energy, and stability analysis calculations.\cite{sharada2015wavefunction} Even though the FD approach approximates analytical methods, it has more limited precision and may require significantly more computational resources. Therefore, analytical derivatives are typically preferred when feasible, and the first purpose of this paper is to report the theory and implementation of VV10 analytical second derivatives.

Nowadays many functionals have been benchmarked extensively for predicting frequencies and zero-point vibrational energies (ZPVEs).\cite{pople1981molecular, merrick2007evaluation, alecu2010computational, biczysko2010harmonic, laury2012vibrational, kesharwani2015frequency, chan2016frequency, katari2017improved, kashinski2017harmonic, hanson2019benchmarking, zapata2022vibfreq1295, unal2021scale, zapata2023model,ravichandran2018performance, howard2015assessing} Most of these works calculated harmonic frequencies with a scaling factor to fit experimental frequencies or ZPVEs, which accords with practical usage, but introduces an additional error in benchmarking.\cite{pople1981molecular, merrick2007evaluation, alecu2010computational, biczysko2010harmonic, laury2012vibrational, kesharwani2015frequency, chan2016frequency, katari2017improved, kashinski2017harmonic, hanson2019benchmarking, unal2021scale, zapata2022vibfreq1295,zapata2023model} In this scenario, B3LYP is good enough to consistently provides the best (or near-best) results among hybrid functionals.\cite{kesharwani2015frequency, chan2016frequency, katari2017improved} For detailed reviews on the topic, please consult refs.~\citenum{palafox2018dft}, \citenum{zapata2021meta}, and \citenum{zapata2023model}. By contrast, relatively few papers have compared harmonic frequencies with theoretical best estimates (TBEs), so as to directly assess the accuracy of functionals themselves.\cite{biczysko2010harmonic, howard2015assessing} Here, as our second purpose, we will benchmark B97M-V, $\omega$B97X-V, and $\omega$B97M-V with other popular functionals mainly against TBEs to avoid errors from other sources. Finally, at the end will we employ experimental reference values to develop scaling factors for recently developed functionals (including r2SCAN, B97M-V, $\omega$B97X-V, M06-SX, and $\omega$B97M-V). 

In this paper, we will describe the theory and implementation of the second analytical derivatives of VV10-containing functionals (Section~\ref{sec:method}), including orbital Hessian, nuclear Hessian, and Fock nuclear derivative contributions. We have already discussed the contribution of VV10 to excitation energies,\cite{liang2022revisiting} so we only benchmark the performance of B97M-V, $\omega$B97X-V, and $\omega$B97M-V for simulating molecular frequencies here. We will benchmark the computational time (Section~\ref{subsec:time}), explore the basis set and quadrature grid convergence (Section~\ref{subsec:convergence}), study the contribution of VV10 to frequency prediction (Section~\ref{subsec:vv10}), compare VV10-containing functionals with other popular functionals against TBEs on various data sets (Section~\ref{subsec:compare}), and finally recommend scaling factors to employ in practice (Section~\ref{subsec:SF}).

\section{Theory and Implementation of VV10} \label{sec:method}

\subsection{VV10 Energy and First Derivative}\label{subsec:1std}
The energy of the nonlocal correlation functional, VV10, can be written as (here we use atomic units)

\begin{equation} \label{eqn:EVV10}
        E^{\rm VV10} = \int d^3\mathbf{r}\, \rho(\mathbf{r})\Big[\beta + 
        \frac{1}{2}\int d^3\mathbf{r}'\, \rho(\mathbf{r}'){\rm \Phi}(\mathbf{r}, \mathbf{r}') \Big], 
\end{equation}
where $\rho(\mathbf{r})$ is the total electron density and ${\rm \Phi}(\mathbf{r}, \mathbf{r}')$ is the correlation kernel defined as
\begin{equation}\label{eqn:Cknernal}
        {\rm \Phi}= - \frac{3}{2gg'(g+g')}
\end{equation}
with
\begin{equation}\label{eqn:g}
\begin{split}
        g(\mathbf{r}, \mathbf{r}') & = \omega(\mathbf{r})|\mathbf{r}-\mathbf{r}'|^2 + \kappa(\mathbf{r}), \\
        g'(\mathbf{r}, \mathbf{r}') & = \omega(\mathbf{r}')|\mathbf{r}-\mathbf{r}'|^2 + \kappa(\mathbf{r}').
\end{split}
\end{equation}
$\omega(\mathbf{r})$ and $\kappa(\mathbf{r})$ are both intermediate single-variable functions, defined as
\begin{equation}
        \omega(\mathbf{r}) = \sqrt{C \frac{\gamma(\mathbf{r})^2}{\rho(\mathbf{r})^4} \ +\  \frac{4\pi\rho(\mathbf{r})}{3} } ,
\end{equation}
\begin{equation}
        \kappa(\mathbf{r}) = b \frac{3\pi}{2} (\frac{\rho(\mathbf{r})}{9\pi})^{\frac{1}{6}},
\end{equation}
where $\gamma =\lvert\nabla\rho\rvert^2$ is the square of the density gradient and $b$ and C are two parameters that control the behavior of VV10. For the detailed physical meaning of each variable, please consult the original paper.\cite{vydrov2010nonlocal} The correction $\beta$ in Eq~\ref{eqn:EVV10} is also determined by $b$ to ensure that the VV10 energy is equal to 0 in the uniform density limit,
\begin{equation}
        \beta = \frac{1}{32} (\frac{3}{b^2})^{\frac{3}{4}}.
\end{equation}

To evaluate the VV10 functional on a quadrature grid (for example the widely used atom-centered method pioneered by Becke\cite{becke1988multicenter}), we adopt the following formalism. The energy can be calculated as
\begin{equation}
E^{\rm VV10} = \sum_i w_i f^0_i,
\end{equation}
\begin{equation}
        f^0_i =\rho_i\big(\beta + \frac{1}{2} \mathrm{E}_i \big),
\end{equation}
\begin{equation}
\mathrm{E}_i = \sum_{j}  w_{j}\rho_{j}{\rm \Phi}_{ij},
\end{equation}
\begin{equation}
{\rm \Phi}_{ij} = - \frac{3}{2g_{ij}g'_{ij}(g_{ij}+g'_{ij})}.
\end{equation}
Here the subscript $i$ indicates that the value is at the $i$-th (outer) grid point. For example, $w_i$ is the quadrature weight of the $i$-th grid point and $f^0_i$ is the VV10 (outer) integrand at this quadrature point. $\mathrm{E}_i$ is an intermediate variable representing the inner integral of Eq~\ref{eqn:EVV10} at point $i$. Similarly, $j$ indicates that the value is at the $j$-th (inner) grid point. $g_{ij} = {\omega}_i {R^2}_{ij} + \kappa_i$, $g'_{ij} = {\omega}_{j}{R^2}_{ij} + \kappa_{j}$, and ${R^2}_{ij} = \lvert \mathbf{r}_i - \mathbf{r}_j \rvert^2$ are corresponding to Eq~\ref{eqn:Cknernal} and \ref{eqn:g}. It is worth noting $g'_{ij} = g_{ji}$ but we keep the index $j$ as the fast index throughout the implementation.

The VV10 contribution to the Fock matrix can be calculated by 
\begin{equation}
    F_{\mu \nu} = \sum_i w_i \Big[f^{\rho}_i \phi_{\mu i} \phi_{\nu i} + 2f^{\gamma}_i \nabla\rho_i \cdot (\phi_{\nu i}\nabla\phi_{\mu i}+ \phi_{\mu i}\nabla\phi_{\nu i})\Big],
\end{equation}
\begin{equation}
f^{\rho}_i = \beta + \mathrm{E}_i + \rho_i\Big[ (\frac{\partial\kappa}{\partial\rho})_i \mathrm{U}_i+ (\frac{\partial\omega}{\partial\rho})_i \mathrm{W}_i\Big],
\end{equation}
\begin{equation}
f^{\gamma}_i =  \rho_i (\frac{\partial\omega}{\partial\gamma})_i \mathrm{W}_i.
\end{equation}
$\phi_{\mu i}$ and $\phi_{\nu i}$ are the values of atomic orbitals $\phi_\mu$ and $\phi_\nu$ at the quadrature point $i$. We use real functions and thus omit complex conjugations for simplicity. $\nabla$ without any subscript stands for the gradient with respect to the electron position. $f^{\rho}_i$ and $f^{\gamma}_i$ are the partial derivatives of the outer integrand at point $i$  with respect to $\rho_i$ and $\gamma_i$ individually. $(\frac{\partial\kappa}{\partial\rho})_i$, $(\frac{\partial\omega}{\partial\rho})_i$, and $(\frac{\partial\omega}{\partial\gamma})_i$ are also the partial derivatives of $\kappa$ and $\omega$. $\mathrm{U}_i$ and $\mathrm{W}_i$ are two intermediate integral values similar to $\mathrm{E}_i$, which are defined as 
\begin{equation}
    \mathrm{U}_i = - \sum_{j} w_{j} \rho_{j}{\rm \Phi}_{ij}(\frac{1}{g_{ij}+g'_{ij}} +\frac{1}{g_{ij}}),
\end{equation}
\begin{equation}
    \mathrm{W}_i = - \sum_{j} w_{j} \rho_{j}{\rm \Phi}_{ij} {R^2}_{ij} (\frac{1}{g_{ij}+g'_{ij}} +\frac{1}{g_{ij}}).
\end{equation}

We split the VV10 gradient into three terms:
\begin{equation}
   E^A= \nabla_A E = E^A_{G} + E^A_{w} + E^A_{gr}.
\end{equation} 
Here $A$ stands for one cartesian coordinate (i.e., $x$, $y$, or $z$) of one atom. $E^A_{G}$, $E^A_{w}$, $E^A_{gr}$ denote the contributions from the change of Gaussian basis functions, quadrature weights, and grid positions respectively.
\begin{equation}
    E^A_{G} = \sum_i w_i [f^{\rho}_i \rho^A_i +  f^{\gamma}_i \gamma^A_i],
\end{equation} 
\begin{equation}
    E^A_{w} = \sum_i  w_i^A \rho_i \big(\beta + \mathrm{E}_i \big),
\end{equation} 
\begin{equation}
\label{grad_gr}
\begin{split}
    E^A_{gr} =&  -\sum_{i\in A} \sum_{j\notin A} w_i w_{j} \rho_i \rho_{j} {\rm \Phi}_{ij} (\frac{\omega_i}{g_{ij}} + \frac{\omega_{j}}{g'_{ij}} + \frac{\omega_i + \omega_{j}}{g_{ij}+ g'_{ij}})  (\mathbf{r}_{i} - \mathbf{r}_{j}) \\  
    &- \sum_{i\notin A} \sum_{j\in A} w_i w_{j} \rho_i \rho_{j} {\rm \Phi}_{ij} (\frac{\omega_i}{g_{ij}} + \frac{\omega_{j}}{g'_{ij}} + \frac{\omega_i + \omega_{j}}{g_{ij}+ g'_{ij}})  (\mathbf{r}_{j} - \mathbf{r}_{i}) \\
    =&  -2\sum_{i\in A} \sum_{j\notin A} w_i w_{j} \rho_i \rho_{j} {\rm \Phi}_{ij} (\frac{\omega_i}{g_{ij}}  + \frac{\omega_i }{g_{ij}+ g'_{ij}})  (\mathbf{r}_{i} - \mathbf{r}_{j}) \\  
    &- 2\sum_{i\notin A} \sum_{j\in A} w_i w_{j} \rho_i \rho_{j} {\rm \Phi}_{ij} (\frac{\omega_{j}}{g'_{ij}} + \frac{\omega_{j}}{g_{ij}+ g'_{ij}})  (\mathbf{r}_{j} - \mathbf{r}_{i}) .
\end{split}
\end{equation} 
Here $w_i^A$, $\rho_i^A$, and $\gamma_i^A$ are the gradient of $w_i$, $\rho_i$, and $\gamma_i$ with respect to the change of $A$. The calculation time of Eq \ref{grad_gr} can be saved half by translational invariance. Eq \ref{grad_gr} can also be written in the form of Eq \ref{EiBgr},
\begin{equation}
    E^A_{gr} =\sum_{i\in A} w_{i} \rho_{i} E^{Agr}_i
\end{equation}

\subsection{Second Derivative}\label{subsec:2nd}
 

To simulate the molecular frequencies by coupled self-consistent field theory (CPSCF), we need several key quantities. The first quantity contains the orbital Hessian contributions (the second derivative of the VV10 energy with respect to the density matrix, which is also needed for time-dependent density functional theory (TDDFT) to predict the excitation energy.\cite{liang2022revisiting}). Usually, we only calculate contracted orbital Hessian $G^t_{\mu\nu}$, to avoid the fourth-rank tensor storage of orbital Hessian. The second quantity is the nuclear Hessian, $E^{AB}$, which is the second derivative of the VV10 energy with respect to nuclear positions. The third quantity is the Fock nuclear derivative $F_{\mu\nu}^A$, which is the derivative of the VV10 contribution to the Fock matrix with respect to nuclear position. 

The contracted orbital Hessian term $G^t_{\mu\nu}$ defined in Eq~9 of  ref~\citenum{liang2022revisiting} can be calculated by
\begin{equation}
        G^t_{\mu\nu} = \sum_i  w_i \Big[ f^{\rho,t}_i
         \phi_{\mu i} \phi_{\nu i} + 2 [f^{\gamma,t}_i
        \nabla\rho_i + f^{\gamma}_i \nabla\rho^t_i ]\cdot  (\phi_{\nu i}\nabla\phi_{\mu.i}+ \phi_{\mu i}\nabla\phi_{\nu i})\Big],
\end{equation}
\begin{equation}\label{eqn:frhot}
\begin{split}
        f^{\rho,t}_i =& \sum_{j}  w_{j}( f^{\rho\rho}_{ij}\rho^t_{j} + 2 f^{\rho\gamma}_{ij}\gamma^t_{j}),\\
        f^{\gamma,t}_i=& \sum_{j}  w_{j}( f^{\gamma\rho}_{ij}\rho^t_{j} + 2 f^{\gamma\gamma}_{ij}\gamma^t_{j} ).
\end{split}
\end{equation}
$\rho^t_i$ and $\gamma^t_i$ are the trial electron density and trial gamma density on $i$-th grid point, which are defined as
\begin{equation}
\rho^t_i = \Sigma_{\mu\nu} P^t_{\mu\nu}\phi_{\mu i} \phi_{\nu i}, \quad
   \gamma^t_i = \nabla\rho_i \cdot \nabla\rho^t_i, \quad
    \nabla\rho^t_i = \Sigma_{\mu\nu} P^t_{\mu\nu} \nabla(\phi_{\mu i} \phi_{\nu i}).
\end{equation}
$P^t_{\mu\nu}$ is the trial density matrix defined in Eq~8 of ref~\citenum{liang2022revisiting}.

$f^{\rho\rho}$, $f^{\rho\gamma}$, $f^{\gamma\rho}$ and $f^{\gamma\gamma}$ in Eq~\ref{eqn:frhot} are all the second derivatives of the VV10 energy, which can be calculated by
\begin{equation}
\begin{split}
f^{\rho\rho}_{ij} &= \frac{\partial^2 E^{\rm VV10}}{\partial \rho_i \partial \rho_j} = {\rm \Phi}_{ij} \Big\{ \rho_i \Big( {R^2}_{ij}(\frac{\partial\omega}{\partial\rho})_i + (\frac{\partial\kappa}{\partial\rho})_i\Big) \rho_{j} \Big( {R^2}_{ij}(\frac{\partial\omega}{\partial\rho})_{j} + (\frac{\partial\kappa}{\partial\rho})_{j}\Big)\Big(\frac{2}{(g_{ij}+g'_{ij})^2} + \frac{2}{g_{ij}g'_{ij}}\Big) \\
& -\rho_i \Big( {R^2}_{ij}(\frac{\partial\omega}{\partial\rho})_i + (\frac{\partial\kappa}{\partial\rho})_i\Big) (\frac{1}{g_{ij}+g'_{ij}} + \frac{1}{g_{ij}}) 
 -\rho_{j} \Big( {R^2}_{ij}(\frac{\partial\omega}{\partial\rho})_{j} + (\frac{\partial\kappa}{\partial\rho})_{j}\Big) (\frac{1}{g_{ij}+g'_{ij}} + \frac{1}{g'_{ij}}) + 1 \Big\} \\
& + \frac{\delta_{ij}}{w_i} \Big\{ \ \Big[ 2(\frac{\partial\omega}{\partial\rho})_i \mathrm{W}_i + 2(\frac{\partial\kappa}{\partial\rho})_i \mathrm{U}_i\Big] + \rho_i \Big[ (\frac{\partial^2\omega}{\partial\rho^2})_i \mathrm{W}_i + (\frac{\partial^2\kappa}{\partial\rho^2})_i \mathrm{U}_i \\
& + (\frac{\partial\kappa}{\partial\rho})_i(\frac{\partial\kappa}{\partial\rho})_i \mathrm{A}_i + (\frac{\partial\omega}{\partial\rho})_i(\frac{\partial\omega}{\partial\rho})_i \mathrm{C}_i 
+ 2(\frac{\partial\omega}{\partial\rho})_i(\frac{\partial\kappa}{\partial\rho})_i \mathrm{B}_i 
\Big] \ \Big\},
\end{split}
\end{equation}
\begin{equation}
\begin{split}
        f^{\gamma\rho}_{ij} &= \frac{\partial^2 E^{\rm VV10}}{\partial \gamma_i \partial \rho_j} =
        \rho_{i} (\frac{\partial\omega}{\partial\gamma})_{i} {R^2}_{ij} {\rm \Phi}_{ij}
        \Big[\rho_{j} \Big( {R^2}_{ij}(\frac{\partial\omega}{\partial\rho})_{j} + (\frac{\partial\kappa}{\partial\rho})_{j}\Big)
        \Big(\frac{2}{(g_{ij}+g'_{ij})^2} + \frac{2}{g_{ij}g'_{ij}}\Big) - (\frac{1}{g_{ij}+g'_{ij}} + \frac{1}{g_{ij}})  \Big] \\
        & + \frac{\delta_{ij}}{w_i} \Big[ (\frac{\partial\omega}{\partial\gamma})_i \mathrm{W}_i 
        + \rho_i (\frac{\partial^2\omega}{\partial\gamma\partial\rho})_i \mathrm{W}_i
        + \rho_i (\frac{\partial\omega}{\partial\gamma})_i(\frac{\partial\kappa}{\partial\rho})_i \mathrm{B}_i
        + \rho_i (\frac{\partial\omega}{\partial\gamma})_i(\frac{\partial\omega}{\partial\rho})_i \mathrm{C}_i  \Big],
\end{split}
\end{equation}
\begin{equation}
\begin{split}
        f^{\rho\gamma}_{ij} &= \frac{\partial^2 E^{\rm VV10}}{\partial \rho_i \partial \gamma_j} = \rho_{j} (\frac{\partial\omega}{\partial\gamma})_{j} {R^2}_{ij} {\rm \Phi}_{ij}
        \Big[\rho_{i} \Big( {R^2}_{ij}(\frac{\partial\omega}{\partial\rho})_{i} + (\frac{\partial\kappa}{\partial\rho})_{i}\Big)
        \Big(\frac{2}{(g'_{ij}+g_{ij})^2} + \frac{2}{g'_{ij}g_{ij}}\Big) - (\frac{1}{g'_{ij}+g_{ij}} + \frac{1}{g'_{ij}})  \Big] \\
        & + \frac{\delta_{ij}}{w_i} \Big[ (\frac{\partial\omega}{\partial\gamma})_i \mathrm{W}_i 
        + \rho_i (\frac{\partial^2\omega}{\partial\gamma\partial\rho})_i \mathrm{W}_i
        + \rho_i (\frac{\partial\omega}{\partial\gamma})_i(\frac{\partial\kappa}{\partial\rho})_i \mathrm{B}_i
        + \rho_i (\frac{\partial\omega}{\partial\gamma})_i(\frac{\partial\omega}{\partial\rho})_i \mathrm{C}_i  \Big],
\end{split}
\end{equation}
\begin{equation}
\begin{split}
f^{\gamma\gamma}_{ij}   &=  \frac{\partial^2 E^{\rm VV10}}{\partial \gamma_i \partial \gamma_j} =
\rho_{i} \rho_{j} (\frac{\partial\omega}{\partial\gamma})_{i}(\frac{\partial\omega}{\partial\gamma})_{j} ({R^2}_{ij})^2 {\rm \Phi}_{ij} (\frac{2}{(g_{ij}+g'_{ij})^2} + \frac{2}{g_{ij}g'_{ij}}) \\
   &+ \frac{\delta_{ij}}{w_i} \rho_{i} ( (\frac{\partial^2\omega}{\partial\gamma^2})_i \mathrm{W}_i + (\frac{\partial\omega}{\partial\gamma})^2 \mathrm{C}_i).
\end{split}
\end{equation}
Here, $\mathrm{A}_i$, $\mathrm{B}_i$ and $\mathrm{C}_i$ are integrals that are similar to $\mathrm{U}_i$ and $\mathrm{W}_i$:
\begin{equation}
        \mathrm{A}_i = \sum_{j} 2 w_{j} \rho_{j}{\rm \Phi}_{ij}
        (\frac{1}{(g_{ij}+g'_{ij})^2} + \frac{1}{(g_{ij}+g'_{ij})g_{ij}} +\frac{1}{g_{ij}^2}),
\end{equation}
\begin{equation}
        \mathrm{B}_i = \sum_{j} 2 w_{j} \rho_{j}{\rm \Phi}_{ij} {R^2}_{ij}
        (\frac{1}{(g_{ij}+g'_{ij})^2} + \frac{1}{(g_{ij}+g'_{ij})g_{ij}} +\frac{1}{g_{ij}^2}),
\end{equation}
\begin{equation}
        \mathrm{C}_i = \sum_{j} 2 w_{j} \rho_{j}{\rm \Phi}_{ij} ({R^2}_{ij})^2 
        (\frac{1}{(g_{ij}+g'_{ij})^2} + \frac{1}{(g_{ij}+g'_{ij})g_{ij}} +\frac{1}{g_{ij}^2}).
\end{equation}

Similarly to the nuclear gradient, the nuclear Hessian can be split into nine terms: 
\begin{equation} \label{E_hess}
\begin{split}
   E^{AB}=& \nabla_B \nabla_A E \\
   =& E^{AB}_{G, G} + E^{AB}_{w, w} + E^{AB}_{gr, gr}  + E^{AB}_{w, G} + E^{AB}_{gr, G} + E^{AB}_{G, w}+ E^{AB}_{gr, w}  + E^{AB}_{G, gr} + E^{AB}_{w, gr} \\
   =& E^{AB}_{G, G} + E^{AB}_{w, w} + E^{AB}_{gr, gr}  + E^{AB}_{w, G} + (E^{AB}_{w, G})^\intercal + E^{AB}_{gr, G} + (E^{AB}_{gr, G})^\intercal + E^{AB}_{gr, w} + (E^{AB}_{gr, w})^\intercal.
\end{split}
\end{equation}

\begin{flalign}
E^{AB}_{w,w}  =& \sum_i (w_i^{AB} ) \rho_i\big(\beta + \mathrm{E}_i \big) + \sum_i w_i^A \otimes \rho_i \mathrm{E}_i^{Bw} ,
\\
E^{AB}_{w,gr} =& \sum_{i} w_i^A \otimes \rho_{i} \mathrm{E}_{i}^{Bgr},
\\
E^{AB}_{G, w} =& \sum_i w_i^B \otimes [f^{\rho}_i \rho^A_i +  f^{\gamma}_i \gamma^A_i] \\
    &+ \sum_i w_i \rho^A_i \otimes \Big[ \mathrm{E}_i^{Bw} +\rho_i[ (\frac{\partial\kappa}{\partial\rho})_i \mathrm{U}_i^{Bw} +  (\frac{\partial\omega}{\partial\rho})_i \otimes \mathrm{W}_i^{Bw}] \Big]\\
    &+\sum_i w_i \gamma^A_i \otimes \rho_i (\frac{\partial\omega}{\partial\gamma})_i \mathrm{W}_i^{Bw},
\\
E^{AB}_{G, gr} =& \sum_{i}  w_i  \rho^A_i \otimes \Big[ \mathrm{E}_{i}^{Bgr} + \rho_i [ (\frac{\partial\kappa}{\partial\rho})_i \mathrm{U}_{i}^{Bgr} + (\frac{\partial\omega}{\partial\rho})_i  \mathrm{W}_i^{Bgr} ]\Big]\\
    &+ \sum_{i}  w_i \gamma^A_i  \otimes \rho_i   (\frac{\partial\omega}{\partial\gamma})_i \mathrm{W}_i^{Bgr},
\\
E^{AB}_{gr,gr} =& \sum_{i\in A} w_{i} \rho_{i} \mathrm{D}_{i}^{B} \text{ if } A \neq B \text{ and } E^{AA}_{gr,gr} = -\sum_{B \neq A} E^{AB}_{gr,gr},
\\
E^{AB}_{G, G} =&\sum_i w_i \Big[f^{\rho}_i \rho^{AB}_i +  f^{\gamma}_i \gamma^{AB}_i +  \rho^A_i \otimes \sum_{j} w_{j} (f^{\rho\rho}_{ij}  \rho^B_{j} + f^{\rho\gamma}_{ij}  \gamma^B_{j}) + \gamma^ A_i \otimes \sum_{j} w_{j} (f^{\gamma\rho}_{ij} \rho^B_{j} +  f^{\gamma\gamma}_{ij} \gamma^B_{j}) \Big].
\end{flalign}
$w_i^{AB}, $ $\rho^{AB}_i$, and $\gamma^{AB}_i$ are the Hessian of $w_i$, $\rho_i$, and $\gamma_i$ with respect to the change of atoms A and B. $\mathrm{E}_i^{Bw}$, $\mathrm{E}_i^{Uw}$, $\mathrm{E}_i^{Ww}$, $\mathrm{E}_{i}^{Bgr}$, $\mathrm{E}_{i}^{Ugr}$, and $\mathrm{E}_{i}^{Wgr}$ are all integrals in the form of three-element vectors (for $x$, $y$, $z$ directions). $\mathrm{D}_{i}$ is a three-by-three matrix. They can be calculated through the following equations.

\begin{flalign}
\mathrm{E}_i^{Bw} =& \sum_{j} w_{j}^B\rho_{j}{\rm \Phi}_{ij}, \\
\mathrm{U}_i^{Bw} =& - \sum_{j}w_{j}^B \rho_{j}{\rm \Phi}_{ij}(\frac{1}{g_{ij}+g'_{ij}} +\frac{1}{g_{ij}}),\\
\mathrm{W}_i^{Bw} =& - \sum_{j}w_{j}^B \rho_{j}{\rm \Phi}_{ij} {R^2}_{ij} (\frac{1}{g_{ij}+g'_{ij}} +\frac{1}{g_{ij}}),
\end{flalign}

\begin{equation}\label{EiBgr}
\mathrm{E}_{i}^{Bgr} =\begin{cases}
          -2\sum_{j\in B} w_{j} \rho_{j} {\rm \Phi}_{ij} (\frac{\omega_i}{g_{ij}} + \frac{\omega_{j}}{g'_{ij}} + \frac{\omega_i + \omega_{j}}{g_{ij}+ g'_{ij}}) (\mathbf{r}_{j} - \mathbf{r}_{i}), \quad i \notin B \\
          -\sum_{C\neq B} \mathrm{E}_{i}^{Cgr} , \quad i \in B \\
     \end{cases}
\end{equation}
\begin{equation}
 \mathrm{U}_{i}^{Bgr} =  \left\{
\begin{array}{ll}
    2\sum_{j\in B}  w_{j} \rho_{j} {\rm \Phi}_{ij}\Big[  (\frac{\omega_i}{g_{ij}} + \frac{\omega_{j}}{g'_{ij}} + \frac{\omega_i + \omega_{j}}{g_{ij}+ g'_{ij}})(\frac{1}{g_{ij}+g'_{ij}} +\frac{1}{g_{ij}}) &\\
    + \Big( \frac{\omega_i}{g_{ij}^2} + \frac{\omega_i + \omega_{j}}{(g_{ij}+ g'_{ij})^2} \Big) \Big](\mathbf{r}_{j} - \mathbf{r}_{i}),  \quad i \notin B &\\
    -\sum_{C\neq B} \mathrm{U}_{i}^{Cgr},  \quad i \in B  &\\
\end{array} 
\right. 
\end{equation}
\begin{equation}
 \mathrm{W}_{i}^{Bgr} =  \left\{
\begin{array}{ll}
    2 \sum_{j\in B} w_{j} \rho_{j} {\rm \Phi}_{ij} \Big[ {R^2}_{ij} (\frac{\omega_i}{g_{ij}} + \frac{\omega_{j}}{g'_{ij}} + \frac{\omega_i + \omega_{j}}{g_{ij}+ g'_{ij}}) (\frac{1}{g_{ij}+g'_{ij}} +\frac{1}{g_{ij}})  \\
        + {R^2}_{ij} \Big( \frac{\omega_i}{g_{ij}^2} + \frac{\omega_i + \omega_{j}}{(g_{ij}+ g'_{ij})^2} \Big)- (\frac{1}{g_{ij}+g'_{ij}} +\frac{1}{g_{ij}}) \Big] (\mathbf{r}_{j} - \mathbf{r}_{i}),  \quad i \notin B &\\
    -\sum_{C\neq B} \mathrm{W}_{i}^{Cgr},  \quad i \in B  &\\
\end{array} 
\right. 
\end{equation}
\begin{equation}
\begin{split}
\mathrm{D}_{i}^{B} = & 
    -2 \sum_{j\in B} w_{j} \rho_{j} {\rm \Phi}_{ij} \Bigg[ 2 \Big[(\frac{\omega_i}{g_{ij}} + \frac{\omega_{j}}{g'_{ij}} + \frac{\omega_i + \omega_{j}}{g_{ij}+ g'_{ij}})^2 + (\frac{\omega_i}{g_{ij}})^2 + (\frac{\omega_{j}}{g'_{ij}})^2  \\
    &+ (\frac{\omega_i + \omega_{j}}{g_{ij} + g'_{ij}})^2 \Big] (\mathbf{r}_{j} - \mathbf{r}_{i})\otimes (\mathbf{r}_{j} - \mathbf{r}_{i})  - (\frac{\omega_i}{g_{ij}} + \frac{\omega_{j}}{g'_{ij}} + \frac{\omega_i + \omega_{j}}{g_{ij}+ g'_{ij}}) \mathbf{I}_3\Bigg], \quad i \notin B 
\end{split}
\end{equation}
Here the symbol $\otimes$ represents the outer product of two vectors and $\mathbf{I}_3$ is the three-by-three identity matrix.

The nuclear derivative of the Fock matrix can be also split into three terms:
\begin{equation}
   F_{\mu\nu}^A = \nabla_A F_{\mu\nu} = F_{\mu\nu,G}^A + F_{\mu\nu,w}^A + F_{\mu\nu,gr}^A ,
\end{equation} 
\begin{equation}
\begin{split}
    F_{\mu\nu,G}^A  &= \sum_i w_i \Big[f^{\rho}_i \nabla_A (\phi_{\mu} \phi_{\nu})_i +  2f^{\gamma}_i \,\nabla_A[\nabla\rho\cdot \nabla(\phi_{\mu}\phi_{\nu})]_i \Big] \\
    &+  \sum_i w_i \Big[ \phi_{\mu i} \phi_{\nu i} \sum_{j} w_{j} [f^{\rho\rho}_{ij}  \rho^A_{j} + f^{\rho\gamma}_{ij}  \gamma^A_{j}] + 2(\nabla\rho)_i \cdot [\nabla(\phi_{\mu}\phi_{\nu})]_i \sum_{j} w_{j} [f^{\gamma\rho}_{ij} \rho^A_{j} + f^{\gamma\gamma}_{ij} \gamma^A_{j}] \Big],
\end{split}
\end{equation}
\begin{equation}
\begin{split}
    F_{\mu\nu,gr}^A &=  \sum_{i}  w_i  \phi_{\mu i} \phi_{\nu i} \Big[ \textbf{E}_{i}^{Bgr} + \rho_i [ (\frac{\partial\kappa}{\partial\rho})_i \textbf{U}_{i}^{Bgr} + (\frac{\partial\omega}{\partial\rho})_i  \textbf{W}_i^{Bgr} ]\Big]\\
    &+ 2\sum_{i}  w_i (\nabla\rho)_i\cdot [\nabla(\phi_{\mu}\phi_{\nu})]_i  \rho_i   (\frac{\partial\omega}{\partial\gamma})_i \textbf{W}_i^{Bgr},
\end{split}
\end{equation} 
\begin{equation}
\begin{split}
F_{\mu\nu,w}^A &= \sum_i (\nabla_{A}w_i) \Big[f^{\rho}_i \phi_{\mu i} \phi_{\nu i} + 2 f^{\gamma}_i (\nabla\rho)_i\cdot [\nabla(\phi_{\mu}\phi_{\nu})]_i \Big] \\
    & + \sum_i w_i \phi_{\mu i} \phi_{\nu i} \Bigg[ \textbf{E}_i^{Aw} + \rho_i\Big[ (\frac{\partial\kappa}{\partial\rho})_i  \textbf{U}_i^{Aw}+ (\frac{\partial\omega}{\partial\rho})_i \textbf{W}_i^{Aw}\Big]\Bigg] \\
    &+ 2\sum_i w_i (\nabla\rho)_i \cdot [\nabla(\phi_{\mu}\phi_{\nu})]_i \Big[ \rho_i (\frac{\partial\omega}{\partial\gamma})_i \textbf{W}_i^{Aw}\Big],\\
\end{split}
\end{equation} 
where the dot symbol $\cdot$ represents the dot product of two vectors.

When implementing the equations above, we have adopted some important optimizations to improve efficiency. For example, translational invariance is employed to avoid or minimize some computationally heavy tasks, like the computations of $w_i^A\, (i\in A)$, $\mathrm{E}_i^{Bgr}\, (i\in B)$, and $E^{AA}_{gr,gr}$. Grid screening technology is also used to avoid unnecessary calculations like $\sum_{j} w_{j} [f^{\rho\rho}_{ij}  \rho^B_{j} + f^{\rho\gamma}_{ij}  \gamma^B_{j}]$ when $\rho^B$ and $\gamma^B$ are under the precision threshold in this batch of $j$.

\section{Computational details and time benchmark} \label{sec:comp}

\subsection{Computational details}\label{subsec:comp}

We have selected a range of widely used density functionals, as well as a set of density functionals containing the VV10 functional for the assessment of predicted frequencies. The tested density functionals include
\begin{enumerate}
    \item GGA (Rung 2): B97-D\cite{grimme2006semiempirical}, PBE\cite{perdew1996generalized};
    \item mGGA (Rung 3): B97M-V\cite{Mardirossian:2015}, SCAN\cite{sun2015strongly}, r2SCAN\cite{furness2020accurate}, M06-L\cite{zhao2006new};
    \item hybrid GGA (Rung 4a): $\omega$B97X-D\cite{chai2008long}, CAM-B3LYP\cite{yanai2004new}, $\omega$B97X-V\cite{Mardirossian:2014}, HSE-HJS\cite{henderson2008generalized,krukau2006influence}, B3LYP\cite{Becke1993,stephens1994ab};
    \item hybrid mGGA (Rung 4b): BMK\cite{boese2004development}, M06-SX\cite{wang2020m06}, M06-2X\cite{zhao2008m06}, $\omega$B97M-V\cite{Mardirossian:2016},  SCAN0\cite{hui2016scan}.
\end{enumerate}
The D3(BJ) empirical correction\cite{grimme2011effect} is added to any functional without its own dispersion component. 

We benchmarked the performance of these functionals for predicting harmonic frequency TBEs on the data sets shown in Table~\ref{tab:dataset} and used their theoretical best geometries (TBGs) as the starting geometries of the optimization.

\begin{table}
\caption{Summary of data sets used for benchmarking the performance of density functionals for predicting harmonic frequencies.
}
\label{tab:dataset}
\resizebox{\textwidth}{!}{
\begin{tabular}{cccccc}
\hline
\textbf{Data set} & \textbf{Subset Name} & \textbf{\begin{tabular}[c]{@{}c@{}}Computational method\\ to generate TBEs\end{tabular}} & \textbf{\begin{tabular}[c]{@{}c@{}}Computational method\\ to generate TBGs\end{tabular}} & \textbf{\begin{tabular}[c]{@{}c@{}}Number of molecules\\ (or conformations)\end{tabular}} & \textbf{\begin{tabular}[c]{@{}c@{}}Number of \\ frequencies\end{tabular}} \\ 
\hline
\multirow{3}{*}{Covalent Set} & HFREQ2014\cite{martin2014assessment}$^,$\footnote{A set of common small molecules with high-quality benchmark values. We discarded the CH\textsubscript{3}OH molecule because its geometry is not reported clearly in the original paper.} & CCSD(T*)-F12c/cc-pVQZ-F12 & CCSD(T*)-F12c/cc-pVQZ-F12 & 30 & 112 \\
 & Diatomic\cite{bertels2021polishing}$^,$\footnote{ A set of diatomic molecules containing more diverse elements than the HFREQ2014 set. We split it into a restricted subset (DiatomicR, which has 25 frequencies) and an unrestricted subset (DiatomicU, which has 45 frequencies).} & \begin{tabular}[c]{@{}c@{}}CCSD(T)(:$\kappa$-OOMP2)\footnote{CCSD(T) based on $\kappa$-OOMP2 orbitals.\cite{lee2018regularized}}/\\ aug-cc-pwCVTZ\cite{peterson2002accurate,prascher2011gaussian}\end{tabular} & \begin{tabular}[c]{@{}c@{}}CCSD(T)(:$\kappa$-OOMP2)/\\ aug-cc-pwCVTZ\end{tabular} & 70 & 70 \\
\hline
\multirow{6}{*}{Noncovalent Set}
& H2O\cite{howard2015benchmark}$^,$\footnote{A set of water clusters ranging in size from the trimer to four different isomers of the hexamer.} & CCSD(T):MP2\footnote{2-body:Many-body CCSD(T):MP2 theory\cite{howard2013n}}/haQZ\footnote{aug-cc-pVQZ basis set for non-hydrogen atoms and cc-pVQZ for hydrogen atoms.} & CCSD(T):MP2/haQZ & 7 & 282 \\
 & V30\cite{hoja2022v30}$^,$\footnote{A set of molecular dimers with different polarity combinations (polar-polar, polar-nonpolar, and nonpolar-nonpolar).} & CCSD(T)/haQZ & CCSD(T)/haQZ & 30 & 331 \\
 & H2S\cite{lemke2017structure}$^,$\footnote{\ce{H2S} dimer.} & CCSD(T)/aug-cc-pVQZ & CCSD(T)/aug-cc-pVQZ & 1 & 12 \\
 & N2O\cite{salmon2016structure}$^,$\footnote{A set of 4 dimers incorporating \ce{N2O}.} & CCSD(T)-F12b/cc-pVQZ-F12\footnote{\ce{O2}-\ce{N2O} is an exception which is calculated by UCCSD(T)-F12b/cc-pVTZ-F12 method.} & CCSD(T)-F12b/cc-pVQZ-F12 & 4 & 36 \\
 & CO2\cite{de2011explicit}$^,$\footnote{\ce{CO2-CO} dimer and \ce{CO2-NH3} dimer that are not in the V30 set} & CCSD(T)-F12b/cc-pVTZ-F12 & CCSD(T)-F12b/cc-pVTZ-F12 & 2 & 24 \\
 & P\cite{van2016big}$^,$\footnote{A set of P\textsubscript{2} dimer and PCCP dimers in different point group symmetries.} & CCSD(T)/aug-cc-pVTZ & CCSD(T)/aug-cc-pVTZ & 5 & 79 \\
\hline
\multicolumn{4}{c}{\textbf{At total}} & 149 & 946 \\ \hline
\end{tabular}}
\end{table}

Q-Chem 5.4 and Q-Chem 6.0 were used to perform all of the calculations.\cite{epifanovsky2021software} In Section~\ref{subsec:convergence}, the def2 basis set family\cite{weigend2005balanced, rappoport2010property} (def2-SVP, def2-SVPD, def2-TZVP, def2-TZVPD, def2-QZVPP, and def2-QZVPPD) is used to study the basis set convergence, and the standard grids\cite{gill1993, dasgupta2017standard} (SG-0/1/2/3) in Q-Chem and their parents [(23, 170), (50, 194), (75, 302), (99, 590)] are used to study the grid convergence. Here (X, Y) means a radial grid with X points and an angular Lebedev grid with Y points. After that, all the calculations employ the (99,590) grid for local functional (XC) integrals and SG-1 for nonlocal VV10 correlation (NL). 

It is worth noting that optimizing the geometries of molecules in the Noncovalent Set proved to be quite challenging due to obtaining imaginary frequencies at a fraction of the optimized geometries. To eliminate imaginary frequencies, we have attempted a variety of techniques to refine the optimized geometries, including altering convergence thresholds, employing the exact Hessian, perturbing the geometries, and so on. The details are described in the Supporting Information. 

\subsection{Time benchmark}\label{subsec:time}

As described in Section~\ref{subsec:2nd}, we can divide a frequency job into different computational tasks: self-consistent field (SCF) energy, orbital Hessian contributions (G), nuclear Hessian (Hess), and Fock nuclear derivative (F\textsubscript{nuc}). For each task, we need to construct matrix contributions from the XC, NL, and Coulomb and HF exchange (JK) parts individually. Here we report the parallel efficiency and time scaling with respect to basis set size and molecule size for $\omega$B97M-V frequency calculations in our implementation. All timing jobs were run on a single Haswell node of the NERSC supercomputer. Each Haswell node (Intel Xeon Processor E5-2698 v3) has two sockets, each populated with a 2.3 GHz 16-core Haswell processor. These timings reflect OpenMP performance up to 32 physical cores, as well as system-size scaling for our current implementation.

\begin{figure}[ht!]
     \centering
     \begin{subfigure}[b]{0.45\textwidth}
         \centering
         \includegraphics[width=\textwidth]{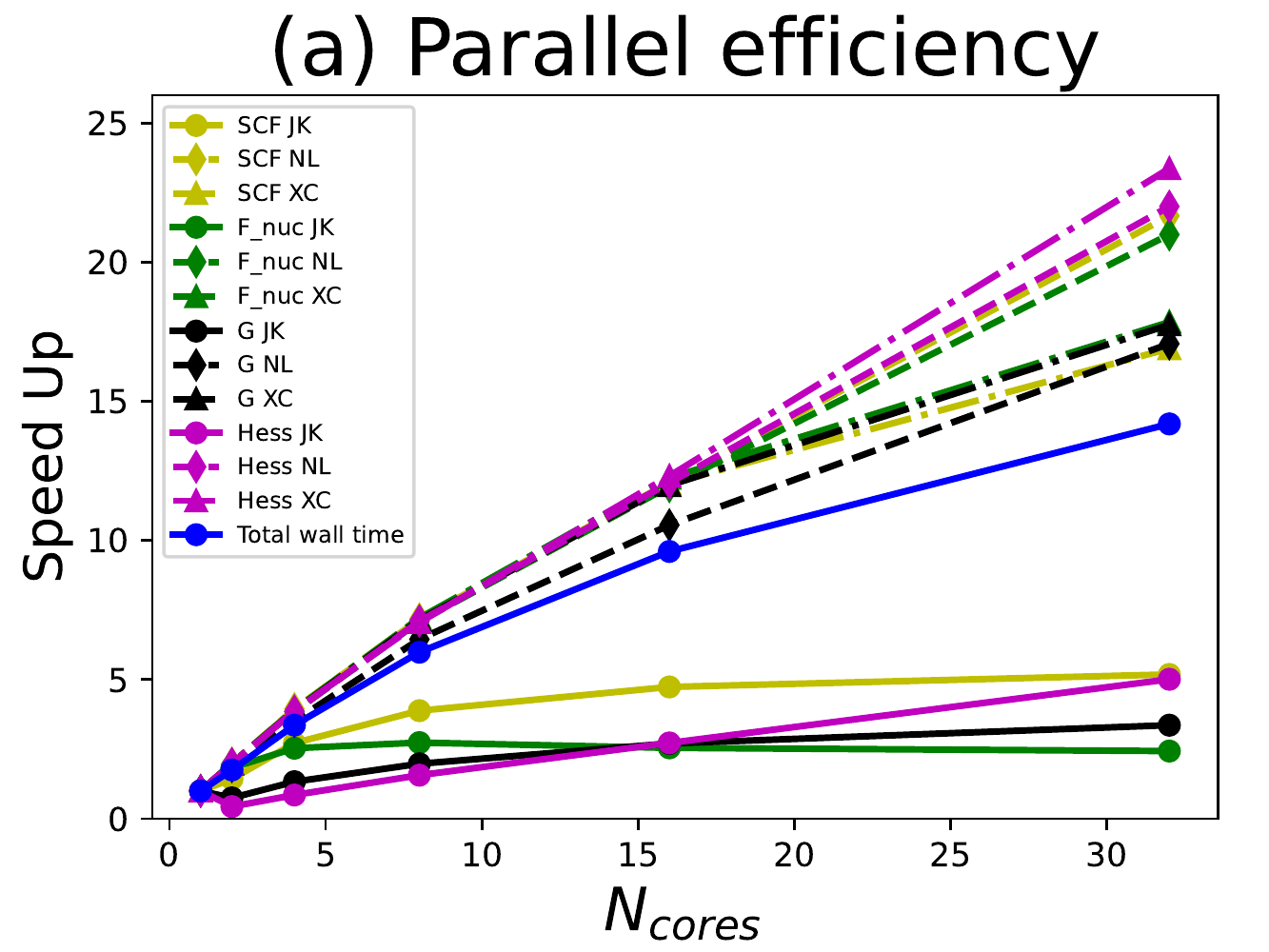}
     \end{subfigure}
     \begin{subfigure}[b]{0.45\textwidth}
         \centering
         \includegraphics[width=\textwidth]{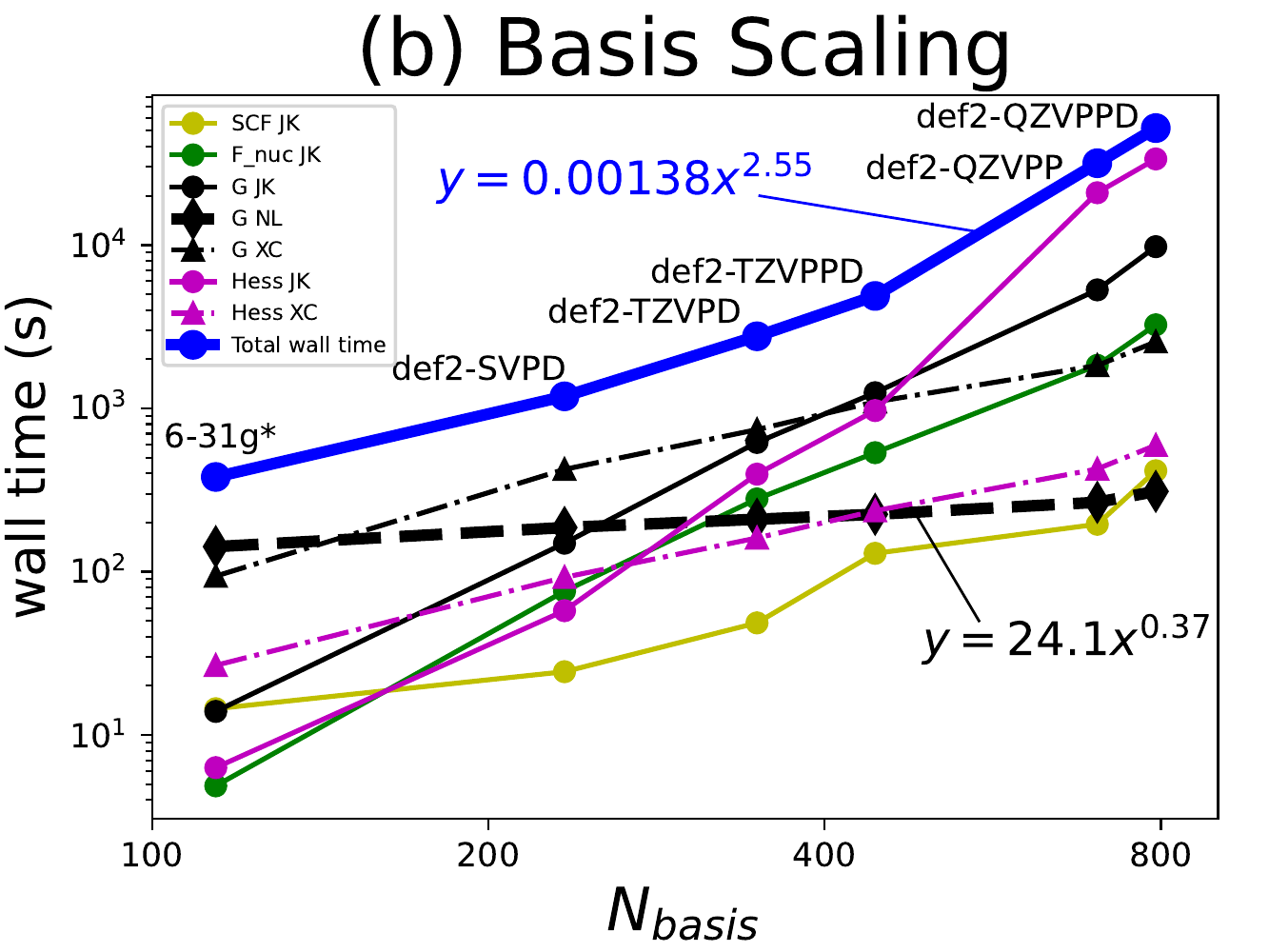}
     \end{subfigure}
     \begin{subfigure}[b]{0.5\textwidth}
         \centering
         \includegraphics[width=\textwidth]{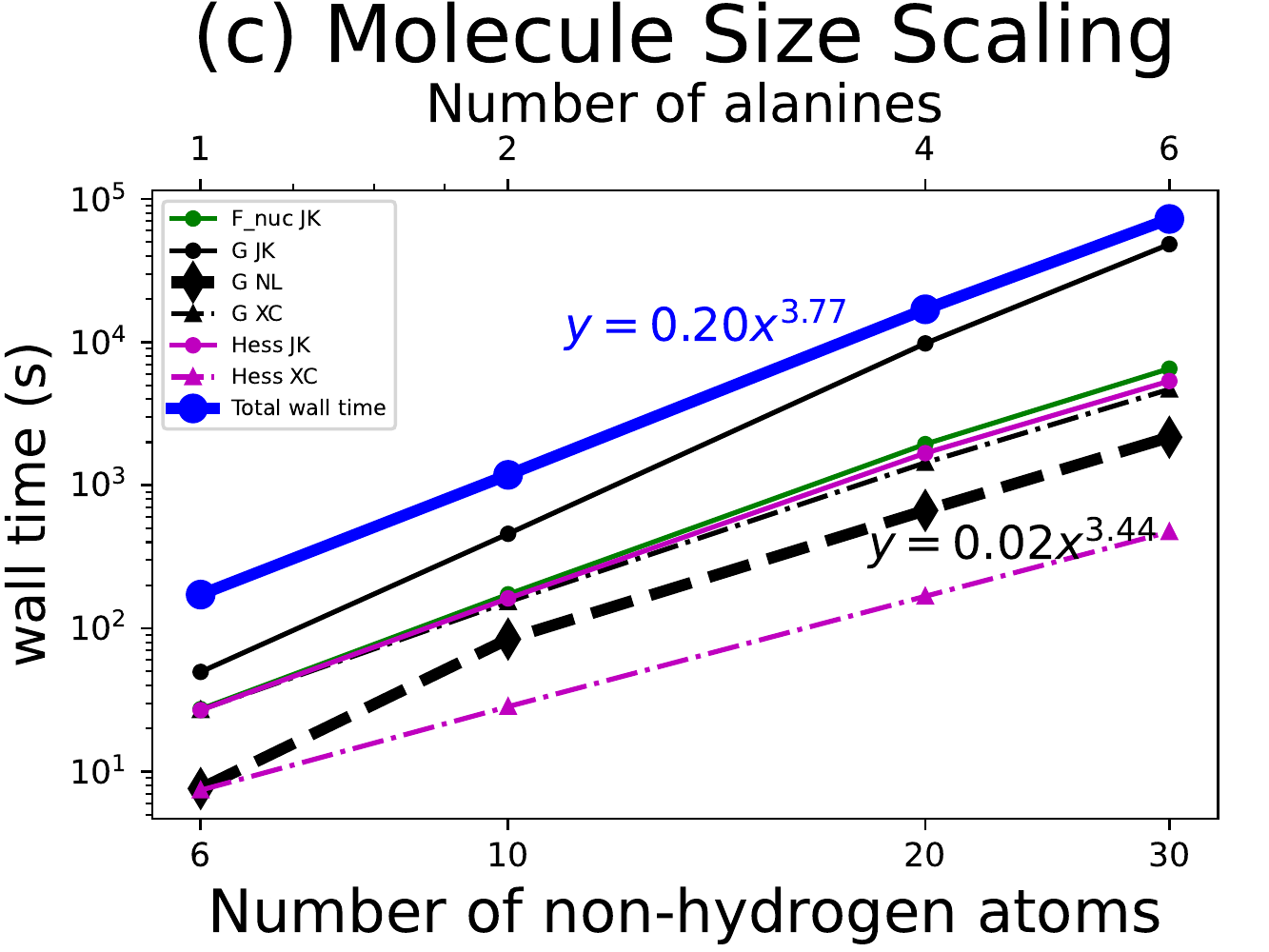}
     \end{subfigure}
        \caption{Wall time for calculating the analytic frequencies with $\omega$B97M-V under different conditions: (a) (H\textsubscript{2}O)\textsubscript{6} using the def2-SVPD basis set as a function of the number of physical cores. (b)  (H\textsubscript{2}O)\textsubscript{6} system using different basis sets (6-31g*, def2-SVPD, def2-TZVPD, def2-TZVPPD, def2-QZVPP, def2-QZVPPD in order of increasing size), in calculations using 32 physical cores. (c) alanine polypeptide systems using the def2-SVPD basis set with 32 physical cores. We use benchmark-level accuracy for (a) and (b), with (99, 590)/SG-1 as the XC/NL grid type and $10^{-14}$ as the integral threshold, and routine production-level protocol for (c), with SG-2/SG-0 as the XC/NL grid type and $10^{-10}$ as the integral threshold.}
        \label{fig:time_bench}
\end{figure}

From Figure~\ref{fig:time_bench}(a), the parallel efficiency of the NL part is as good as for the XC part and we can still get an appreciable speedup between 16 and 32 physical cores. While it is not the topic of this work, we see that the parallel efficiency of the JK terms is not good, and suggests that these codes would benefit from restructuring. Figure~\ref{fig:time_bench}(b) and \ref{fig:time_bench}(c) displays the elapsed time of those tasks which take more than 1\% of total wall time. It is clear that only the orbital Hessian (G) of the NL part takes more than 1\% of the total wall time. Encouragingly, this is still less than the XC part and likewise is less than 10\% of total wall time except for the smallest basis set, 6-31G*. The total wall time and time scaling are both dominated by the JK part, which appears to reflect its relatively poor parallel scaling as already discussed.

From Figure~\ref{fig:time_bench}(b) and Table~S1.1, we can see that the elapsed time for the NL part of all tasks grows  sublinearly with respect to the number of basis functions ($N_{basis}$) (i.e. increasing basis set size for fixed molecule size). This is because the dominant step of the NL calculations is mainly the double integral, which is not related to the basis set. The scalings of wall time with respect to the molecule size ($M$, the number of non-hydrogen atoms) in alanine poly-peptide systems are around $O(M^{3.0})$, $O(M^{3.44})$, and $O(M^{2.77})$ for F\textsubscript{nuc}, G, and Hess respectively. They can be further reduced to $O(M^{2.84})$, $O(M^{2.90})$, and $O(M^{2.73})$ if only fitting the data with 20 and 30 non-hydrogen atoms, indicating the optimized code performs very well with the SG-0 grid. 

\section{Results and discussion} \label{sec:result}

\subsection{Basis set and grid convergence}\label{subsec:convergence}

\begin{figure}[ht!]
    \centering
     \begin{subfigure}[b]{0.45\textwidth}
         \centering
         \includegraphics[width=\textwidth]{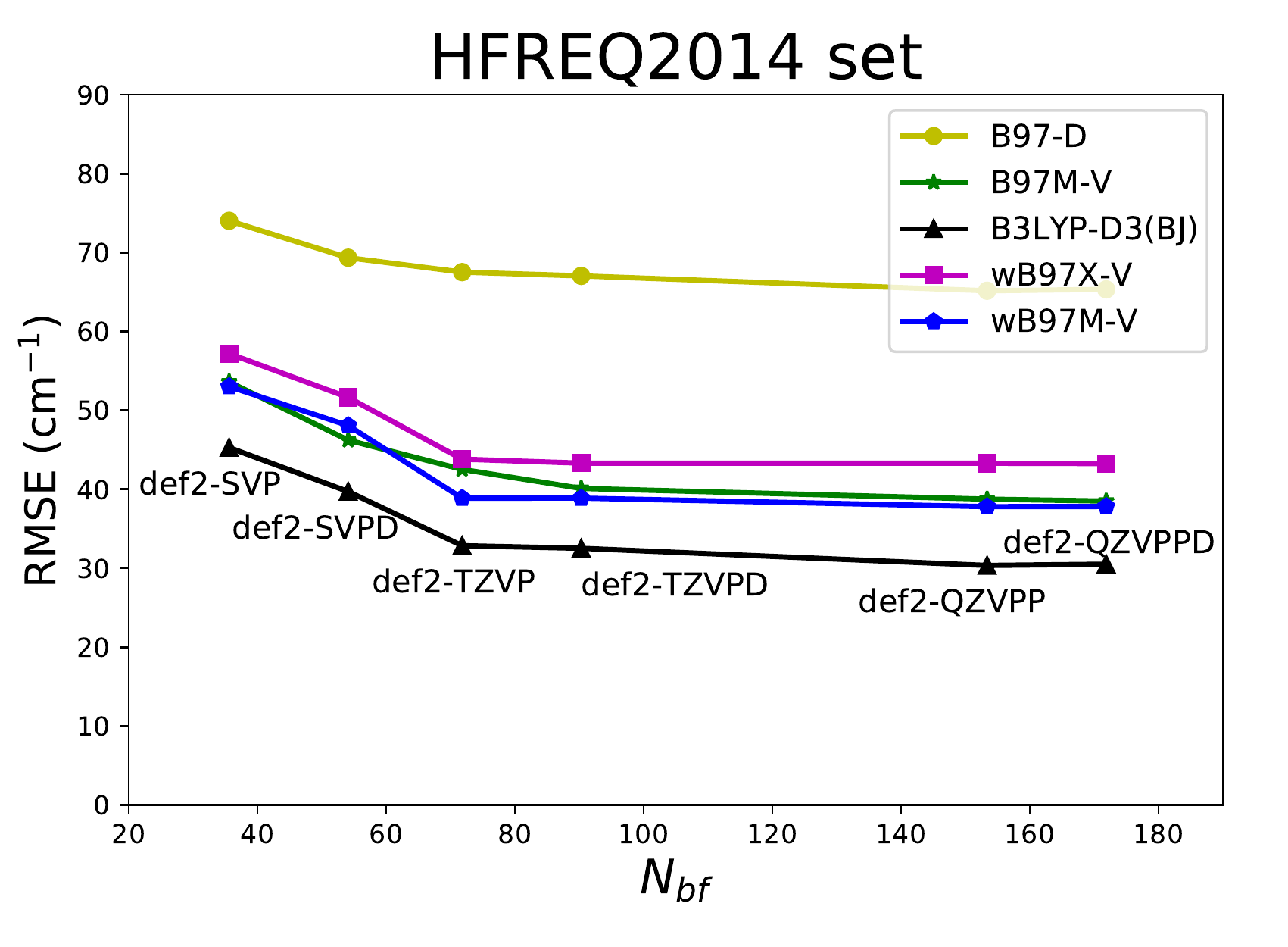}
     \end{subfigure}
     \begin{subfigure}[b]{0.45\textwidth}
         \centering
         \includegraphics[width=\textwidth]{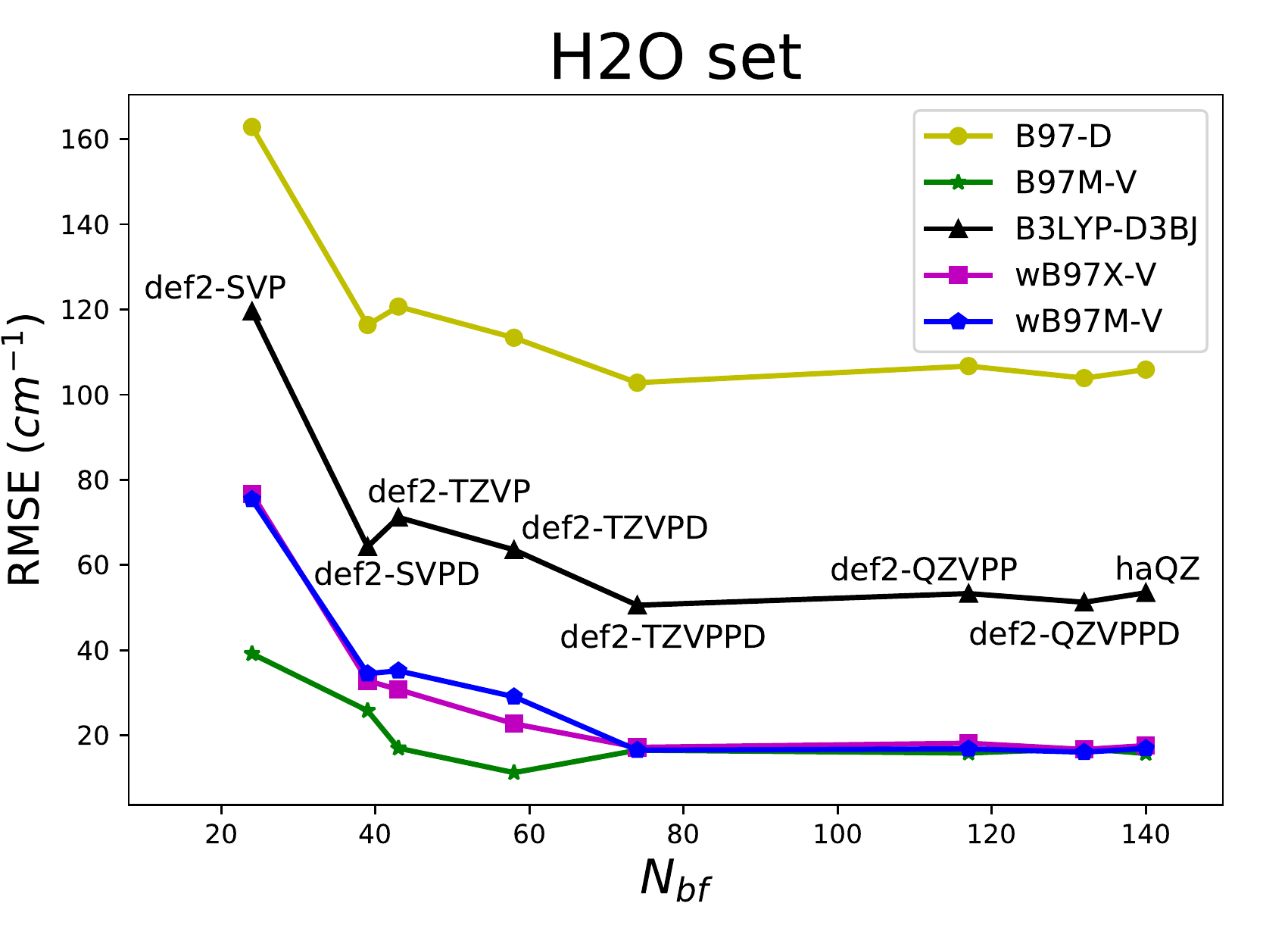}
     \end{subfigure}
    \caption{Comparison of the RMSEs (cm$^{-1}$) of several basis sets relative to TBEs using the optimized geometries against their average number of basis functions (for the  HFREQ2014 data set) or the number of functions on a single water molecule (for the H2O cluster data set) ($N_{bf}$)
    Tested basis sets include def2-SVP, def2-SVPD, def2-TZVP, def2-TZVPD, def2-TZVPPD, def2-QZVPP, def2-QZVPPD, and haQZ (only for the H2O set).}
    \label{fig:F12_basis}
\end{figure}

We explored the basis set and quadrature grid convergence of harmonic frequencies evaluated with B97M-V, $\omega$B97X-V, and $\omega$B97M-V functionals at their optimized geometries (OptGs) and compared them with B97-D and B3LYP-D3(BJ). These calculations used the (99, 590) XC integral grid, and the SG-1 NL grid.
From Figure~\ref{fig:F12_basis}, it is evident that basis set convergence for water clusters is more difficult than for the HFREQ2014 set. The difference in root-mean-squared errors (RMSEs) obtained with def2-TZVP and def2-QZVPPD basis sets for the HFREQ2014 set is at most 4 cm$^{-1}$, around $10\%$ of the method error. In contrast, the larger def2-TZVPPD basis is needed to achieve the same accuracy for the H2O set. As a result, we recommend def2-TZVP for common chemically-bonded systems and def2-TZVPPD for systems with noncovalent interactions such as water clusters.

\begin{figure}[ht!]
     \centering
     \begin{subfigure}[b]{0.45\textwidth}
         \centering
         \includegraphics[width=\textwidth]{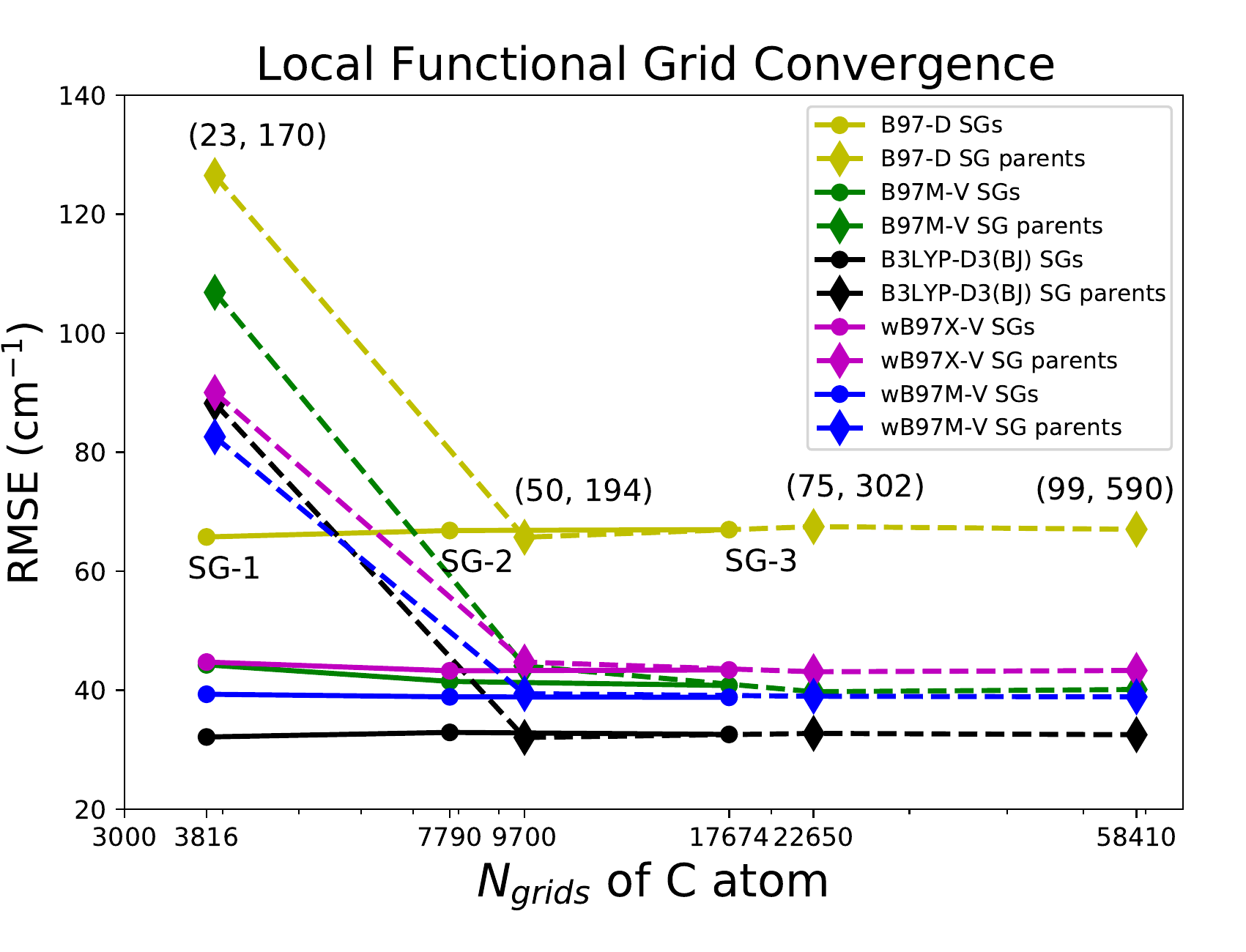}
     \end{subfigure}
     \begin{subfigure}[b]{0.45\textwidth}
         \centering
         \includegraphics[width=\textwidth]{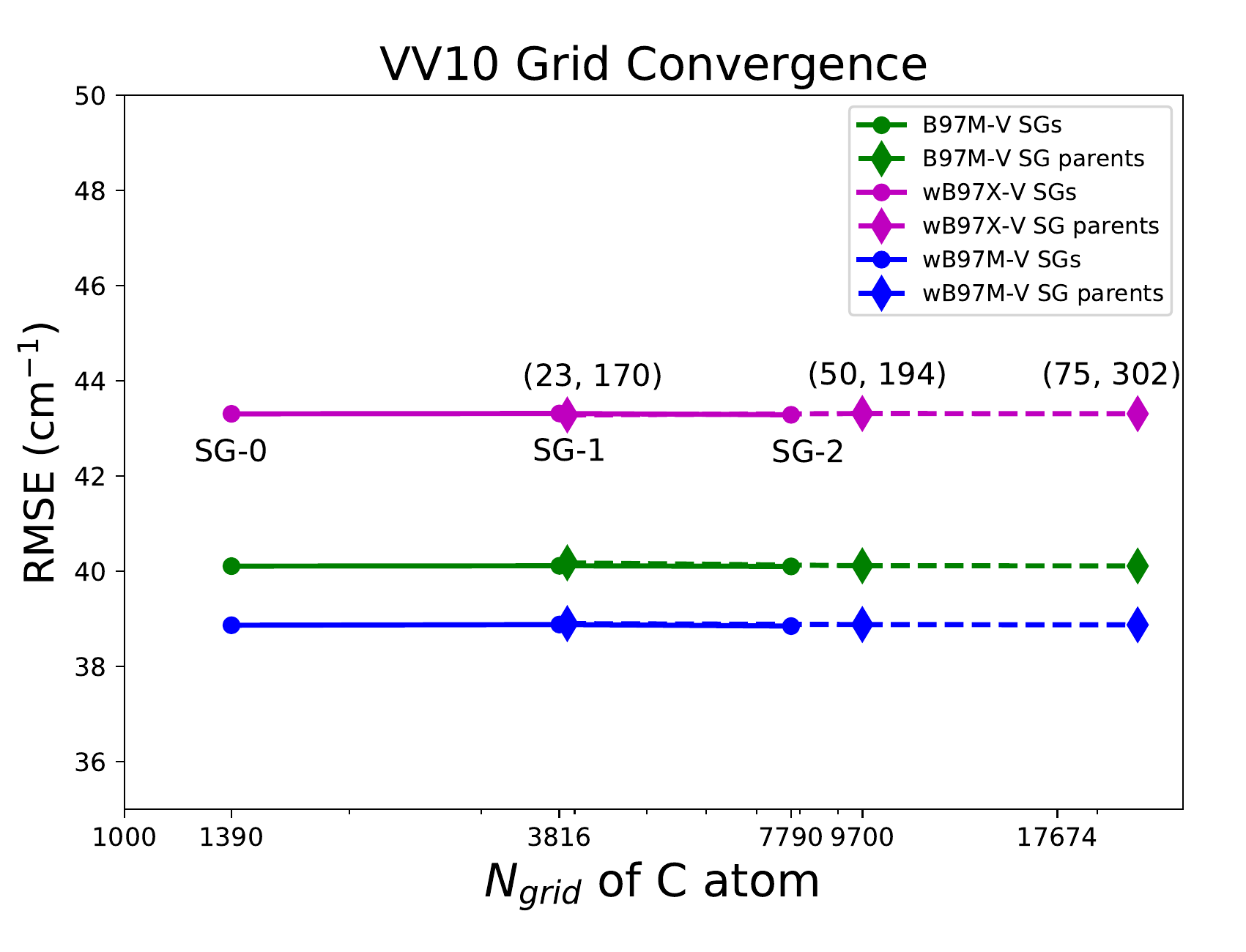}
     \end{subfigure}
        \caption{Comparison of the RMSEs (cm$^{-1}$) of several grid types for XC and NL functionals relative to TBEs using the optimized geometries against the number of grid points for one carbon atom on the HFREQ2014 set. Tested grid types include the standard grids (SG-0/1/2/3) in Q-Chem and their parents [(23, 170), (50, 194), (75, 302), (99, 590)].}
        \label{fig:F12_grid}
\end{figure}

Figure~\ref{fig:F12_grid} displays the RMSEs produced with different grids on the HFREQ2014 data set. It turns out that quadrature grid convergence is quite easy to achieve for simulating frequencies. If we regard 1 cm$^{-1}$ of RMSE difference with the biggest tested grid as the accepted threshold, then SG-2 is large enough for semi-local XC integrals and SG-0 is large enough for the nonlocal (NL) integrals of VV10. The smaller SG-1 grid for XC integration can also offer an RMSE difference of at most 4 cm$^{-1}$. Table~\ref{tab:H20_grid} and Table~S1.3 further validate this conclusion on the H2O set and the DiatomicU set. We only found that the XC part of B97M-V is a little more sensitive to the grid type, especially the pruning associated with the use of standard grids. For example, SG-3 is a pruned subset of (99, 590) grid and its performance is expected to be fairly similar to (99, 590) and better than (75, 302). However, the difference between RMSEs obtained with SG-3/SG-1 and (99, 590)/SG-1 on the H20 data set is 2.42 cm$^{-1}$, much bigger than the difference between (75, 302)/SG-1 and (99, 590)/SG-1 (0.66 cm$^{-1}$). One possible reason is that SG-3 removes some grid points far away from the nucleus, which are important to describe hydrogen bonds. From Table~S1.3, we can see that the low frequencies ($<$ 1000 cm$^{-1}$) are more affected than the high frequencies ($\geq$ 1000 cm$^{-1}$). Therefore, we recommend the SG-2/SG-0 grid combination for common functionals on chemically-bonded systems and (75, 302)/SG-1 for sensitive functionals on hydrogen-bonding systems, provided that SCF and CPSCF can be converged. It's worth noting that purely dispersion-bound systems such as Ar\textsubscript{2} and He\textsubscript{2} may require a larger grid according to Ref~\citenum{sitkiewicz2022reliable}. Additionally, some functionals, like SCAN, exhibit poor grid convergence behavior on all systems, as shown in Figure S18 of Ref~\citenum{sitkiewicz2022reliable}. In fact, we also tested SCAN on the HFREQ2014 dataset using (99, 590) and (250, 974) grids, and the RMSE difference is as high as 8 cm$^{-1}$.

\begin{table}[hbt!]
\caption{Comparison of the RMSEs (cm$^{-1}$) of several grid types for some representative functionals relative to TBEs using the optimized geometries on the H2O set. For B97-D and B3LYP-D3(BJ), only one entry is displayed at each XC grid level since they have no NL component.}
\label{tab:H20_grid}
\begin{tabular}{lccccc}
\hline
\textbf{XC/NL Grid Type}      & \textbf{B97-D} & \textbf{B97M-V} & \textbf{B3LYP} & \textbf{$\omega$B97X-V} & \textbf{$\omega$B97M-V} \\ \hline
\textbf{SG-2/SG-0}            & 112.72         & 16.92           & 63.31          & 22.72            & 29.31            \\ 
\textbf{SG-3/SG-0}            & 113.19         & 13.71           & 63.71          & 22.72            & 29.48            \\ 
\textbf{SG-3/SG-1}            &                & 13.67           &                & 22.70            & 29.22            \\ 
\textbf{(75, 302)/SG-1}            &                & 11.91           &                &             &             \\ 
\textbf{(99, 590)/SG-1}       & 113.35         & 11.25           & 63.57          & 22.74            & 29.08            \\ 
\textbf{(99, 590)/(50,194)} &                & 11.26           &                & 22.75            & 29.08            \\ \hline
\end{tabular}
\end{table}

\subsection{Effect of VV10} \label{subsec:vv10}

\begin{figure}[htb!]
     \centering
     \begin{subfigure}[b]{0.45\textwidth}
         \centering
         \includegraphics[width=\textwidth]{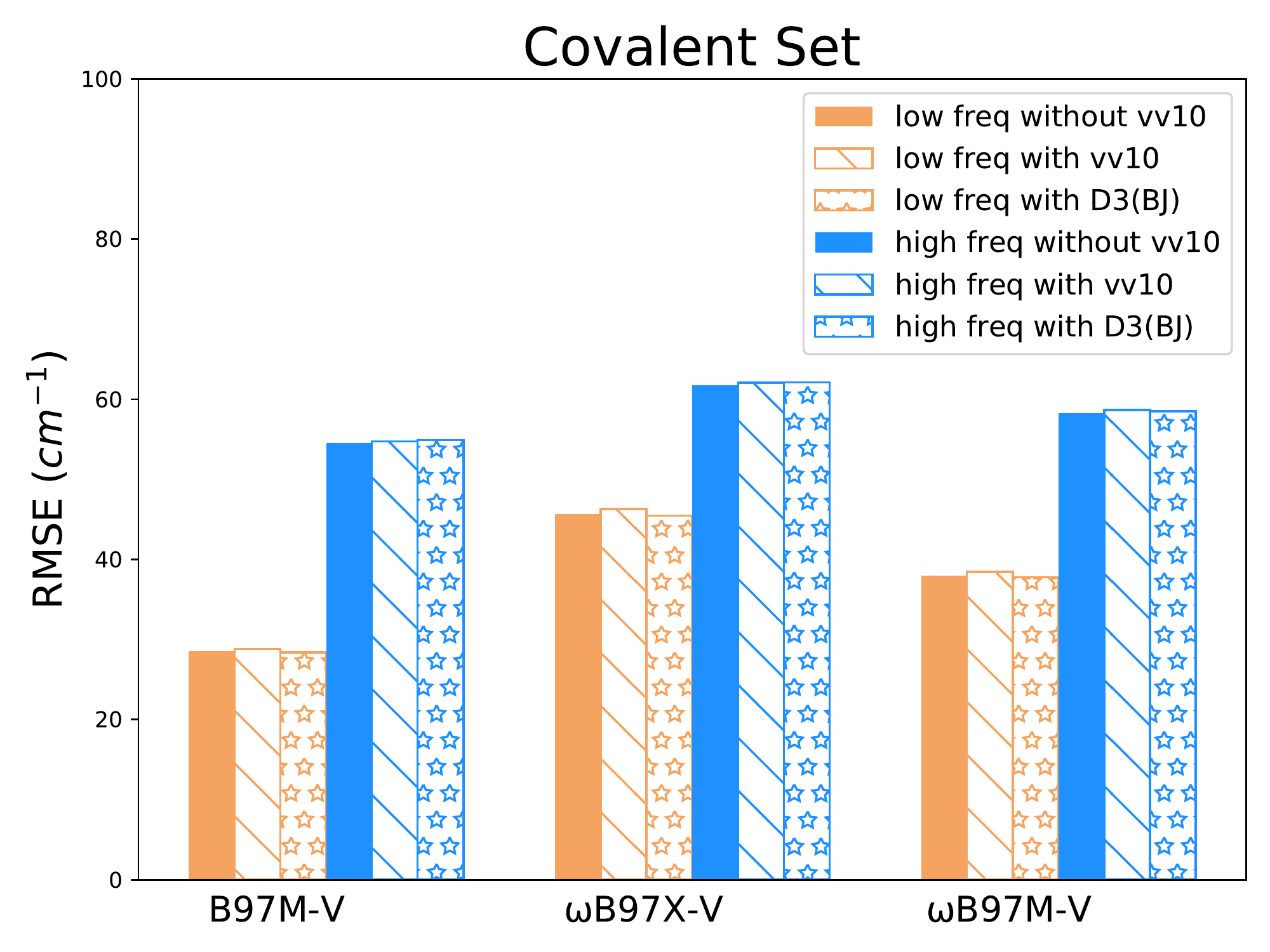}
     \end{subfigure}
     \begin{subfigure}[b]{0.45\textwidth}
         \centering
         \includegraphics[width=\textwidth]{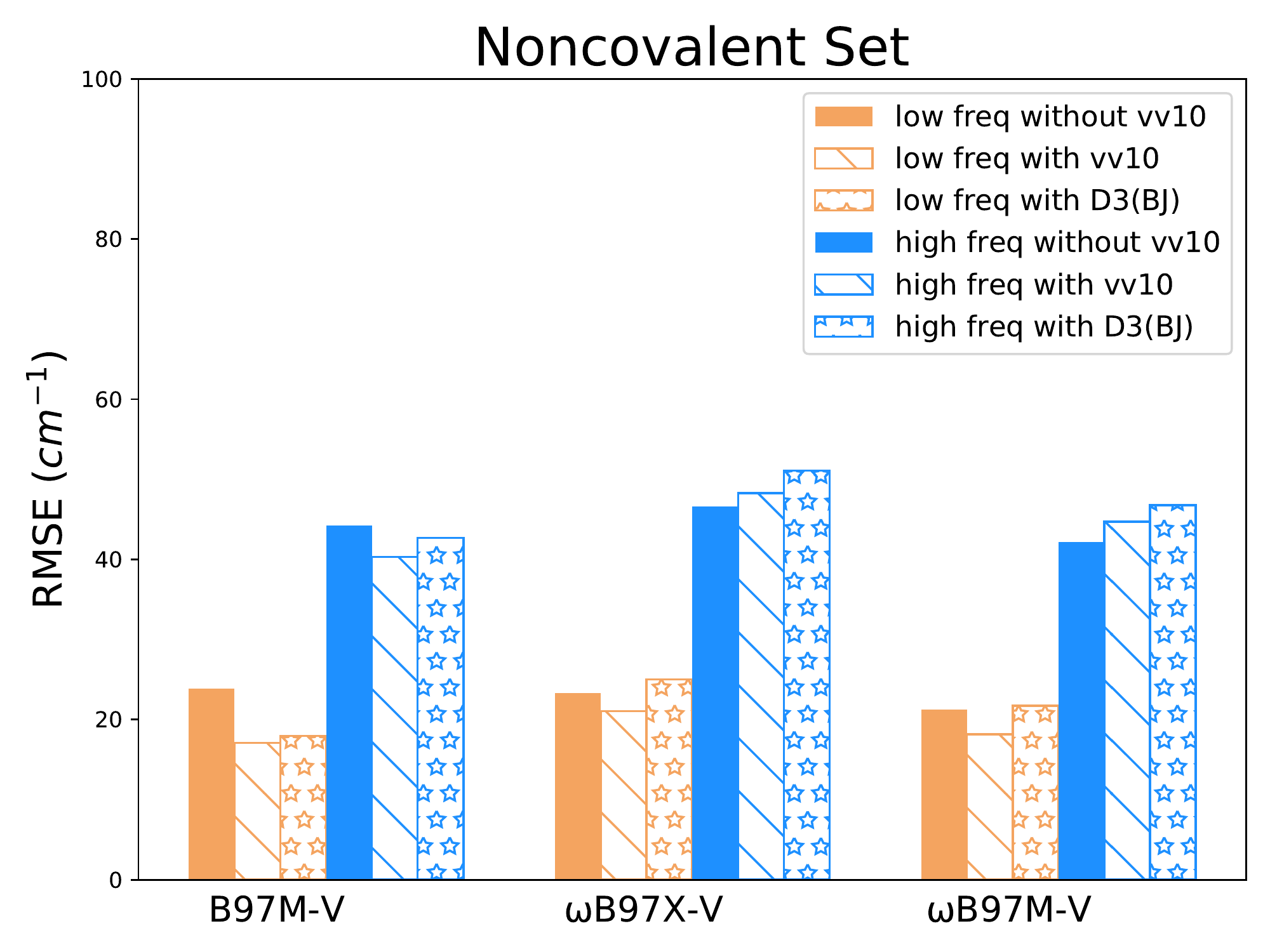}
     \end{subfigure}
        \caption{Comparison of the RMSEs (cm$^{-1}$) of low frequencies ($<$ 1000 cm$^{-1}$) and high frequencies ($\geq$ 1000 cm$^{-1}$) of B97M-V, $\omega$B97X-V, and $\omega$B97M-V with VV10, without VV10, and with D3(BJ) [replacing VV10 with D3(BJ)] relative to TBEs at the optimized geometries on the Covalent set (with def2-TZVP for the HFREQ2014 subset and aug-cc-pwCVTZ for the Diatomic subsets) and Noncovalent Set (with aug-cc-pVTZ for the P subset, the def2-QZVPP for the H2O and V30 sets, aug-cc-pVQZ for the H2S set, and the def2-QZVPPD for the N2O and CO2 sets) }
        \label{fig:vv10}
\end{figure}

Figure~\ref{fig:vv10} compares the RMSEs of B97M-V, $\omega$B97X-V, and $\omega$B97M-V with VV10, without VV10, and with D3(BJ) [namely replacing VV10 with D3(BJ)]. The inclusion of the dispersion correction [both D3(BJ) and VV10] has a minimal impact on the Covalent Set and a slightly bigger impact on the Noncovalent Set. Therefore, the following analysis is based on the Noncovalent Set and should be suitable for molecules with non-covalent interactions. For B97M-V, both D3(BJ) and VV10 can lower the RMSEs of low frequencies and high frequencies, implying the importance of dispersion correction on a semi-local functional such as B97M-V. For $\omega$B97X-V and $\omega$B97M-V, the high-frequency RMSEs are increased with both D3(BJ) and VV10, while the low-frequency RMSEs are reduced with VV10. If we compare the RMSEs on each subset (Figure~S1), we can see VV10 is able to improve the low-frequency predictions on all Noncovalent subsets, which is consistent with VV10's goal of better describing long-range correlation predictions, namely weak interactions. However, the consistent improvement is small and only exists at the optimized geometries in this work. 

\subsection{Comparison of functionals on all data sets} \label{subsec:compare}

\begin{figure}[hbt!]
    \centering
     \begin{subfigure}[b]{0.45\textwidth}
         \centering
         \includegraphics[width=\textwidth]{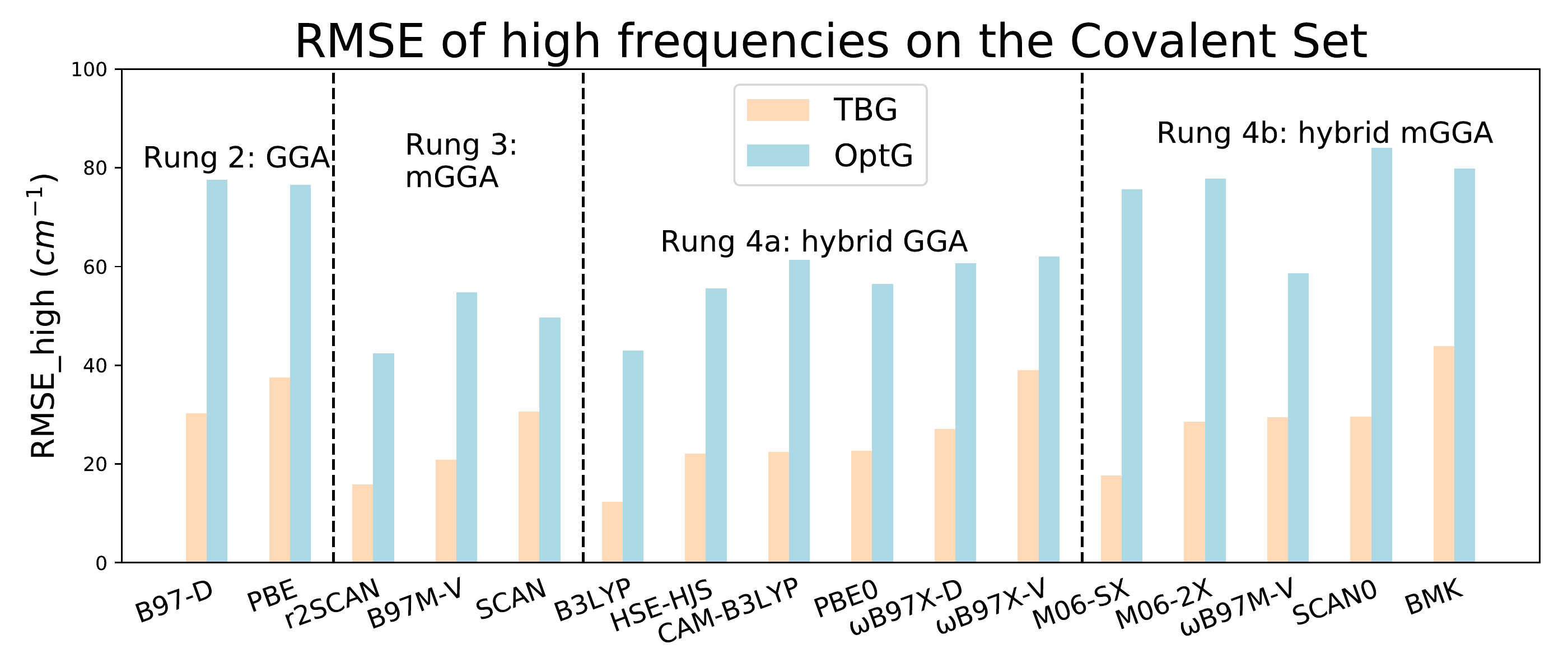}
     \end{subfigure}
    \centering
     \begin{subfigure}[b]{0.45\textwidth}
         \centering
         \includegraphics[width=\textwidth]{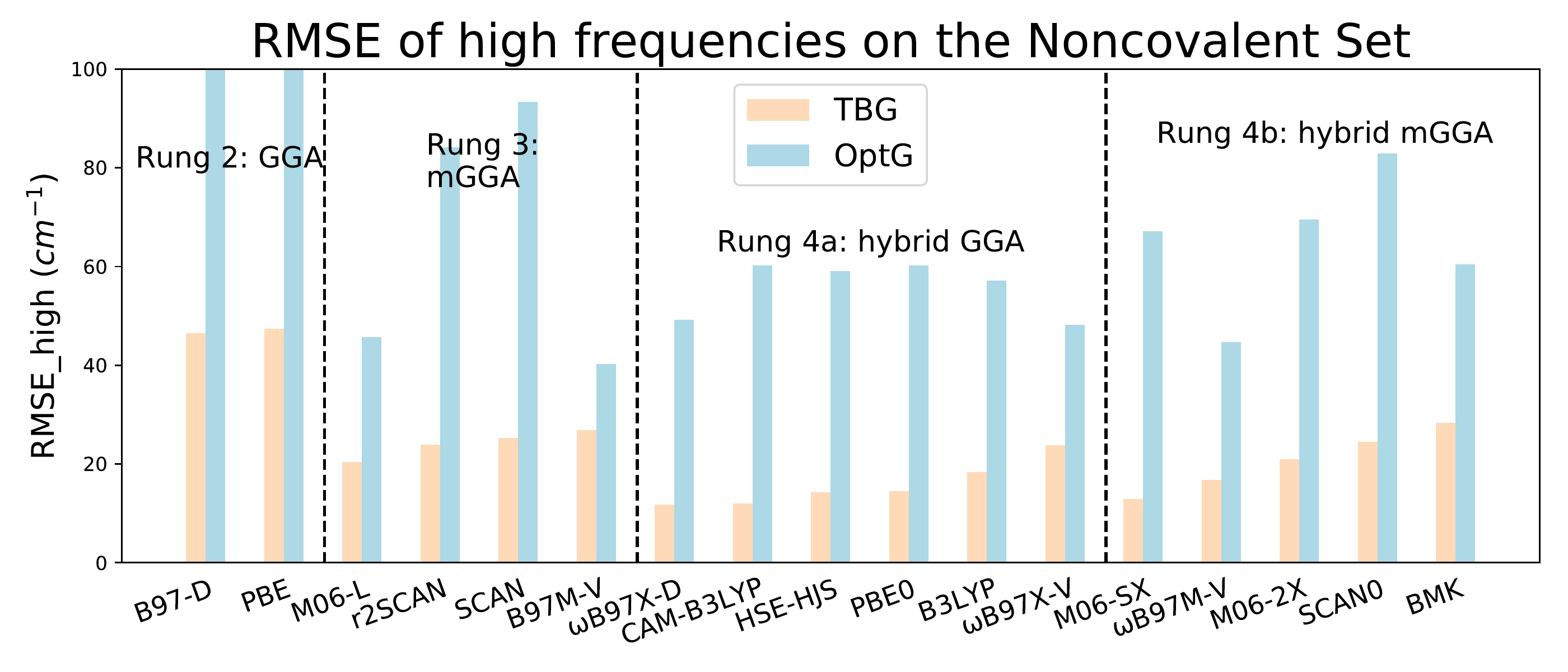}
     \end{subfigure}
    \centering
     \begin{subfigure}[b]{0.45\textwidth}
         \centering
         \includegraphics[width=\textwidth]{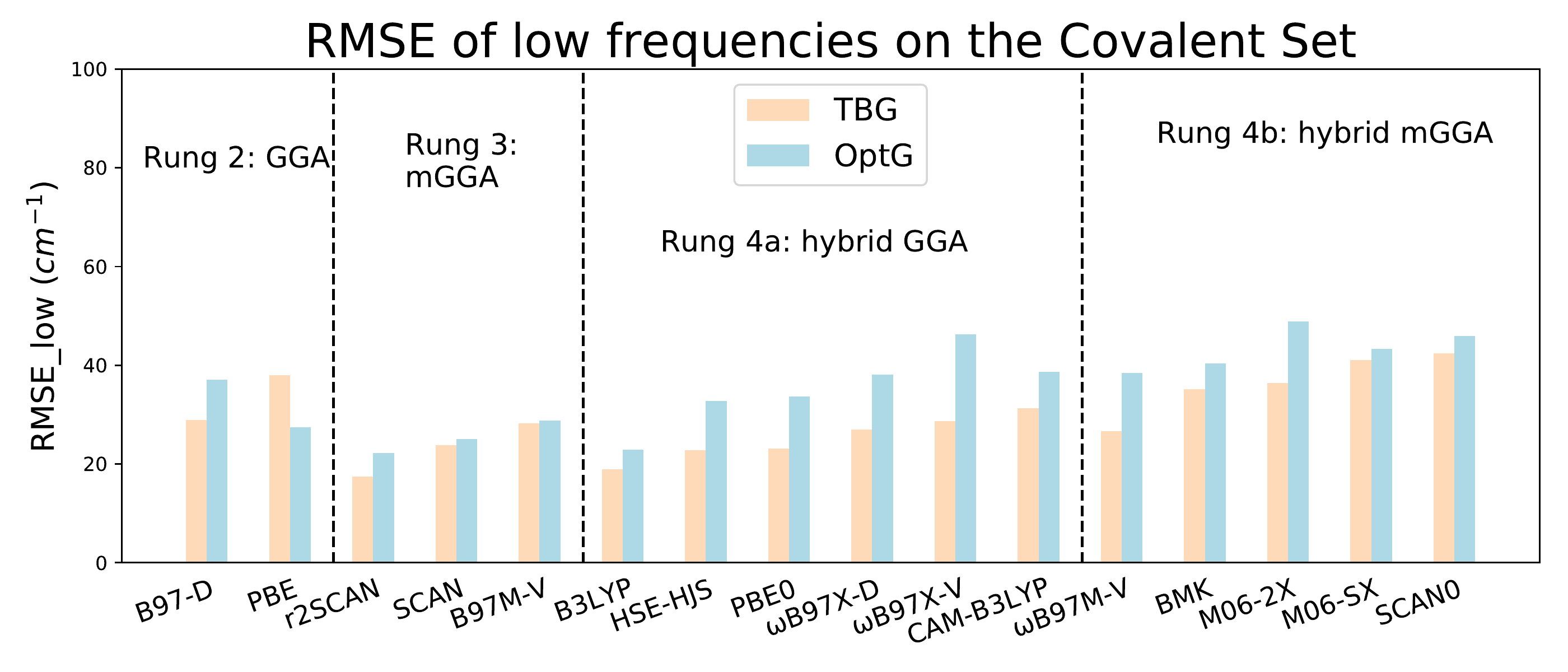}
     \end{subfigure}
    \centering
     \begin{subfigure}[b]{0.45\textwidth}
         \centering
         \includegraphics[width=\textwidth]{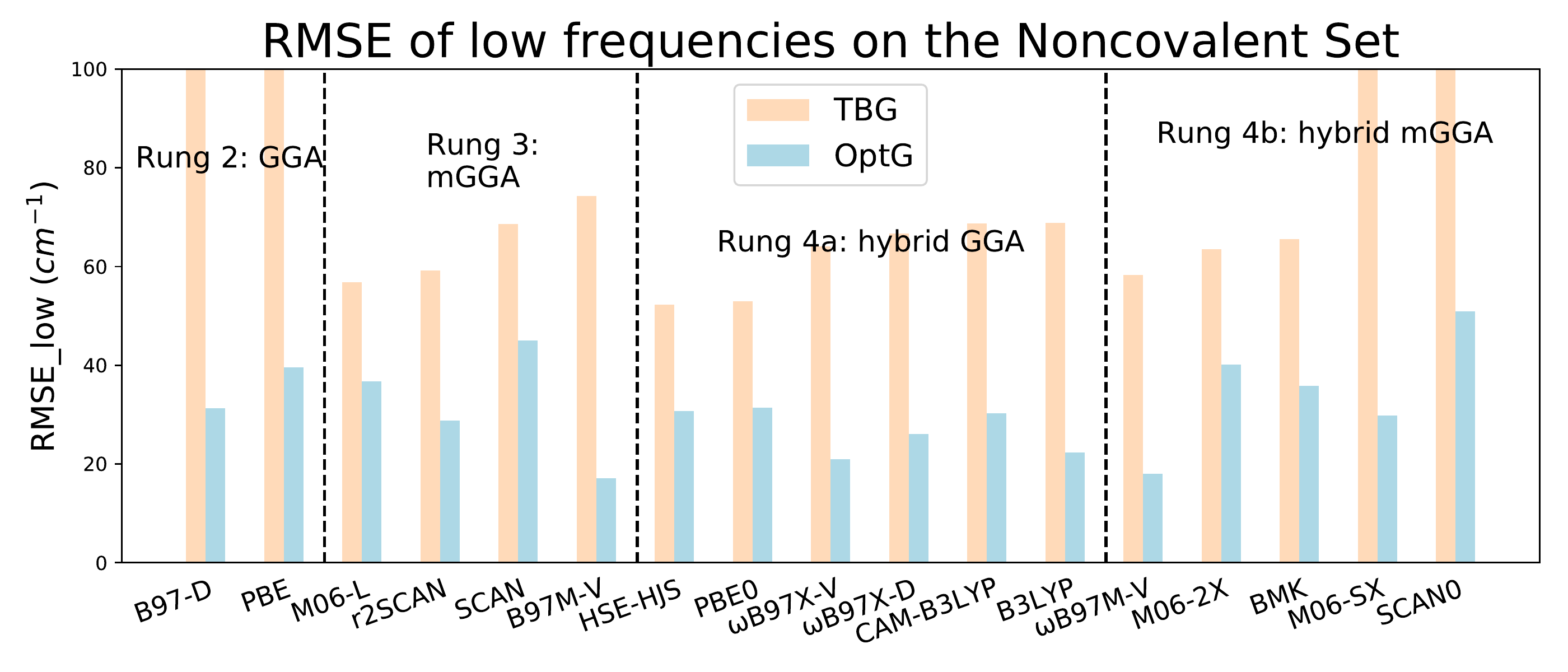}
     \end{subfigure}
    \caption{Comparison of the RMSEs (cm$^{-1}$) of low frequencies ($<$ 1000 cm$^{-1}$) and high frequencies ($\geq$ 1000 cm$^{-1}$) of different functionals relative to TBEs using TBGs and optimized geometries (OptGs) on the Covalent Set and Noncovalent Set (with the same basis sets as Figure~\ref{fig:vv10}). The results are organized left to right in order of increasing RMSE using TBGs on each rung of Jacob's Ladder. The D3(BJ) empirical correction is added to any functional without its own dispersion component.}
    \label{fig:func}
\end{figure}

Figure~\ref{fig:func} shows some recent or popular functionals' performance at theoretical best geometries (TBGs) and optimized geometries (OptGs). The SCF for M06-L cannot be converged for some radicals, and thus M06-L results for the Covalent Set are not presented. The D3(BJ) empirical correction is added to any functional without its own dispersion component.

The first striking outcome of Figure~\ref{fig:func} is that the RMSEs of all functionals at TBGs are significantly lower than those at OptGs for predicting high frequencies. We tend to always perform frequency calculations at the optimized geometry because this is essential to reliably characterize stationary points and avoid imaginary frequencies. One might also believe that there exists cancellation of method error and geometry error at the optimized (i.e., equilibrium) geometry of the same-level method and thus frequency prediction could be more accurate. This is consistent with our results for low frequencies, especially on the Noncovalent Set, where a lot of imaginary frequencies appear at TBGs. However, this is clearly not true for simulating high frequencies. One possible explanation is that the high frequencies imply that the potential energy surface (PES) curve is steep and sensitive to even very small structure changes. Therefore, a better geometry will improve frequency prediction. Given that geometry optimization consumes less computational resources than frequency calculations, it can therefore be useful to geometry optimize with a more accurate (and more expensive) method than that used for frequency evaluation when high frequencies ($\geq$ 1000 cm$^{-1}$) are of primary interest.

Turning to the performance of individual functionals, the conclusions are very different for the Covalent Set and Noncovalent Set. It is evident from Figure~\ref{fig:func} and Figure~S2 that B3LYP offers the overall smallest RMSEs at TBGs and nearly the smallest RMSEs at OptGs on the Covalent Set. While we do notice a significant improvement of all mGGAs relative to GGAs (B97-D and PBE), Jacob's Ladder\cite{perdew2005prescription} is partially violated since the best hybrid mGGAs ($\omega$B97M-V and M06-SX) are not better than the best hybrid GGA (B3LYP). In fact, r2SCAN, a semi-local mGGA, is the second-best functional at TBGs and can slightly outperform B3LYP at OptGs. It is possible that present functionals like B3LYP or r2SCAN might have reached the maximum capability of semi-local/hybrid DFT to predict the frequencies of common small molecules. It is also possible that the hybrid mGGAs can be further improved because of their enormous functional flexibility.\cite{mardirossian2017thirty}

The molecules included in the Covalent Set are too small for intramolecular dispersion to be significant, and therefore the effect of the VV10 component of the tested functionals is very small as discussed in the previous subsection. However, the situation changes for the intermolecular interactions associated with the Noncovalent Set. Firstly, the inclusion of many low-frequency vibrations causes the overall RMSE at OptG to be smaller than the RMSE at TBG. Secondly, we observe dramatic improvements in frequency prediction in the VV10-containing functionals using OptGs. Figure~\ref{fig:func} shows that B97M-V (overall RMSE of 29.4 cm$^{-1}$) dramatically outperforms r2SCAN (RMSE of 59.3 cm$^{-1}$) at the mGGA level and both $\omega$B97M-V (RMSE of 32.3 cm$^{-1}$) and $\omega$B97X-V (RMSE of 35.4 cm$^{-1}$) outperform B3LYP (RMSE of 41.0 cm$^{-1}$) significantly at the hybrid level. When we look at each subset separately in Figure~S3, we can observe that the main advantage of the VV10-containing functionals comes from the H2O set. Figure~S4 also shows that VV10-containing functionals describe dipole-dipole interactions better than pure dispersion. This is a surprising result that seems at odds with the design goal of using VV10 to better describe weak interactions, and could merit further studies, and/or support the idea of using (finite difference) frequency information in the development of future functionals. 

\subsection{Scaling Factors for New Developed Functionals} \label{subsec:SF}

The data presented in previous sections are all based on the comparison of calculated harmonic frequencies and corresponding TBEs. However, in practice, many researchers prefer to employ scaled  harmonic frequencies to approximate (anharmonic) experimental fundamental frequencies and to evaluate the zero-point vibrational energy (ZPVE). Therefore, we present the scaling factors and after-scaling RMSEs of some recent functionals (i.e., SCAN, r2SCAN, B97M-V, $\omega$B97X-V, $\omega$B97M-V, SCAN0, M06-SX, and HSE-HJS) with the def2-TZVP basis set in Table~\ref{tab:SF}. For a detailed error display, please see Table~S1.6. The data set used here is from ref~\citenum{kesharwani2015frequency} and most TBGs are from the HFREQ2014 set. Only CH\textsubscript{3}OH TBG is from CCCBDB, optimized by CCSD(T)/aug-cc-pVTZ, and unrestricted species TBGs are from the DiatomicU set.

\begin{table}[htb!]
\caption{The scaling factors ($\lambda$) of some recent functionals with the def2-TZVP basis set for predicting fundamental frequencies and zero-point vibrational energy (ZPVE), and their RMSEs (cm$^{-1}$ for frequency and kcal/mol for ZPVE) after scaling. The D3(BJ) empirical correction is added to any functional without its own dispersion component. Note that SCAN is tested at the same grid type (99,590) as other functionals rather than at its own converged grid type.}
\label{tab:SF}
\resizebox{\textwidth}{!}{
\begin{tabular}{llcrrrrrrrrr}
\hline
\textbf{Property}                                                                         & \multicolumn{2}{l}{\textbf{Result}}             & \textbf{SCAN} & \textbf{r2SCAN} & \textbf{B97M-V} & \textbf{$\omega$B97X-V} & \textbf{SCAN0} & \textbf{M06-SX} & \textbf{HSE-HJS} & \textbf{$\omega$B97M-V}  & \textbf{B3LYP} \\ \hline
\multirow{4}{*}{\textbf{\begin{tabular}[c]{@{}l@{}}fundamental\\ frequency\end{tabular}}} & \multirow{2}{*}{\textbf{TBG}} & \textbf{$\lambda$}    & 0.962   & 0.962    & 0.963           & 0.952            & 0.954          & 0.959           & 0.957            & 0.959     & 0.964        \\
&         & \textbf{RMSE} & 30.9      & 28.4    & 28.1            & 30.7             & 35.6           & 31.2            & 32.4             & 28.5  & 25.6            
\\ 
\cline{2-12} 
& \multirow{2}{*}{\textbf{OptG}}   & \textbf{$\lambda$}    & 0.963    & 0.967       & 0.957           & 0.954            & 0.941          & 0.949           & 0.958            & 0.958   & 0.967         \\
&              & \textbf{RMSE} & 27.6      & 21.5     & 27.8            & 25.4             & 25.3           & 28.3            & 25.5             & 24.5  & 19.4          \\ \hline
\multirow{4}{*}{\textbf{ZPVE}}                                                            & \multirow{2}{*}{\textbf{TBG}} & \textbf{$\lambda$}    & 0.988    & 0.989      & 0.984           & 0.977            & 0.973          & 0.979           & 0.984            & 0.984     & 0.989        \\
&                                 & \textbf{RMSE} & 0.105   &0.057      & 0.099           & 0.080            & 0.114          & 0.101           & 0.073            & 0.084 &0.056           \\ \cline{2-12} 
& \multirow{2}{*}{\textbf{OptG}}   & \textbf{$\lambda$}    & 0.986    & 0.991      & 0.980           & 0.978            & 0.965          & 0.972           & 0.983            & 0.983  & 0.990       \\
&                                 & \textbf{RMSE} & 0.123     & 0.085    & 0.133           & 0.128            & 0.148          & 0.148           & 0.117            & 0.115  & 0.092          \\ \hline
\end{tabular}}
\end{table}

As shown in Table~\ref{tab:SF}, RMSEs at TBGs are larger than those at OptGs for predicting fundamental frequencies but smaller for predicting ZPVEs. Although no modern functional can outperform B3LYP (RMSE of 19.44 cm$^{-1}$ at OptGs), scaling significantly lowers functional sensitivity for fundamentals (the biggest RMSE at OptGs is just 28.3 cm$^{-1}$ here). Similar findings hold true for ZPVEs, however, r2SCAN can provide RMSE that is comparable to B3LYP at TBGs and even smaller at OptGs. Compared with the result of ref~\citenum{kesharwani2015frequency}, the optimal scaling factors of most functionals for fundamental frequencies and ZPVE are still around 0.96 and 0.98 respectively.
Note that the data set only involves small molecules here so it is not clear how well these conclusions will transfer to larger systems with significant noncovalent interactions.

\section{Conclusions} \label{sec:conclusion}

In this work, we have formulated and efficiently implemented the analytical second derivatives of the nonlocal (NL) correlation functional, VV10. Tests of our OpenMP NL implementation establish that its parallel performance is quite good out to at least 32 cores. Furthermore, aside from the smallest basis sets, we have shown that the computational time for the NL part is negligible compared with that of the entire job.

As an application of the new analytical derivative code, we have also examined the performance of some recent or popular functionals for molecular vibrational frequencies. The assessment work builds on a number of previous efforts to establish benchmark data sets, which we have compiled into a Covalent Set and a Noncovalent Set. The Covalent Set contains small molecules without weak interactions while the Noncovalent Set can represent chemical systems containing weak intermolecular interactions. 
As shown in Table~\ref{tab:conclusion}, our conclusions are different for these two kinds of systems. While VV10-containing functionals provide little or no advantage in the accuracy of harmonic frequencies for small molecules, recent functionals such as B97M-V, $\omega$B97X-V, and $\omega$B97M-V provide very significant improvements for harmonic frequencies in polar molecular complexes. It is a fascinating question as to whether those same advantages will hold for vibrational frequencies in larger molecules. 
At the moment, a lack of suitable benchmarks for such molecules prevents an answer.

Finally, we also examined scaling factors for harmonic vibrational frequencies to permit direct comparison against (anharmonic) experimental fundamental vibrational frequencies and to evaluate the ZPVEs. B3LYP-D3(BJ) provides the smallest RMSE after scaling although the importance of functional choice is greatly reduced by scaling, at least for the small molecules which we have assessed.  

\begin{table}[ht!]
\caption{Broad conclusions and recommendations for simulating  molecular harmonic frequencies (and ZPVE) for small molecules and weakly interacting systems. Note that choice of functional is less important if frequencies are scaled to compare against experiments, at least for small molecules.}
\label{tab:conclusion}
\begin{tabular}{|l|l|l|}
\hline
Conclusions           & Small molecules & Weak-interaction systems \\ \hline
\begin{tabular}[c]{@{}l@{}}Recommended \\ basis set\end{tabular} &
  def2-TZVP &
  def2-TZVPPD (or larger) \\ \hline
\begin{tabular}[c]{@{}l@{}}Recommended \\ XC/NL grid \end{tabular} &
  SG-2/SG-0 &
  \begin{tabular}[c]{@{}l@{}}SG-2/SG-0 typically but (75, 302)/SG-1 \\ for sensitive functionals\end{tabular} \\ \hline
Effect of VV10        & minimal                 & helpful for low-frequencies                    \\ \hline
Recommended functional & B3LYP-D3(BJ), r2SCAN-D3(BJ)            &      B97M-V, $\omega$B97X-V, $\omega$B97M-V                     \\ \hline
\end{tabular}
\end{table}

\section*{Supporting Information}

Additional figures (SI.pdf)

S1-data\_analysis.xlsx

S2-raw\_data.xlsx

\section*{Author Declarations}
Martin Head-Gordon is a part-owner of Q-Chem, which is the software platform used to perform the developments and calculations described in this work. The other authors declare no competing financial interests.

\begin{acknowledgments}
This work was supported by the Director, Office of Science, Office of Basic Energy Sciences, of the U.S. Department of Energy through the Gas Phase Chemical Physics Program, under Contract No. DE-AC02-05CH11231. Additional support to X.F. and M.H.-G. was provided through NIH grant R44GM121126-02. This research used computational resources of the National Energy Research Scientific Computing Center, a DOE Office of Science User Facility supported by the Office of Science of the U.S. Department of Energy under Contract No. DE-AC02-05CH11231.
\end{acknowledgments}

\clearpage

\bibliography{aipsamp}

\begin{thebibliography}{99}%
\makeatletter
\providecommand \@ifxundefined [1]{%
 \@ifx{#1\undefined}
}%
\providecommand \@ifnum [1]{%
 \ifnum #1\expandafter \@firstoftwo
 \else \expandafter \@secondoftwo
 \fi
}%
\providecommand \@ifx [1]{%
 \ifx #1\expandafter \@firstoftwo
 \else \expandafter \@secondoftwo
 \fi
}%
\providecommand \natexlab [1]{#1}%
\providecommand \enquote  [1]{``#1''}%
\providecommand \bibnamefont  [1]{#1}%
\providecommand \bibfnamefont [1]{#1}%
\providecommand \citenamefont [1]{#1}%
\providecommand \href@noop [0]{\@secondoftwo}%
\providecommand \href [0]{\begingroup \@sanitize@url \@href}%
\providecommand \@href[1]{\@@startlink{#1}\@@href}%
\providecommand \@@href[1]{\endgroup#1\@@endlink}%
\providecommand \@sanitize@url [0]{\catcode `\\12\catcode `\$12\catcode
  `\&12\catcode `\#12\catcode `\^12\catcode `\_12\catcode `\%12\relax}%
\providecommand \@@startlink[1]{}%
\providecommand \@@endlink[0]{}%
\providecommand \url  [0]{\begingroup\@sanitize@url \@url }%
\providecommand \@url [1]{\endgroup\@href {#1}{\urlprefix }}%
\providecommand \urlprefix  [0]{URL }%
\providecommand \Eprint [0]{\href }%
\providecommand \doibase [0]{http://dx.doi.org/}%
\providecommand \selectlanguage [0]{\@gobble}%
\providecommand \bibinfo  [0]{\@secondoftwo}%
\providecommand \bibfield  [0]{\@secondoftwo}%
\providecommand \translation [1]{[#1]}%
\providecommand \BibitemOpen [0]{}%
\providecommand \bibitemStop [0]{}%
\providecommand \bibitemNoStop [0]{.\EOS\space}%
\providecommand \EOS [0]{\spacefactor3000\relax}%
\providecommand \BibitemShut  [1]{\csname bibitem#1\endcsname}%
\let\auto@bib@innerbib\@empty
\bibitem [{\citenamefont {Mardirossian}\ and\ \citenamefont
  {Head-Gordon}(2017)}]{mardirossian2017thirty}%
  \BibitemOpen
  \bibfield  {author} {\bibinfo {author} {\bibfnamefont {N.}~\bibnamefont
  {Mardirossian}}\ and\ \bibinfo {author} {\bibfnamefont {M.}~\bibnamefont
  {Head-Gordon}},\ }\bibfield  {title} {\enquote {\bibinfo {title} {Thirty
  years of density functional theory in computational chemistry: {An} overview
  and extensive assessment of 200 density functionals},}\ }\href {\doibase
  10.1080/00268976.2017.1333644} {\bibfield  {journal} {\bibinfo  {journal}
  {Mol. Phys.}\ }\textbf {\bibinfo {volume} {115}},\ \bibinfo {pages}
  {2315--2372} (\bibinfo {year} {2017})}\BibitemShut {NoStop}%
\bibitem [{\citenamefont {Palafox}(2018)}]{palafox2018dft}%
  \BibitemOpen
  \bibfield  {author} {\bibinfo {author} {\bibfnamefont {M.~A.}\ \bibnamefont
  {Palafox}},\ }\bibfield  {title} {\enquote {\bibinfo {title} {Dft
  computations on vibrational spectra: Scaling procedures to improve the
  wavenumbers},}\ }\href {\doibase doi:10.1515/psr-2017-0184} {\bibfield
  {journal} {\bibinfo  {journal} {Physical Sciences Reviews}\ }\textbf
  {\bibinfo {volume} {3}},\ \bibinfo {pages} {20170184} (\bibinfo {year}
  {2018})}\BibitemShut {NoStop}%
\bibitem [{\citenamefont {Zapata~Trujillo}\ and\ \citenamefont
  {McKemmish}(2021)}]{zapata2021meta}%
  \BibitemOpen
  \bibfield  {author} {\bibinfo {author} {\bibfnamefont {J.~C.}\ \bibnamefont
  {Zapata~Trujillo}}\ and\ \bibinfo {author} {\bibfnamefont {L.~K.}\
  \bibnamefont {McKemmish}},\ }\bibfield  {title} {\enquote {\bibinfo {title}
  {Meta-analysis of uniform scaling factors for harmonic frequency
  calculations},}\ }\href {\doibase 10.1002/wcms.1584} {\bibfield  {journal}
  {\bibinfo  {journal} {WIREs Comput. Mol. Sci.}\ }\textbf {\bibinfo {volume}
  {12}},\ \bibinfo {pages} {e1584} (\bibinfo {year} {2021})}\BibitemShut
  {NoStop}%
\bibitem [{\citenamefont {Liang}\ \emph {et~al.}(2022)\citenamefont {Liang},
  \citenamefont {Feng}, \citenamefont {Hait},\ and\ \citenamefont
  {Head-Gordon}}]{liang2022revisiting}%
  \BibitemOpen
  \bibfield  {author} {\bibinfo {author} {\bibfnamefont {J.}~\bibnamefont
  {Liang}}, \bibinfo {author} {\bibfnamefont {X.}~\bibnamefont {Feng}},
  \bibinfo {author} {\bibfnamefont {D.}~\bibnamefont {Hait}}, \ and\ \bibinfo
  {author} {\bibfnamefont {M.}~\bibnamefont {Head-Gordon}},\ }\bibfield
  {title} {\enquote {\bibinfo {title} {Revisiting the performance of
  time-dependent density functional theory for electronic excitations:
  {Assessment} of 43 popular and recently developed functionals from rungs one
  to four},}\ }\href {\doibase 10.1021/acs.jctc.2c00160} {\bibfield  {journal}
  {\bibinfo  {journal} {J. Chem. Theory Comput.}\ } (\bibinfo {year} {2022}),\
  10.1021/acs.jctc.2c00160}\BibitemShut {NoStop}%
\bibitem [{\citenamefont {Kristy\'an}\ and\ \citenamefont
  {Pulay}(1994)}]{kristyan1994can}%
  \BibitemOpen
  \bibfield  {author} {\bibinfo {author} {\bibfnamefont {S.}~\bibnamefont
  {Kristy\'an}}\ and\ \bibinfo {author} {\bibfnamefont {P.}~\bibnamefont
  {Pulay}},\ }\bibfield  {title} {\enquote {\bibinfo {title} {Can (semi)local
  density functional theory account for the {London} dispersion forces?}}\
  }\href {\doibase 10.1016/0009-2614(94)01027-7} {\bibfield  {journal}
  {\bibinfo  {journal} {Chem. Phys. Lett.}\ }\textbf {\bibinfo {volume}
  {229}},\ \bibinfo {pages} {175--180} (\bibinfo {year} {1994})}\BibitemShut
  {NoStop}%
\bibitem [{\citenamefont {P\'erez-Jord\'a}\ and\ \citenamefont
  {Becke}(1995)}]{perez1995density}%
  \BibitemOpen
  \bibfield  {author} {\bibinfo {author} {\bibfnamefont {J.}~\bibnamefont
  {P\'erez-Jord\'a}}\ and\ \bibinfo {author} {\bibfnamefont {A.}~\bibnamefont
  {Becke}},\ }\bibfield  {title} {\enquote {\bibinfo {title} {A
  density-functional study of van der {Waals} forces: {Rare} gas diatomics},}\
  }\href {\doibase 10.1016/0009-2614(94)01402-h} {\bibfield  {journal}
  {\bibinfo  {journal} {Chem. Phys. Lett.}\ }\textbf {\bibinfo {volume}
  {233}},\ \bibinfo {pages} {134--137} (\bibinfo {year} {1995})}\BibitemShut
  {NoStop}%
\bibitem [{\citenamefont {Hobza}, \citenamefont {{\v{s}}poner},\ and\
  \citenamefont {Reschel}(1995)}]{hobza1995density}%
  \BibitemOpen
  \bibfield  {author} {\bibinfo {author} {\bibfnamefont {P.}~\bibnamefont
  {Hobza}}, \bibinfo {author} {\bibfnamefont {J.}~\bibnamefont {{\v{s}}poner}},
  \ and\ \bibinfo {author} {\bibfnamefont {T.}~\bibnamefont {Reschel}},\
  }\bibfield  {title} {\enquote {\bibinfo {title} {Density functional theory
  and molecular clusters},}\ }\href {\doibase 10.1002/jcc.540161102} {\bibfield
   {journal} {\bibinfo  {journal} {J. Comput. Chem.}\ }\textbf {\bibinfo
  {volume} {16}},\ \bibinfo {pages} {1315--1325} (\bibinfo {year}
  {1995})}\BibitemShut {NoStop}%
\bibitem [{\citenamefont {Najibi}\ and\ \citenamefont
  {Goerigk}(2018)}]{najibi2018nonlocal}%
  \BibitemOpen
  \bibfield  {author} {\bibinfo {author} {\bibfnamefont {A.}~\bibnamefont
  {Najibi}}\ and\ \bibinfo {author} {\bibfnamefont {L.}~\bibnamefont
  {Goerigk}},\ }\bibfield  {title} {\enquote {\bibinfo {title} {The nonlocal
  kernel in van der {Waals} density functionals as an additive correction: {An}
  extensive analysis with special emphasis on the b97m-v and
  \ensuremath{\omega}b97m-v approaches},}\ }\href {\doibase
  10.1021/acs.jctc.8b00842} {\bibfield  {journal} {\bibinfo  {journal} {J.
  Chem. Theory Comput.}\ }\textbf {\bibinfo {volume} {14}},\ \bibinfo {pages}
  {5725--5738} (\bibinfo {year} {2018})}\BibitemShut {NoStop}%
\bibitem [{\citenamefont {Klime{\v{s}}}\ and\ \citenamefont
  {Michaelides}(2012)}]{klimevs2012perspective}%
  \BibitemOpen
  \bibfield  {author} {\bibinfo {author} {\bibfnamefont {J.}~\bibnamefont
  {Klime{\v{s}}}}\ and\ \bibinfo {author} {\bibfnamefont {A.}~\bibnamefont
  {Michaelides}},\ }\bibfield  {title} {\enquote {\bibinfo {title}
  {Perspective: Advances and challenges in treating van der waals dispersion
  forces in density functional theory},}\ }\href@noop {} {\bibfield  {journal}
  {\bibinfo  {journal} {J. Chem. Phys.}\ }\textbf {\bibinfo {volume} {137}},\
  \bibinfo {pages} {120901} (\bibinfo {year} {2012})}\BibitemShut {NoStop}%
\bibitem [{\citenamefont {St{\"o}hr}, \citenamefont {Van~Voorhis},\ and\
  \citenamefont {Tkatchenko}(2019)}]{stohr2019theory}%
  \BibitemOpen
  \bibfield  {author} {\bibinfo {author} {\bibfnamefont {M.}~\bibnamefont
  {St{\"o}hr}}, \bibinfo {author} {\bibfnamefont {T.}~\bibnamefont
  {Van~Voorhis}}, \ and\ \bibinfo {author} {\bibfnamefont {A.}~\bibnamefont
  {Tkatchenko}},\ }\bibfield  {title} {\enquote {\bibinfo {title} {Theory and
  practice of modeling van der waals interactions in electronic-structure
  calculations},}\ }\href@noop {} {\bibfield  {journal} {\bibinfo  {journal}
  {Chem. Soc. Rev.}\ }\textbf {\bibinfo {volume} {48}},\ \bibinfo {pages}
  {4118--4154} (\bibinfo {year} {2019})}\BibitemShut {NoStop}%
\bibitem [{\citenamefont {Becke}\ and\ \citenamefont
  {Johnson}(2007)}]{becke2007exchange}%
  \BibitemOpen
  \bibfield  {author} {\bibinfo {author} {\bibfnamefont {A.~D.}\ \bibnamefont
  {Becke}}\ and\ \bibinfo {author} {\bibfnamefont {E.~R.}\ \bibnamefont
  {Johnson}},\ }\bibfield  {title} {\enquote {\bibinfo {title} {Exchange-hole
  dipole moment and the dispersion interaction revisited},}\ }\href@noop {}
  {\bibfield  {journal} {\bibinfo  {journal} {J. Chem. Phys.}\ }\textbf
  {\bibinfo {volume} {127}},\ \bibinfo {pages} {154108} (\bibinfo {year}
  {2007})}\BibitemShut {NoStop}%
\bibitem [{\citenamefont {Tkatchenko}\ and\ \citenamefont
  {Scheffler}(2009)}]{tkatchenko2009accurate}%
  \BibitemOpen
  \bibfield  {author} {\bibinfo {author} {\bibfnamefont {A.}~\bibnamefont
  {Tkatchenko}}\ and\ \bibinfo {author} {\bibfnamefont {M.}~\bibnamefont
  {Scheffler}},\ }\bibfield  {title} {\enquote {\bibinfo {title} {Accurate
  molecular van der waals interactions from ground-state electron density and
  free-atom reference data},}\ }\href@noop {} {\bibfield  {journal} {\bibinfo
  {journal} {Phys. Rev. Lett.}\ }\textbf {\bibinfo {volume} {102}},\ \bibinfo
  {pages} {073005} (\bibinfo {year} {2009})}\BibitemShut {NoStop}%
\bibitem [{\citenamefont {Grimme}(2006)}]{grimme2006semiempirical}%
  \BibitemOpen
  \bibfield  {author} {\bibinfo {author} {\bibfnamefont {S.}~\bibnamefont
  {Grimme}},\ }\bibfield  {title} {\enquote {\bibinfo {title} {Semiempirical
  gga-type density functional constructed with a long-range dispersion
  correction},}\ }\href {\doibase 10.1002/jcc.20495} {\bibfield  {journal}
  {\bibinfo  {journal} {J. Comput. Chem.}\ }\textbf {\bibinfo {volume} {27}},\
  \bibinfo {pages} {1787--1799} (\bibinfo {year} {2006})}\BibitemShut {NoStop}%
\bibitem [{\citenamefont {Grimme}\ \emph {et~al.}(2010)\citenamefont {Grimme},
  \citenamefont {Antony}, \citenamefont {Ehrlich},\ and\ \citenamefont
  {Krieg}}]{grimme2010consistent}%
  \BibitemOpen
  \bibfield  {author} {\bibinfo {author} {\bibfnamefont {S.}~\bibnamefont
  {Grimme}}, \bibinfo {author} {\bibfnamefont {J.}~\bibnamefont {Antony}},
  \bibinfo {author} {\bibfnamefont {S.}~\bibnamefont {Ehrlich}}, \ and\
  \bibinfo {author} {\bibfnamefont {H.}~\bibnamefont {Krieg}},\ }\bibfield
  {title} {\enquote {\bibinfo {title} {A consistent and accurate ab initio
  parametrization of density functional dispersion correction (dft-d) for the
  94 elements h-p},}\ }\href {\doibase 10.1063/1.3382344} {\bibfield  {journal}
  {\bibinfo  {journal} {J. Chem. Phys.}\ }\textbf {\bibinfo {volume} {132}},\
  \bibinfo {pages} {154104} (\bibinfo {year} {2010})}\BibitemShut {NoStop}%
\bibitem [{\citenamefont {Grimme}, \citenamefont {Ehrlich},\ and\ \citenamefont
  {Goerigk}(2011)}]{grimme2011effect}%
  \BibitemOpen
  \bibfield  {author} {\bibinfo {author} {\bibfnamefont {S.}~\bibnamefont
  {Grimme}}, \bibinfo {author} {\bibfnamefont {S.}~\bibnamefont {Ehrlich}}, \
  and\ \bibinfo {author} {\bibfnamefont {L.}~\bibnamefont {Goerigk}},\
  }\bibfield  {title} {\enquote {\bibinfo {title} {Effect of the damping
  function in dispersion corrected density functional theory},}\ }\href
  {\doibase 10.1002/jcc.21759} {\bibfield  {journal} {\bibinfo  {journal} {J.
  Comput. Chem.}\ }\textbf {\bibinfo {volume} {32}},\ \bibinfo {pages}
  {1456--1465} (\bibinfo {year} {2011})}\BibitemShut {NoStop}%
\bibitem [{\citenamefont {Grimme}\ \emph {et~al.}(2016)\citenamefont {Grimme},
  \citenamefont {Hansen}, \citenamefont {Brandenburg},\ and\ \citenamefont
  {Bannwarth}}]{grimme2016dispersion}%
  \BibitemOpen
  \bibfield  {author} {\bibinfo {author} {\bibfnamefont {S.}~\bibnamefont
  {Grimme}}, \bibinfo {author} {\bibfnamefont {A.}~\bibnamefont {Hansen}},
  \bibinfo {author} {\bibfnamefont {J.~G.}\ \bibnamefont {Brandenburg}}, \ and\
  \bibinfo {author} {\bibfnamefont {C.}~\bibnamefont {Bannwarth}},\ }\bibfield
  {title} {\enquote {\bibinfo {title} {Dispersion-corrected mean-field
  electronic structure methods},}\ }\href {\doibase
  10.1021/acs.chemrev.5b00533} {\bibfield  {journal} {\bibinfo  {journal}
  {Chem. Rev.}\ }\textbf {\bibinfo {volume} {116}},\ \bibinfo {pages}
  {5105--5154} (\bibinfo {year} {2016})}\BibitemShut {NoStop}%
\bibitem [{\citenamefont {Nguyen}\ and\ \citenamefont
  {Galli}(2010)}]{nguyen2010first}%
  \BibitemOpen
  \bibfield  {author} {\bibinfo {author} {\bibfnamefont {H.-V.}\ \bibnamefont
  {Nguyen}}\ and\ \bibinfo {author} {\bibfnamefont {G.}~\bibnamefont {Galli}},\
  }\bibfield  {title} {\enquote {\bibinfo {title} {A first-principles study of
  weakly bound molecules using exact exchange and the random phase
  approximation},}\ }\href {\doibase 10.1063/1.3299247} {\bibfield  {journal}
  {\bibinfo  {journal} {J. Chem. Phys.}\ }\textbf {\bibinfo {volume} {132}},\
  \bibinfo {pages} {044109} (\bibinfo {year} {2010})}\BibitemShut {NoStop}%
\bibitem [{\citenamefont {Zhu}\ \emph {et~al.}(2010)\citenamefont {Zhu},
  \citenamefont {Toulouse}, \citenamefont {Savin},\ and\ \citenamefont
  {\'Angy\'an}}]{zhu2010range}%
  \BibitemOpen
  \bibfield  {author} {\bibinfo {author} {\bibfnamefont {W.}~\bibnamefont
  {Zhu}}, \bibinfo {author} {\bibfnamefont {J.}~\bibnamefont {Toulouse}},
  \bibinfo {author} {\bibfnamefont {A.}~\bibnamefont {Savin}}, \ and\ \bibinfo
  {author} {\bibfnamefont {J.~G.}\ \bibnamefont {\'Angy\'an}},\ }\bibfield
  {title} {\enquote {\bibinfo {title} {Range-separated density-functional
  theory with random phase approximation applied to noncovalent intermolecular
  interactions},}\ }\href {\doibase 10.1063/1.3431616} {\bibfield  {journal}
  {\bibinfo  {journal} {J. Chem. Phys.}\ }\textbf {\bibinfo {volume} {132}},\
  \bibinfo {pages} {244108} (\bibinfo {year} {2010})}\BibitemShut {NoStop}%
\bibitem [{\citenamefont {Chen}\ \emph {et~al.}(2017)\citenamefont {Chen},
  \citenamefont {Voora}, \citenamefont {Agee}, \citenamefont {Balasubramani},\
  and\ \citenamefont {Furche}}]{chen2017random}%
  \BibitemOpen
  \bibfield  {author} {\bibinfo {author} {\bibfnamefont {G.~P.}\ \bibnamefont
  {Chen}}, \bibinfo {author} {\bibfnamefont {V.~K.}\ \bibnamefont {Voora}},
  \bibinfo {author} {\bibfnamefont {M.~M.}\ \bibnamefont {Agee}}, \bibinfo
  {author} {\bibfnamefont {S.~G.}\ \bibnamefont {Balasubramani}}, \ and\
  \bibinfo {author} {\bibfnamefont {F.}~\bibnamefont {Furche}},\ }\bibfield
  {title} {\enquote {\bibinfo {title} {Random-phase approximation methods},}\
  }\href@noop {} {\bibfield  {journal} {\bibinfo  {journal} {Ann. Rev. Phys.
  Chem.}\ }\textbf {\bibinfo {volume} {68}},\ \bibinfo {pages} {421--445}
  (\bibinfo {year} {2017})}\BibitemShut {NoStop}%
\bibitem [{\citenamefont {Goerigk}\ and\ \citenamefont
  {Grimme}(2014)}]{goerigk2014double}%
  \BibitemOpen
  \bibfield  {author} {\bibinfo {author} {\bibfnamefont {L.}~\bibnamefont
  {Goerigk}}\ and\ \bibinfo {author} {\bibfnamefont {S.}~\bibnamefont
  {Grimme}},\ }\bibfield  {title} {\enquote {\bibinfo {title} {Double-hybrid
  density functionals},}\ }\href@noop {} {\bibfield  {journal} {\bibinfo
  {journal} {WIRes: Comput. Mol. Sci.}\ }\textbf {\bibinfo {volume} {4}},\
  \bibinfo {pages} {576--600} (\bibinfo {year} {2014})}\BibitemShut {NoStop}%
\bibitem [{\citenamefont {Martin}\ and\ \citenamefont
  {Santra}(2020)}]{martin2020empirical}%
  \BibitemOpen
  \bibfield  {author} {\bibinfo {author} {\bibfnamefont {J.~M.}\ \bibnamefont
  {Martin}}\ and\ \bibinfo {author} {\bibfnamefont {G.}~\bibnamefont
  {Santra}},\ }\bibfield  {title} {\enquote {\bibinfo {title} {Empirical
  double-hybrid density functional theory: A ‘third way’in between wft and
  dft},}\ }\href@noop {} {\bibfield  {journal} {\bibinfo  {journal} {Isr. J.
  Chem.}\ }\textbf {\bibinfo {volume} {60}},\ \bibinfo {pages} {787--804}
  (\bibinfo {year} {2020})}\BibitemShut {NoStop}%
\bibitem [{\citenamefont {Dion}\ \emph {et~al.}(2004)\citenamefont {Dion},
  \citenamefont {Rydberg}, \citenamefont {Schr\"oder}, \citenamefont
  {Langreth},\ and\ \citenamefont {Lundqvist}}]{dion2004van}%
  \BibitemOpen
  \bibfield  {author} {\bibinfo {author} {\bibfnamefont {M.}~\bibnamefont
  {Dion}}, \bibinfo {author} {\bibfnamefont {H.}~\bibnamefont {Rydberg}},
  \bibinfo {author} {\bibfnamefont {E.}~\bibnamefont {Schr\"oder}}, \bibinfo
  {author} {\bibfnamefont {D.~C.}\ \bibnamefont {Langreth}}, \ and\ \bibinfo
  {author} {\bibfnamefont {B.~I.}\ \bibnamefont {Lundqvist}},\ }\bibfield
  {title} {\enquote {\bibinfo {title} {Van der {Waals} density functional for
  general geometries},}\ }\href {\doibase 10.1103/physrevlett.92.246401}
  {\bibfield  {journal} {\bibinfo  {journal} {Phys. Rev. Lett.}\ }\textbf
  {\bibinfo {volume} {92}},\ \bibinfo {pages} {246401} (\bibinfo {year}
  {2004})}\BibitemShut {NoStop}%
\bibitem [{\citenamefont {Lee}\ \emph {et~al.}(2010)\citenamefont {Lee},
  \citenamefont {Murray}, \citenamefont {Kong}, \citenamefont {Lundqvist},\
  and\ \citenamefont {Langreth}}]{lee2010higher}%
  \BibitemOpen
  \bibfield  {author} {\bibinfo {author} {\bibfnamefont {K.}~\bibnamefont
  {Lee}}, \bibinfo {author} {\bibfnamefont {E.~D.}\ \bibnamefont {Murray}},
  \bibinfo {author} {\bibfnamefont {L.}~\bibnamefont {Kong}}, \bibinfo {author}
  {\bibfnamefont {B.~I.}\ \bibnamefont {Lundqvist}}, \ and\ \bibinfo {author}
  {\bibfnamefont {D.~C.}\ \bibnamefont {Langreth}},\ }\bibfield  {title}
  {\enquote {\bibinfo {title} {Higher-accuracy van der {Waals} density
  functional},}\ }\href {\doibase 10.1103/physrevb.82.081101} {\bibfield
  {journal} {\bibinfo  {journal} {Phys. Rev. B}\ }\textbf {\bibinfo {volume}
  {82}},\ \bibinfo {pages} {081101} (\bibinfo {year} {2010})}\BibitemShut
  {NoStop}%
\bibitem [{\citenamefont {Vydrov}\ and\ \citenamefont
  {Van~Voorhis}(2009)}]{vydrov2009nonlocal}%
  \BibitemOpen
  \bibfield  {author} {\bibinfo {author} {\bibfnamefont {O.~A.}\ \bibnamefont
  {Vydrov}}\ and\ \bibinfo {author} {\bibfnamefont {T.}~\bibnamefont
  {Van~Voorhis}},\ }\bibfield  {title} {\enquote {\bibinfo {title} {Nonlocal
  van der {Waals} density functional made simple},}\ }\href {\doibase
  10.1103/physrevlett.103.063004} {\bibfield  {journal} {\bibinfo  {journal}
  {Phys. Rev. Lett.}\ }\textbf {\bibinfo {volume} {103}},\ \bibinfo {pages}
  {063004} (\bibinfo {year} {2009})}\BibitemShut {NoStop}%
\bibitem [{\citenamefont {Vydrov}\ and\ \citenamefont
  {Van~Voorhis}(2010)}]{vydrov2010nonlocal}%
  \BibitemOpen
  \bibfield  {author} {\bibinfo {author} {\bibfnamefont {O.~A.}\ \bibnamefont
  {Vydrov}}\ and\ \bibinfo {author} {\bibfnamefont {T.}~\bibnamefont
  {Van~Voorhis}},\ }\bibfield  {title} {\enquote {\bibinfo {title} {Nonlocal
  van der {Waals} density functional: {The} simpler the better},}\ }\href
  {\doibase 10.1063/1.3521275} {\bibfield  {journal} {\bibinfo  {journal} {J.
  Chem. Phys.}\ }\textbf {\bibinfo {volume} {133}},\ \bibinfo {pages} {244103}
  (\bibinfo {year} {2010})}\BibitemShut {NoStop}%
\bibitem [{\citenamefont {Calbo}\ \emph {et~al.}(2015)\citenamefont {Calbo},
  \citenamefont {Ort{\'\i}}, \citenamefont {Sancho-Garc{\'\i}a},\ and\
  \citenamefont {Arag{\'o}}}]{calbo2015nonlocal}%
  \BibitemOpen
  \bibfield  {author} {\bibinfo {author} {\bibfnamefont {J.}~\bibnamefont
  {Calbo}}, \bibinfo {author} {\bibfnamefont {E.}~\bibnamefont {Ort{\'\i}}},
  \bibinfo {author} {\bibfnamefont {J.~C.}\ \bibnamefont {Sancho-Garc{\'\i}a}},
  \ and\ \bibinfo {author} {\bibfnamefont {J.}~\bibnamefont {Arag{\'o}}},\
  }\bibfield  {title} {\enquote {\bibinfo {title} {The nonlocal correlation
  density functional vv10: a successful attempt to accurately capture
  noncovalent interactions},}\ }in\ \href@noop {} {\emph {\bibinfo {booktitle}
  {Ann. Rep. Comput. Chem.}}},\ Vol.~\bibinfo {volume} {11}\ (\bibinfo
  {publisher} {Elsevier},\ \bibinfo {year} {2015})\ pp.\ \bibinfo {pages}
  {37--102}\BibitemShut {NoStop}%
\bibitem [{\citenamefont {Mardirossian}\ and\ \citenamefont
  {Head-Gordon}(2014)}]{Mardirossian:2014}%
  \BibitemOpen
  \bibfield  {author} {\bibinfo {author} {\bibfnamefont {N.}~\bibnamefont
  {Mardirossian}}\ and\ \bibinfo {author} {\bibfnamefont {M.}~\bibnamefont
  {Head-Gordon}},\ }\bibfield  {title} {\enquote {\bibinfo {title}
  {\ensuremath{\omega}b97x-v: A 10-parameter, range-separated hybrid,
  generalized gradient approximation density functional with nonlocal
  correlation, designed by a survival-of-the-fittest strategy},}\ }\href
  {\doibase 10.1039/c3cp54374a} {\bibfield  {journal} {\bibinfo  {journal}
  {Phys. Chem. Chem. Phys.}\ }\textbf {\bibinfo {volume} {16}},\ \bibinfo
  {pages} {9904} (\bibinfo {year} {2014})}\BibitemShut {NoStop}%
\bibitem [{\citenamefont {Mardirossian}\ and\ \citenamefont
  {Head-Gordon}(2015)}]{Mardirossian:2015}%
  \BibitemOpen
  \bibfield  {author} {\bibinfo {author} {\bibfnamefont {N.}~\bibnamefont
  {Mardirossian}}\ and\ \bibinfo {author} {\bibfnamefont {M.}~\bibnamefont
  {Head-Gordon}},\ }\bibfield  {title} {\enquote {\bibinfo {title} {Mapping the
  genome of meta-generalized gradient approximation density functionals: The
  search for b97m-v},}\ }\href {\doibase 10.1063/1.4907719} {\bibfield
  {journal} {\bibinfo  {journal} {J. Chem. Phys.}\ }\textbf {\bibinfo {volume}
  {142}},\ \bibinfo {pages} {074111} (\bibinfo {year} {2015})}\BibitemShut
  {NoStop}%
\bibitem [{\citenamefont {Mardirossian}\ and\ \citenamefont
  {Head-Gordon}(2016)}]{Mardirossian:2016}%
  \BibitemOpen
  \bibfield  {author} {\bibinfo {author} {\bibfnamefont {N.}~\bibnamefont
  {Mardirossian}}\ and\ \bibinfo {author} {\bibfnamefont {M.}~\bibnamefont
  {Head-Gordon}},\ }\bibfield  {title} {\enquote {\bibinfo {title}
  {\ensuremath{\omega}b97m-v: A combinatorially optimized, range-separated
  hybrid, meta-gga density functional with vv10 nonlocal correlation},}\ }\href
  {\doibase 10.1063/1.4952647} {\bibfield  {journal} {\bibinfo  {journal} {J.
  Chem. Phys.}\ }\textbf {\bibinfo {volume} {144}},\ \bibinfo {pages} {214110}
  (\bibinfo {year} {2016})}\BibitemShut {NoStop}%
\bibitem [{\citenamefont {Sabatini}, \citenamefont {Gorni},\ and\ \citenamefont
  {De~Gironcoli}(2013)}]{sabatini2013nonlocal}%
  \BibitemOpen
  \bibfield  {author} {\bibinfo {author} {\bibfnamefont {R.}~\bibnamefont
  {Sabatini}}, \bibinfo {author} {\bibfnamefont {T.}~\bibnamefont {Gorni}}, \
  and\ \bibinfo {author} {\bibfnamefont {S.}~\bibnamefont {De~Gironcoli}},\
  }\bibfield  {title} {\enquote {\bibinfo {title} {Nonlocal van der waals
  density functional made simple and efficient},}\ }\href@noop {} {\bibfield
  {journal} {\bibinfo  {journal} {Phys. Rev. B}\ }\textbf {\bibinfo {volume}
  {87}},\ \bibinfo {pages} {041108} (\bibinfo {year} {2013})}\BibitemShut
  {NoStop}%
\bibitem [{\citenamefont {Pulay}(1987)}]{pulay1987analytical}%
  \BibitemOpen
  \bibfield  {author} {\bibinfo {author} {\bibfnamefont {P.}~\bibnamefont
  {Pulay}},\ }\bibfield  {title} {\enquote {\bibinfo {title} {Analytical
  derivative methods in quantum chemistry},}\ }\href@noop {} {\bibfield
  {journal} {\bibinfo  {journal} {Adv. Chem. Phys.}\ }\textbf {\bibinfo
  {volume} {67}},\ \bibinfo {pages} {241--286} (\bibinfo {year}
  {1987})}\BibitemShut {NoStop}%
\bibitem [{\citenamefont {Amos}\ and\ \citenamefont
  {Rice}(1989)}]{amos1989implementation}%
  \BibitemOpen
  \bibfield  {author} {\bibinfo {author} {\bibfnamefont {R.~D.}\ \bibnamefont
  {Amos}}\ and\ \bibinfo {author} {\bibfnamefont {J.~E.}\ \bibnamefont
  {Rice}},\ }\bibfield  {title} {\enquote {\bibinfo {title} {Implementation of
  analytic derivative methods in quantum chemistry},}\ }\href@noop {}
  {\bibfield  {journal} {\bibinfo  {journal} {Comput. Phys. Rep.}\ }\textbf
  {\bibinfo {volume} {10}},\ \bibinfo {pages} {147--187} (\bibinfo {year}
  {1989})}\BibitemShut {NoStop}%
\bibitem [{\citenamefont {Pople}\ \emph {et~al.}(1979)\citenamefont {Pople},
  \citenamefont {Krishnan}, \citenamefont {Schlegel},\ and\ \citenamefont
  {Binkley}}]{pople1979derivative}%
  \BibitemOpen
  \bibfield  {author} {\bibinfo {author} {\bibfnamefont {J.}~\bibnamefont
  {Pople}}, \bibinfo {author} {\bibfnamefont {R.}~\bibnamefont {Krishnan}},
  \bibinfo {author} {\bibfnamefont {H.}~\bibnamefont {Schlegel}}, \ and\
  \bibinfo {author} {\bibfnamefont {J.~S.}\ \bibnamefont {Binkley}},\
  }\bibfield  {title} {\enquote {\bibinfo {title} {Derivative studies in
  hartree-fock and m{\o}ller-plesset theories},}\ }\href@noop {} {\bibfield
  {journal} {\bibinfo  {journal} {Int. J. Quantum Chem.}\ }\textbf {\bibinfo
  {volume} {16}},\ \bibinfo {pages} {225--241} (\bibinfo {year}
  {1979})}\BibitemShut {NoStop}%
\bibitem [{\citenamefont {Frisch}, \citenamefont {Head-Gordon},\ and\
  \citenamefont {Pople}(1990)}]{frisch1990direct}%
  \BibitemOpen
  \bibfield  {author} {\bibinfo {author} {\bibfnamefont {M.}~\bibnamefont
  {Frisch}}, \bibinfo {author} {\bibfnamefont {M.}~\bibnamefont {Head-Gordon}},
  \ and\ \bibinfo {author} {\bibfnamefont {J.}~\bibnamefont {Pople}},\
  }\bibfield  {title} {\enquote {\bibinfo {title} {Direct analytic scf second
  derivatives and electric field properties},}\ }\href@noop {} {\bibfield
  {journal} {\bibinfo  {journal} {Chem. Phys.}\ }\textbf {\bibinfo {volume}
  {141}},\ \bibinfo {pages} {189--196} (\bibinfo {year} {1990})}\BibitemShut
  {NoStop}%
\bibitem [{\citenamefont {Johnson}\ and\ \citenamefont
  {Fisch}(1994)}]{johnson1994implementation}%
  \BibitemOpen
  \bibfield  {author} {\bibinfo {author} {\bibfnamefont {B.~G.}\ \bibnamefont
  {Johnson}}\ and\ \bibinfo {author} {\bibfnamefont {M.~J.}\ \bibnamefont
  {Fisch}},\ }\bibfield  {title} {\enquote {\bibinfo {title} {An implementation
  of analytic second derivatives of the gradient-corrected density functional
  energy},}\ }\href@noop {} {\bibfield  {journal} {\bibinfo  {journal} {J.
  Chem. Phys.}\ }\textbf {\bibinfo {volume} {100}},\ \bibinfo {pages}
  {7429--7442} (\bibinfo {year} {1994})}\BibitemShut {NoStop}%
\bibitem [{\citenamefont {Deglmann}, \citenamefont {Furche},\ and\
  \citenamefont {Ahlrichs}(2002)}]{deglmann2002efficient}%
  \BibitemOpen
  \bibfield  {author} {\bibinfo {author} {\bibfnamefont {P.}~\bibnamefont
  {Deglmann}}, \bibinfo {author} {\bibfnamefont {F.}~\bibnamefont {Furche}}, \
  and\ \bibinfo {author} {\bibfnamefont {R.}~\bibnamefont {Ahlrichs}},\
  }\bibfield  {title} {\enquote {\bibinfo {title} {An efficient implementation
  of second analytical derivatives for density functional methods},}\
  }\href@noop {} {\bibfield  {journal} {\bibinfo  {journal} {Chem. Phys.
  Lett.}\ }\textbf {\bibinfo {volume} {362}},\ \bibinfo {pages} {511--518}
  (\bibinfo {year} {2002})}\BibitemShut {NoStop}%
\bibitem [{\citenamefont {Wolff}(2005)}]{wolff2005analytical}%
  \BibitemOpen
  \bibfield  {author} {\bibinfo {author} {\bibfnamefont {S.~K.}\ \bibnamefont
  {Wolff}},\ }\bibfield  {title} {\enquote {\bibinfo {title} {Analytical second
  derivatives in the amsterdam density functional package},}\ }\href@noop {}
  {\bibfield  {journal} {\bibinfo  {journal} {Int. J. Quantum Chem.}\ }\textbf
  {\bibinfo {volume} {104}},\ \bibinfo {pages} {645--659} (\bibinfo {year}
  {2005})}\BibitemShut {NoStop}%
\bibitem [{\citenamefont {Bykov}\ \emph {et~al.}(2015)\citenamefont {Bykov},
  \citenamefont {Petrenko}, \citenamefont {Izs{\'a}k}, \citenamefont
  {Kossmann}, \citenamefont {Becker}, \citenamefont {Valeev},\ and\
  \citenamefont {Neese}}]{bykov2015efficient}%
  \BibitemOpen
  \bibfield  {author} {\bibinfo {author} {\bibfnamefont {D.}~\bibnamefont
  {Bykov}}, \bibinfo {author} {\bibfnamefont {T.}~\bibnamefont {Petrenko}},
  \bibinfo {author} {\bibfnamefont {R.}~\bibnamefont {Izs{\'a}k}}, \bibinfo
  {author} {\bibfnamefont {S.}~\bibnamefont {Kossmann}}, \bibinfo {author}
  {\bibfnamefont {U.}~\bibnamefont {Becker}}, \bibinfo {author} {\bibfnamefont
  {E.}~\bibnamefont {Valeev}}, \ and\ \bibinfo {author} {\bibfnamefont
  {F.}~\bibnamefont {Neese}},\ }\bibfield  {title} {\enquote {\bibinfo {title}
  {Efficient implementation of the analytic second derivatives of hartree--fock
  and hybrid dft energies: a detailed analysis of different approximations},}\
  }\href@noop {} {\bibfield  {journal} {\bibinfo  {journal} {Mol. Phys.}\
  }\textbf {\bibinfo {volume} {113}},\ \bibinfo {pages} {1961--1977} (\bibinfo
  {year} {2015})}\BibitemShut {NoStop}%
\bibitem [{\citenamefont {Delgado-Venegas}\ \emph {et~al.}(2016)\citenamefont
  {Delgado-Venegas}, \citenamefont {Mej{\'\i}a-Rodr{\'\i}guez}, \citenamefont
  {Flores-Moreno}, \citenamefont {Calaminici},\ and\ \citenamefont
  {K{\"o}ster}}]{delgado2016analytic}%
  \BibitemOpen
  \bibfield  {author} {\bibinfo {author} {\bibfnamefont {R.~I.}\ \bibnamefont
  {Delgado-Venegas}}, \bibinfo {author} {\bibfnamefont {D.}~\bibnamefont
  {Mej{\'\i}a-Rodr{\'\i}guez}}, \bibinfo {author} {\bibfnamefont
  {R.}~\bibnamefont {Flores-Moreno}}, \bibinfo {author} {\bibfnamefont
  {P.}~\bibnamefont {Calaminici}}, \ and\ \bibinfo {author} {\bibfnamefont
  {A.~M.}\ \bibnamefont {K{\"o}ster}},\ }\bibfield  {title} {\enquote {\bibinfo
  {title} {Analytic second derivatives from auxiliary density perturbation
  theory},}\ }\href@noop {} {\bibfield  {journal} {\bibinfo  {journal} {J.
  Chem. Phys.}\ }\textbf {\bibinfo {volume} {145}},\ \bibinfo {pages} {224103}
  (\bibinfo {year} {2016})}\BibitemShut {NoStop}%
\bibitem [{\citenamefont {Gu}, \citenamefont {Zhu},\ and\ \citenamefont
  {Xu}(2021)}]{gu2021second}%
  \BibitemOpen
  \bibfield  {author} {\bibinfo {author} {\bibfnamefont {Y.}~\bibnamefont
  {Gu}}, \bibinfo {author} {\bibfnamefont {Z.}~\bibnamefont {Zhu}}, \ and\
  \bibinfo {author} {\bibfnamefont {X.}~\bibnamefont {Xu}},\ }\bibfield
  {title} {\enquote {\bibinfo {title} {Second-order analytic derivatives for
  xyg3 type of doubly hybrid density functionals: theory, implementation, and
  application to harmonic and anharmonic vibrational frequency calculations},}\
  }\href@noop {} {\bibfield  {journal} {\bibinfo  {journal} {J. Chem. Theory
  Comput.}\ }\textbf {\bibinfo {volume} {17}},\ \bibinfo {pages} {4860--4871}
  (\bibinfo {year} {2021})}\BibitemShut {NoStop}%
\bibitem [{\citenamefont {Abbott}\ \emph {et~al.}(2021)\citenamefont {Abbott},
  \citenamefont {Abbott}, \citenamefont {Turney},\ and\ \citenamefont
  {Schaefer~III}}]{abbott2021arbitrary}%
  \BibitemOpen
  \bibfield  {author} {\bibinfo {author} {\bibfnamefont {A.~S.}\ \bibnamefont
  {Abbott}}, \bibinfo {author} {\bibfnamefont {B.~Z.}\ \bibnamefont {Abbott}},
  \bibinfo {author} {\bibfnamefont {J.~M.}\ \bibnamefont {Turney}}, \ and\
  \bibinfo {author} {\bibfnamefont {H.~F.}\ \bibnamefont {Schaefer~III}},\
  }\bibfield  {title} {\enquote {\bibinfo {title} {Arbitrary-order derivatives
  of quantum chemical methods via automatic differentiation},}\ }\href@noop {}
  {\bibfield  {journal} {\bibinfo  {journal} {J. Phys. Chem. Lett.}\ }\textbf
  {\bibinfo {volume} {12}},\ \bibinfo {pages} {3232--3239} (\bibinfo {year}
  {2021})}\BibitemShut {NoStop}%
\bibitem [{\citenamefont {Kasim}, \citenamefont {Lehtola},\ and\ \citenamefont
  {Vinko}(2022)}]{kasim2022dqc}%
  \BibitemOpen
  \bibfield  {author} {\bibinfo {author} {\bibfnamefont {M.~F.}\ \bibnamefont
  {Kasim}}, \bibinfo {author} {\bibfnamefont {S.}~\bibnamefont {Lehtola}}, \
  and\ \bibinfo {author} {\bibfnamefont {S.~M.}\ \bibnamefont {Vinko}},\
  }\bibfield  {title} {\enquote {\bibinfo {title} {Dqc: A python program
  package for differentiable quantum chemistry},}\ }\href@noop {} {\bibfield
  {journal} {\bibinfo  {journal} {J. Chem. Phys.}\ }\textbf {\bibinfo {volume}
  {156}},\ \bibinfo {pages} {084801} (\bibinfo {year} {2022})}\BibitemShut
  {NoStop}%
\bibitem [{\citenamefont {Sa{\l}ek}\ and\ \citenamefont
  {Hesselmann}(2007)}]{salek2007self}%
  \BibitemOpen
  \bibfield  {author} {\bibinfo {author} {\bibfnamefont {P.}~\bibnamefont
  {Sa{\l}ek}}\ and\ \bibinfo {author} {\bibfnamefont {A.}~\bibnamefont
  {Hesselmann}},\ }\bibfield  {title} {\enquote {\bibinfo {title} {A
  self-contained and portable density functional theory library for use in ab
  initio quantum chemistry programs},}\ }\href@noop {} {\bibfield  {journal}
  {\bibinfo  {journal} {Journal of computational chemistry}\ }\textbf {\bibinfo
  {volume} {28}},\ \bibinfo {pages} {2569--2575} (\bibinfo {year}
  {2007})}\BibitemShut {NoStop}%
\bibitem [{\citenamefont {Mazur}\ \emph {et~al.}(2016)\citenamefont {Mazur},
  \citenamefont {Makowski}, \citenamefont {{\L}azarski}, \citenamefont
  {W{\l}odarczyk}, \citenamefont {Czajkowska},\ and\ \citenamefont
  {Glanowski}}]{mazur2016automatic}%
  \BibitemOpen
  \bibfield  {author} {\bibinfo {author} {\bibfnamefont {G.}~\bibnamefont
  {Mazur}}, \bibinfo {author} {\bibfnamefont {M.}~\bibnamefont {Makowski}},
  \bibinfo {author} {\bibfnamefont {R.}~\bibnamefont {{\L}azarski}}, \bibinfo
  {author} {\bibfnamefont {R.}~\bibnamefont {W{\l}odarczyk}}, \bibinfo {author}
  {\bibfnamefont {E.}~\bibnamefont {Czajkowska}}, \ and\ \bibinfo {author}
  {\bibfnamefont {M.}~\bibnamefont {Glanowski}},\ }\bibfield  {title} {\enquote
  {\bibinfo {title} {Automatic code generation for quantum chemistry
  applications},}\ }\href@noop {} {\bibfield  {journal} {\bibinfo  {journal}
  {International Journal of Quantum Chemistry}\ }\textbf {\bibinfo {volume}
  {116}},\ \bibinfo {pages} {1370--1381} (\bibinfo {year} {2016})}\BibitemShut
  {NoStop}%
\bibitem [{\citenamefont {van Dam}(2016)}]{van2016comparison}%
  \BibitemOpen
  \bibfield  {author} {\bibinfo {author} {\bibfnamefont {H.~J.}\ \bibnamefont
  {van Dam}},\ }\href@noop {} {\enquote {\bibinfo {title} {A comparison of
  different methods to implement higher order derivatives of density
  functionals},}\ }\bibinfo {type} {Tech. Rep.}\ (\bibinfo  {institution}
  {Brookhaven National Lab.(BNL), Upton, NY (United States)},\ \bibinfo {year}
  {2016})\BibitemShut {NoStop}%
\bibitem [{\citenamefont {Lehtola}\ \emph {et~al.}(2018)\citenamefont
  {Lehtola}, \citenamefont {Steigemann}, \citenamefont {Oliveira},\ and\
  \citenamefont {Marques}}]{lehtola2018recent}%
  \BibitemOpen
  \bibfield  {author} {\bibinfo {author} {\bibfnamefont {S.}~\bibnamefont
  {Lehtola}}, \bibinfo {author} {\bibfnamefont {C.}~\bibnamefont {Steigemann}},
  \bibinfo {author} {\bibfnamefont {M.~J.}\ \bibnamefont {Oliveira}}, \ and\
  \bibinfo {author} {\bibfnamefont {M.~A.}\ \bibnamefont {Marques}},\
  }\bibfield  {title} {\enquote {\bibinfo {title} {Recent developments in
  libxc—a comprehensive library of functionals for density functional
  theory},}\ }\href@noop {} {\bibfield  {journal} {\bibinfo  {journal}
  {SoftwareX}\ }\textbf {\bibinfo {volume} {7}},\ \bibinfo {pages} {1--5}
  (\bibinfo {year} {2018})}\BibitemShut {NoStop}%
\bibitem [{\citenamefont {Sabatini}\ \emph {et~al.}(2016)\citenamefont
  {Sabatini}, \citenamefont {K\"u\c{c}\"ukbenli}, \citenamefont {Pham},\ and\
  \citenamefont {de~Gironcoli}}]{sabatini2016phonons}%
  \BibitemOpen
  \bibfield  {author} {\bibinfo {author} {\bibfnamefont {R.}~\bibnamefont
  {Sabatini}}, \bibinfo {author} {\bibfnamefont {E.}~\bibnamefont
  {K\"u\c{c}\"ukbenli}}, \bibinfo {author} {\bibfnamefont {C.~H.}\ \bibnamefont
  {Pham}}, \ and\ \bibinfo {author} {\bibfnamefont {S.}~\bibnamefont
  {de~Gironcoli}},\ }\bibfield  {title} {\enquote {\bibinfo {title} {Phonons in
  nonlocal van der {Waals} density functional theory},}\ }\href {\doibase
  10.1103/physrevb.93.235120} {\bibfield  {journal} {\bibinfo  {journal} {Phys.
  Rev. B}\ }\textbf {\bibinfo {volume} {93}},\ \bibinfo {pages} {235120}
  (\bibinfo {year} {2016})}\BibitemShut {NoStop}%
\bibitem [{\citenamefont {Miwa}(2022)}]{miwa2022linear}%
  \BibitemOpen
  \bibfield  {author} {\bibinfo {author} {\bibfnamefont {K.}~\bibnamefont
  {Miwa}},\ }\bibfield  {title} {\enquote {\bibinfo {title} {Linear response
  calculation with nonlocal van der {Waals} density functionals},}\ }\href
  {\doibase 10.1103/physrevb.105.024109} {\bibfield  {journal} {\bibinfo
  {journal} {Phys. Rev. B}\ }\textbf {\bibinfo {volume} {105}},\ \bibinfo
  {pages} {024109} (\bibinfo {year} {2022})}\BibitemShut {NoStop}%
\bibitem [{\citenamefont {Liu}, \citenamefont {Liu},\ and\ \citenamefont
  {Herbert}(2017)}]{liu2017accuracy}%
  \BibitemOpen
  \bibfield  {author} {\bibinfo {author} {\bibfnamefont {K.-Y.}\ \bibnamefont
  {Liu}}, \bibinfo {author} {\bibfnamefont {J.}~\bibnamefont {Liu}}, \ and\
  \bibinfo {author} {\bibfnamefont {J.~M.}\ \bibnamefont {Herbert}},\
  }\bibfield  {title} {\enquote {\bibinfo {title} {Accuracy of
  finite-difference harmonic frequencies in density functional theory},}\
  }\href {\doibase 10.1002/jcc.24811} {\bibfield  {journal} {\bibinfo
  {journal} {J. Comput. Chem.}\ }\textbf {\bibinfo {volume} {38}},\ \bibinfo
  {pages} {1678--1684} (\bibinfo {year} {2017})}\BibitemShut {NoStop}%
\bibitem [{\citenamefont {Sharada}\ \emph {et~al.}(2015)\citenamefont
  {Sharada}, \citenamefont {St\"uck}, \citenamefont {Sundstrom}, \citenamefont
  {Bell},\ and\ \citenamefont {Head-Gordon}}]{sharada2015wavefunction}%
  \BibitemOpen
  \bibfield  {author} {\bibinfo {author} {\bibfnamefont {S.~M.}\ \bibnamefont
  {Sharada}}, \bibinfo {author} {\bibfnamefont {D.}~\bibnamefont {St\"uck}},
  \bibinfo {author} {\bibfnamefont {E.~J.}\ \bibnamefont {Sundstrom}}, \bibinfo
  {author} {\bibfnamefont {A.~T.}\ \bibnamefont {Bell}}, \ and\ \bibinfo
  {author} {\bibfnamefont {M.}~\bibnamefont {Head-Gordon}},\ }\bibfield
  {title} {\enquote {\bibinfo {title} {Wavefunction stability analysis without
  analytical electronic hessians: {Application} to orbital-optimised
  second-order {M{\o}ller{\textendash}Plesset} theory and vv10-containing
  density functionals},}\ }\href {\doibase 10.1080/00268976.2015.1014442}
  {\bibfield  {journal} {\bibinfo  {journal} {Mol. Phys.}\ }\textbf {\bibinfo
  {volume} {113}},\ \bibinfo {pages} {1802--1808} (\bibinfo {year}
  {2015})}\BibitemShut {NoStop}%
\bibitem [{\citenamefont {Pople}\ \emph {et~al.}(2009)\citenamefont {Pople},
  \citenamefont {Schlegel}, \citenamefont {Krishnan}, \citenamefont {Defrees},
  \citenamefont {Binkley}, \citenamefont {Frisch}, \citenamefont {Whiteside},
  \citenamefont {Hout},\ and\ \citenamefont {Hehre}}]{pople1981molecular}%
  \BibitemOpen
  \bibfield  {author} {\bibinfo {author} {\bibfnamefont {J.~A.}\ \bibnamefont
  {Pople}}, \bibinfo {author} {\bibfnamefont {H.~B.}\ \bibnamefont {Schlegel}},
  \bibinfo {author} {\bibfnamefont {R.}~\bibnamefont {Krishnan}}, \bibinfo
  {author} {\bibfnamefont {D.~J.}\ \bibnamefont {Defrees}}, \bibinfo {author}
  {\bibfnamefont {J.~S.}\ \bibnamefont {Binkley}}, \bibinfo {author}
  {\bibfnamefont {M.~J.}\ \bibnamefont {Frisch}}, \bibinfo {author}
  {\bibfnamefont {R.~A.}\ \bibnamefont {Whiteside}}, \bibinfo {author}
  {\bibfnamefont {R.~F.}\ \bibnamefont {Hout}}, \ and\ \bibinfo {author}
  {\bibfnamefont {W.~J.}\ \bibnamefont {Hehre}},\ }\bibfield  {title} {\enquote
  {\bibinfo {title} {Molecular orbital studies of vibrational frequencies},}\
  }\href {\doibase 10.1002/qua.560200829} {\bibfield  {journal} {\bibinfo
  {journal} {Int. J. Quantum Chem.}\ }\textbf {\bibinfo {volume} {20}},\
  \bibinfo {pages} {269--278} (\bibinfo {year} {2009})}\BibitemShut {NoStop}%
\bibitem [{\citenamefont {Merrick}, \citenamefont {Moran},\ and\ \citenamefont
  {Radom}(2007)}]{merrick2007evaluation}%
  \BibitemOpen
  \bibfield  {author} {\bibinfo {author} {\bibfnamefont {J.~P.}\ \bibnamefont
  {Merrick}}, \bibinfo {author} {\bibfnamefont {D.}~\bibnamefont {Moran}}, \
  and\ \bibinfo {author} {\bibfnamefont {L.}~\bibnamefont {Radom}},\ }\bibfield
   {title} {\enquote {\bibinfo {title} {An evaluation of harmonic vibrational
  frequency scale factors},}\ }\href {\doibase 10.1021/jp073974n} {\bibfield
  {journal} {\bibinfo  {journal} {J. Phys. Chem. A}\ }\textbf {\bibinfo
  {volume} {111}},\ \bibinfo {pages} {11683--11700} (\bibinfo {year}
  {2007})}\BibitemShut {NoStop}%
\bibitem [{\citenamefont {Alecu}\ \emph {et~al.}(2010)\citenamefont {Alecu},
  \citenamefont {Zheng}, \citenamefont {Zhao},\ and\ \citenamefont
  {Truhlar}}]{alecu2010computational}%
  \BibitemOpen
  \bibfield  {author} {\bibinfo {author} {\bibfnamefont {I.~M.}\ \bibnamefont
  {Alecu}}, \bibinfo {author} {\bibfnamefont {J.}~\bibnamefont {Zheng}},
  \bibinfo {author} {\bibfnamefont {Y.}~\bibnamefont {Zhao}}, \ and\ \bibinfo
  {author} {\bibfnamefont {D.~G.}\ \bibnamefont {Truhlar}},\ }\bibfield
  {title} {\enquote {\bibinfo {title} {Computational thermochemistry: {Scale}
  factor databases and scale factors for vibrational frequencies obtained from
  electronic model chemistries},}\ }\href {\doibase 10.1021/ct100326h}
  {\bibfield  {journal} {\bibinfo  {journal} {J. Chem. Theory Comput.}\
  }\textbf {\bibinfo {volume} {6}},\ \bibinfo {pages} {2872--2887} (\bibinfo
  {year} {2010})}\BibitemShut {NoStop}%
\bibitem [{\citenamefont {Biczysko}\ \emph {et~al.}(2010)\citenamefont
  {Biczysko}, \citenamefont {Panek}, \citenamefont {Scalmani}, \citenamefont
  {Bloino},\ and\ \citenamefont {Barone}}]{biczysko2010harmonic}%
  \BibitemOpen
  \bibfield  {author} {\bibinfo {author} {\bibfnamefont {M.}~\bibnamefont
  {Biczysko}}, \bibinfo {author} {\bibfnamefont {P.}~\bibnamefont {Panek}},
  \bibinfo {author} {\bibfnamefont {G.}~\bibnamefont {Scalmani}}, \bibinfo
  {author} {\bibfnamefont {J.}~\bibnamefont {Bloino}}, \ and\ \bibinfo {author}
  {\bibfnamefont {V.}~\bibnamefont {Barone}},\ }\bibfield  {title} {\enquote
  {\bibinfo {title} {Harmonic and anharmonic vibrational frequency calculations
  with the double-hybrid {B2PLYP} method: {Analytic} second derivatives and
  benchmark studies},}\ }\href {\doibase 10.1021/ct100212p} {\bibfield
  {journal} {\bibinfo  {journal} {J. Chem. Theory Comput.}\ }\textbf {\bibinfo
  {volume} {6}},\ \bibinfo {pages} {2115--2125} (\bibinfo {year}
  {2010})}\BibitemShut {NoStop}%
\bibitem [{\citenamefont {Laury}, \citenamefont {Carlson},\ and\ \citenamefont
  {Wilson}(2012)}]{laury2012vibrational}%
  \BibitemOpen
  \bibfield  {author} {\bibinfo {author} {\bibfnamefont {M.~L.}\ \bibnamefont
  {Laury}}, \bibinfo {author} {\bibfnamefont {M.~J.}\ \bibnamefont {Carlson}},
  \ and\ \bibinfo {author} {\bibfnamefont {A.~K.}\ \bibnamefont {Wilson}},\
  }\bibfield  {title} {\enquote {\bibinfo {title} {Vibrational frequency scale
  factors for density functional theory and the polarization consistent basis
  sets},}\ }\href {\doibase 10.1002/jcc.23073} {\bibfield  {journal} {\bibinfo
  {journal} {J. Comput. Chem.}\ }\textbf {\bibinfo {volume} {33}},\ \bibinfo
  {pages} {2380--2387} (\bibinfo {year} {2012})}\BibitemShut {NoStop}%
\bibitem [{\citenamefont {Kesharwani}, \citenamefont {Brauer},\ and\
  \citenamefont {Martin}(2014)}]{kesharwani2015frequency}%
  \BibitemOpen
  \bibfield  {author} {\bibinfo {author} {\bibfnamefont {M.~K.}\ \bibnamefont
  {Kesharwani}}, \bibinfo {author} {\bibfnamefont {B.}~\bibnamefont {Brauer}},
  \ and\ \bibinfo {author} {\bibfnamefont {J.~M.~L.}\ \bibnamefont {Martin}},\
  }\bibfield  {title} {\enquote {\bibinfo {title} {Frequency and zero-point
  vibrational energy scale factors for double-hybrid density functionals (and
  other selected methods): {Can} anharmonic force fields be avoided?}}\ }\href
  {\doibase 10.1021/jp508422u} {\bibfield  {journal} {\bibinfo  {journal} {J.
  Phys. Chem. A}\ }\textbf {\bibinfo {volume} {119}},\ \bibinfo {pages}
  {1701--1714} (\bibinfo {year} {2014})}\BibitemShut {NoStop}%
\bibitem [{\citenamefont {Chan}\ and\ \citenamefont
  {Radom}(2016)}]{chan2016frequency}%
  \BibitemOpen
  \bibfield  {author} {\bibinfo {author} {\bibfnamefont {B.}~\bibnamefont
  {Chan}}\ and\ \bibinfo {author} {\bibfnamefont {L.}~\bibnamefont {Radom}},\
  }\bibfield  {title} {\enquote {\bibinfo {title} {Frequency scale factors for
  some double-hybrid density functional theory procedures: {Accurate}
  thermochemical components for high-level composite protocols},}\ }\href
  {\doibase 10.1021/acs.jctc.6b00554} {\bibfield  {journal} {\bibinfo
  {journal} {J. Chem. Theory Comput.}\ }\textbf {\bibinfo {volume} {12}},\
  \bibinfo {pages} {3774--3780} (\bibinfo {year} {2016})}\BibitemShut {NoStop}%
\bibitem [{\citenamefont {Katari}\ \emph {et~al.}(2017)\citenamefont {Katari},
  \citenamefont {Nicol}, \citenamefont {Steinmetz}, \citenamefont {van~der
  Rest}, \citenamefont {Carmichael},\ and\ \citenamefont
  {Frison}}]{katari2017improved}%
  \BibitemOpen
  \bibfield  {author} {\bibinfo {author} {\bibfnamefont {M.}~\bibnamefont
  {Katari}}, \bibinfo {author} {\bibfnamefont {E.}~\bibnamefont {Nicol}},
  \bibinfo {author} {\bibfnamefont {V.}~\bibnamefont {Steinmetz}}, \bibinfo
  {author} {\bibfnamefont {G.}~\bibnamefont {van~der Rest}}, \bibinfo {author}
  {\bibfnamefont {D.}~\bibnamefont {Carmichael}}, \ and\ \bibinfo {author}
  {\bibfnamefont {G.}~\bibnamefont {Frison}},\ }\bibfield  {title} {\enquote
  {\bibinfo {title} {Improved infrared spectra prediction by {DFT} from a new
  experimental database},}\ }\href {\doibase 10.1002/chem.201700340} {\bibfield
   {journal} {\bibinfo  {journal} {Chem. Eur. J.}\ }\textbf {\bibinfo {volume}
  {23}},\ \bibinfo {pages} {8414--8423} (\bibinfo {year} {2017})}\BibitemShut
  {NoStop}%
\bibitem [{\citenamefont {Kashinski}\ \emph {et~al.}(2017)\citenamefont
  {Kashinski}, \citenamefont {Chase}, \citenamefont {Nelson}, \citenamefont
  {Di~Nallo}, \citenamefont {Scales}, \citenamefont {VanderLey},\ and\
  \citenamefont {Byrd}}]{kashinski2017harmonic}%
  \BibitemOpen
  \bibfield  {author} {\bibinfo {author} {\bibfnamefont {D.~O.}\ \bibnamefont
  {Kashinski}}, \bibinfo {author} {\bibfnamefont {G.~M.}\ \bibnamefont
  {Chase}}, \bibinfo {author} {\bibfnamefont {R.~G.}\ \bibnamefont {Nelson}},
  \bibinfo {author} {\bibfnamefont {O.~E.}\ \bibnamefont {Di~Nallo}}, \bibinfo
  {author} {\bibfnamefont {A.~N.}\ \bibnamefont {Scales}}, \bibinfo {author}
  {\bibfnamefont {D.~L.}\ \bibnamefont {VanderLey}}, \ and\ \bibinfo {author}
  {\bibfnamefont {E.~F.~C.}\ \bibnamefont {Byrd}},\ }\bibfield  {title}
  {\enquote {\bibinfo {title} {Harmonic vibrational frequencies: {Approximate}
  global scaling factors for {TPSS,} m06, and m11 functional families using
  several common basis sets},}\ }\href {\doibase 10.1021/acs.jpca.6b12147}
  {\bibfield  {journal} {\bibinfo  {journal} {J. Phys. Chem. A}\ }\textbf
  {\bibinfo {volume} {121}},\ \bibinfo {pages} {2265--2273} (\bibinfo {year}
  {2017})}\BibitemShut {NoStop}%
\bibitem [{\citenamefont {Hanson-Heine}(2019)}]{hanson2019benchmarking}%
  \BibitemOpen
  \bibfield  {author} {\bibinfo {author} {\bibfnamefont {M.~W.~D.}\
  \bibnamefont {Hanson-Heine}},\ }\bibfield  {title} {\enquote {\bibinfo
  {title} {Benchmarking dft-d dispersion corrections for anharmonic vibrational
  frequencies and harmonic scaling factors},}\ }\href {\doibase
  10.1021/acs.jpca.9b07886} {\bibfield  {journal} {\bibinfo  {journal} {J.
  Phys. Chem. A}\ }\textbf {\bibinfo {volume} {123}},\ \bibinfo {pages}
  {9800--9808} (\bibinfo {year} {2019})}\BibitemShut {NoStop}%
\bibitem [{\citenamefont {Zapata~Trujillo}\ and\ \citenamefont
  {McKemmish}(2022)}]{zapata2022vibfreq1295}%
  \BibitemOpen
  \bibfield  {author} {\bibinfo {author} {\bibfnamefont {J.~C.}\ \bibnamefont
  {Zapata~Trujillo}}\ and\ \bibinfo {author} {\bibfnamefont {L.~K.}\
  \bibnamefont {McKemmish}},\ }\bibfield  {title} {\enquote {\bibinfo {title}
  {Vibfreq1295: A new database for vibrational frequency calculations},}\
  }\href@noop {} {\bibfield  {journal} {\bibinfo  {journal} {J. Phys. Chem. A}\
  }\textbf {\bibinfo {volume} {126}},\ \bibinfo {pages} {4100--4122} (\bibinfo
  {year} {2022})}\BibitemShut {NoStop}%
\bibitem [{\citenamefont {\"Unal}\ \emph {et~al.}(2021)\citenamefont {\"Unal},
  \citenamefont {Nassif}, \citenamefont {\"Ozaydin},\ and\ \citenamefont
  {Sayin}}]{unal2021scale}%
  \BibitemOpen
  \bibfield  {author} {\bibinfo {author} {\bibfnamefont {Y.}~\bibnamefont
  {\"Unal}}, \bibinfo {author} {\bibfnamefont {W.}~\bibnamefont {Nassif}},
  \bibinfo {author} {\bibfnamefont {B.~C.}\ \bibnamefont {\"Ozaydin}}, \ and\
  \bibinfo {author} {\bibfnamefont {K.}~\bibnamefont {Sayin}},\ }\bibfield
  {title} {\enquote {\bibinfo {title} {Scale factor database for the vibration
  frequencies calculated in m06-2x, one of the {DFT} methods},}\ }\href
  {\doibase 10.1016/j.vibspec.2020.103189} {\bibfield  {journal} {\bibinfo
  {journal} {Vib. Spectrosc.}\ }\textbf {\bibinfo {volume} {112}},\ \bibinfo
  {pages} {103189} (\bibinfo {year} {2021})}\BibitemShut {NoStop}%
\bibitem [{\citenamefont {Zapata~Trujillo}\ and\ \citenamefont
  {McKemmish}(2023)}]{zapata2023model}%
  \BibitemOpen
  \bibfield  {author} {\bibinfo {author} {\bibfnamefont {J.~C.}\ \bibnamefont
  {Zapata~Trujillo}}\ and\ \bibinfo {author} {\bibfnamefont {L.~K.}\
  \bibnamefont {McKemmish}},\ }\bibfield  {title} {\enquote {\bibinfo {title}
  {Model chemistry recommendations for scaled harmonic frequency calculations:
  A benchmark study},}\ }\href@noop {} {\bibfield  {journal} {\bibinfo
  {journal} {The Journal of Physical Chemistry A}\ } (\bibinfo {year}
  {2023})}\BibitemShut {NoStop}%
\bibitem [{\citenamefont {Ravichandran}\ and\ \citenamefont
  {Banik}(2017)}]{ravichandran2018performance}%
  \BibitemOpen
  \bibfield  {author} {\bibinfo {author} {\bibfnamefont {L.}~\bibnamefont
  {Ravichandran}}\ and\ \bibinfo {author} {\bibfnamefont {S.}~\bibnamefont
  {Banik}},\ }\bibfield  {title} {\enquote {\bibinfo {title} {Performance of
  different density functionals for the calculation of vibrational frequencies
  with vibrational coupled cluster method in bosonic representation},}\ }\href
  {\doibase 10.1007/s00214-017-2177-9} {\bibfield  {journal} {\bibinfo
  {journal} {Theor. Chem. Acc.}\ }\textbf {\bibinfo {volume} {137}},\ \bibinfo
  {pages} {1--14} (\bibinfo {year} {2017})}\BibitemShut {NoStop}%
\bibitem [{\citenamefont {Howard}, \citenamefont {Enyard},\ and\ \citenamefont
  {Tschumper}(2015)}]{howard2015assessing}%
  \BibitemOpen
  \bibfield  {author} {\bibinfo {author} {\bibfnamefont {J.~C.}\ \bibnamefont
  {Howard}}, \bibinfo {author} {\bibfnamefont {J.~D.}\ \bibnamefont {Enyard}},
  \ and\ \bibinfo {author} {\bibfnamefont {G.~S.}\ \bibnamefont {Tschumper}},\
  }\bibfield  {title} {\enquote {\bibinfo {title} {Assessing the accuracy of
  some popular {DFT} methods for computing harmonic vibrational frequencies of
  water clusters},}\ }\href {\doibase 10.1063/1.4936654} {\bibfield  {journal}
  {\bibinfo  {journal} {J. Chem. Phys.}\ }\textbf {\bibinfo {volume} {143}},\
  \bibinfo {pages} {214103} (\bibinfo {year} {2015})}\BibitemShut {NoStop}%
\bibitem [{\citenamefont {Becke}(1988)}]{becke1988multicenter}%
  \BibitemOpen
  \bibfield  {author} {\bibinfo {author} {\bibfnamefont {A.~D.}\ \bibnamefont
  {Becke}},\ }\bibfield  {title} {\enquote {\bibinfo {title} {A multicenter
  numerical integration scheme for polyatomic molecules},}\ }\href@noop {}
  {\bibfield  {journal} {\bibinfo  {journal} {J. Chem. Phys.}\ }\textbf
  {\bibinfo {volume} {88}},\ \bibinfo {pages} {2547--2553} (\bibinfo {year}
  {1988})}\BibitemShut {NoStop}%
\bibitem [{\citenamefont {Perdew}, \citenamefont {Burke},\ and\ \citenamefont
  {Ernzerhof}(1996)}]{perdew1996generalized}%
  \BibitemOpen
  \bibfield  {author} {\bibinfo {author} {\bibfnamefont {J.~P.}\ \bibnamefont
  {Perdew}}, \bibinfo {author} {\bibfnamefont {K.}~\bibnamefont {Burke}}, \
  and\ \bibinfo {author} {\bibfnamefont {M.}~\bibnamefont {Ernzerhof}},\
  }\bibfield  {title} {\enquote {\bibinfo {title} {Generalized gradient
  approximation made simple},}\ }\href@noop {} {\bibfield  {journal} {\bibinfo
  {journal} {Phys. Rev. Lett.}\ }\textbf {\bibinfo {volume} {77}},\ \bibinfo
  {pages} {3865} (\bibinfo {year} {1996})}\BibitemShut {NoStop}%
\bibitem [{\citenamefont {Sun}, \citenamefont {Ruzsinszky},\ and\ \citenamefont
  {Perdew}(2015)}]{sun2015strongly}%
  \BibitemOpen
  \bibfield  {author} {\bibinfo {author} {\bibfnamefont {J.}~\bibnamefont
  {Sun}}, \bibinfo {author} {\bibfnamefont {A.}~\bibnamefont {Ruzsinszky}}, \
  and\ \bibinfo {author} {\bibfnamefont {J.~P.}\ \bibnamefont {Perdew}},\
  }\bibfield  {title} {\enquote {\bibinfo {title} {Strongly constrained and
  appropriately normed semilocal density functional},}\ }\href {\doibase
  10.1103/physrevlett.115.036402} {\bibfield  {journal} {\bibinfo  {journal}
  {Phys. Rev. Lett.}\ }\textbf {\bibinfo {volume} {115}},\ \bibinfo {pages}
  {036402} (\bibinfo {year} {2015})}\BibitemShut {NoStop}%
\bibitem [{\citenamefont {Furness}\ \emph {et~al.}(2020)\citenamefont
  {Furness}, \citenamefont {Kaplan}, \citenamefont {Ning}, \citenamefont
  {Perdew},\ and\ \citenamefont {Sun}}]{furness2020accurate}%
  \BibitemOpen
  \bibfield  {author} {\bibinfo {author} {\bibfnamefont {J.~W.}\ \bibnamefont
  {Furness}}, \bibinfo {author} {\bibfnamefont {A.~D.}\ \bibnamefont {Kaplan}},
  \bibinfo {author} {\bibfnamefont {J.}~\bibnamefont {Ning}}, \bibinfo {author}
  {\bibfnamefont {J.~P.}\ \bibnamefont {Perdew}}, \ and\ \bibinfo {author}
  {\bibfnamefont {J.}~\bibnamefont {Sun}},\ }\bibfield  {title} {\enquote
  {\bibinfo {title} {Accurate and numerically efficient r2scan meta-generalized
  gradient approximation},}\ }\href@noop {} {\bibfield  {journal} {\bibinfo
  {journal} {J. Phys. Chem. Lett.}\ }\textbf {\bibinfo {volume} {11}},\
  \bibinfo {pages} {8208--8215} (\bibinfo {year} {2020})}\BibitemShut {NoStop}%
\bibitem [{\citenamefont {Zhao}\ and\ \citenamefont
  {Truhlar}(2006)}]{zhao2006new}%
  \BibitemOpen
  \bibfield  {author} {\bibinfo {author} {\bibfnamefont {Y.}~\bibnamefont
  {Zhao}}\ and\ \bibinfo {author} {\bibfnamefont {D.~G.}\ \bibnamefont
  {Truhlar}},\ }\bibfield  {title} {\enquote {\bibinfo {title} {A new local
  density functional for main-group thermochemistry, transition metal bonding,
  thermochemical kinetics, and noncovalent interactions},}\ }\href {\doibase
  10.1063/1.2370993} {\bibfield  {journal} {\bibinfo  {journal} {J. Chem.
  Phys.}\ }\textbf {\bibinfo {volume} {125}},\ \bibinfo {pages} {194101}
  (\bibinfo {year} {2006})}\BibitemShut {NoStop}%
\bibitem [{\citenamefont {Chai}\ and\ \citenamefont
  {Head-Gordon}(2008)}]{chai2008long}%
  \BibitemOpen
  \bibfield  {author} {\bibinfo {author} {\bibfnamefont {J.-D.}\ \bibnamefont
  {Chai}}\ and\ \bibinfo {author} {\bibfnamefont {M.}~\bibnamefont
  {Head-Gordon}},\ }\bibfield  {title} {\enquote {\bibinfo {title} {Long-range
  corrected hybrid density functionals with damped atom{\textendash}atom
  dispersion corrections},}\ }\href {\doibase 10.1039/b810189b} {\bibfield
  {journal} {\bibinfo  {journal} {Phys. Chem. Chem. Phys.}\ }\textbf {\bibinfo
  {volume} {10}},\ \bibinfo {pages} {6615} (\bibinfo {year}
  {2008})}\BibitemShut {NoStop}%
\bibitem [{\citenamefont {Yanai}, \citenamefont {Tew},\ and\ \citenamefont
  {Handy}(2004)}]{yanai2004new}%
  \BibitemOpen
  \bibfield  {author} {\bibinfo {author} {\bibfnamefont {T.}~\bibnamefont
  {Yanai}}, \bibinfo {author} {\bibfnamefont {D.~P.}\ \bibnamefont {Tew}}, \
  and\ \bibinfo {author} {\bibfnamefont {N.~C.}\ \bibnamefont {Handy}},\
  }\bibfield  {title} {\enquote {\bibinfo {title} {A new hybrid
  exchange{\textendash}correlation functional using the coulomb-attenuating
  method (cam-b3lyp)},}\ }\href {\doibase 10.1016/j.cplett.2004.06.011}
  {\bibfield  {journal} {\bibinfo  {journal} {Chem. Phys. Lett.}\ }\textbf
  {\bibinfo {volume} {393}},\ \bibinfo {pages} {51--57} (\bibinfo {year}
  {2004})}\BibitemShut {NoStop}%
\bibitem [{\citenamefont {Henderson}, \citenamefont {Janesko},\ and\
  \citenamefont {Scuseria}(2008)}]{henderson2008generalized}%
  \BibitemOpen
  \bibfield  {author} {\bibinfo {author} {\bibfnamefont {T.~M.}\ \bibnamefont
  {Henderson}}, \bibinfo {author} {\bibfnamefont {B.~G.}\ \bibnamefont
  {Janesko}}, \ and\ \bibinfo {author} {\bibfnamefont {G.~E.}\ \bibnamefont
  {Scuseria}},\ }\bibfield  {title} {\enquote {\bibinfo {title} {Generalized
  gradient approximation model exchange holes for range-separated hybrids},}\
  }\href {\doibase 10.1063/1.2921797} {\bibfield  {journal} {\bibinfo
  {journal} {J. Chem. Phys.}\ }\textbf {\bibinfo {volume} {128}},\ \bibinfo
  {pages} {194105} (\bibinfo {year} {2008})}\BibitemShut {NoStop}%
\bibitem [{\citenamefont {Krukau}\ \emph {et~al.}(2006)\citenamefont {Krukau},
  \citenamefont {Vydrov}, \citenamefont {Izmaylov},\ and\ \citenamefont
  {Scuseria}}]{krukau2006influence}%
  \BibitemOpen
  \bibfield  {author} {\bibinfo {author} {\bibfnamefont {A.~V.}\ \bibnamefont
  {Krukau}}, \bibinfo {author} {\bibfnamefont {O.~A.}\ \bibnamefont {Vydrov}},
  \bibinfo {author} {\bibfnamefont {A.~F.}\ \bibnamefont {Izmaylov}}, \ and\
  \bibinfo {author} {\bibfnamefont {G.~E.}\ \bibnamefont {Scuseria}},\
  }\bibfield  {title} {\enquote {\bibinfo {title} {Influence of the exchange
  screening parameter on the performance of screened hybrid functionals},}\
  }\href {\doibase 10.1063/1.2404663} {\bibfield  {journal} {\bibinfo
  {journal} {J. Chem. Phys.}\ }\textbf {\bibinfo {volume} {125}},\ \bibinfo
  {pages} {224106} (\bibinfo {year} {2006})}\BibitemShut {NoStop}%
\bibitem [{\citenamefont {Becke}(1993)}]{Becke1993}%
  \BibitemOpen
  \bibfield  {author} {\bibinfo {author} {\bibfnamefont {A.~D.}\ \bibnamefont
  {Becke}},\ }\bibfield  {title} {\enquote {\bibinfo {title}
  {Density-functional thermochemistry. iii. the role of exact exchange},}\
  }\href {\doibase 10.1063/1.464913} {\bibfield  {journal} {\bibinfo  {journal}
  {J. Chem. Phys.}\ }\textbf {\bibinfo {volume} {98}},\ \bibinfo {pages}
  {5648--5652} (\bibinfo {year} {1993})}\BibitemShut {NoStop}%
\bibitem [{\citenamefont {Stephens}\ \emph {et~al.}(1994)\citenamefont
  {Stephens}, \citenamefont {Devlin}, \citenamefont {Chabalowski},\ and\
  \citenamefont {Frisch}}]{stephens1994ab}%
  \BibitemOpen
  \bibfield  {author} {\bibinfo {author} {\bibfnamefont {P.~J.}\ \bibnamefont
  {Stephens}}, \bibinfo {author} {\bibfnamefont {F.~J.}\ \bibnamefont
  {Devlin}}, \bibinfo {author} {\bibfnamefont {C.~F.}\ \bibnamefont
  {Chabalowski}}, \ and\ \bibinfo {author} {\bibfnamefont {M.~J.}\ \bibnamefont
  {Frisch}},\ }\bibfield  {title} {\enquote {\bibinfo {title} {Ab initio
  calculation of vibrational absorption and circular dichroism spectra using
  density functional force fields},}\ }\href {\doibase 10.1021/j100096a001}
  {\bibfield  {journal} {\bibinfo  {journal} {J. Phys. Chem.}\ }\textbf
  {\bibinfo {volume} {98}},\ \bibinfo {pages} {11623--11627} (\bibinfo {year}
  {1994})}\BibitemShut {NoStop}%
\bibitem [{\citenamefont {Boese}\ and\ \citenamefont
  {Martin}(2004)}]{boese2004development}%
  \BibitemOpen
  \bibfield  {author} {\bibinfo {author} {\bibfnamefont {A.~D.}\ \bibnamefont
  {Boese}}\ and\ \bibinfo {author} {\bibfnamefont {J.~M.~L.}\ \bibnamefont
  {Martin}},\ }\bibfield  {title} {\enquote {\bibinfo {title} {Development of
  density functionals for thermochemical kinetics},}\ }\href {\doibase
  10.1063/1.1774975} {\bibfield  {journal} {\bibinfo  {journal} {J. Chem.
  Phys.}\ }\textbf {\bibinfo {volume} {121}},\ \bibinfo {pages} {3405--3416}
  (\bibinfo {year} {2004})}\BibitemShut {NoStop}%
\bibitem [{\citenamefont {Wang}\ \emph {et~al.}(2020)\citenamefont {Wang},
  \citenamefont {Verma}, \citenamefont {Zhang}, \citenamefont {Li},
  \citenamefont {Liu}, \citenamefont {Truhlar},\ and\ \citenamefont
  {He}}]{wang2020m06}%
  \BibitemOpen
  \bibfield  {author} {\bibinfo {author} {\bibfnamefont {Y.}~\bibnamefont
  {Wang}}, \bibinfo {author} {\bibfnamefont {P.}~\bibnamefont {Verma}},
  \bibinfo {author} {\bibfnamefont {L.}~\bibnamefont {Zhang}}, \bibinfo
  {author} {\bibfnamefont {Y.}~\bibnamefont {Li}}, \bibinfo {author}
  {\bibfnamefont {Z.}~\bibnamefont {Liu}}, \bibinfo {author} {\bibfnamefont
  {D.~G.}\ \bibnamefont {Truhlar}}, \ and\ \bibinfo {author} {\bibfnamefont
  {X.}~\bibnamefont {He}},\ }\bibfield  {title} {\enquote {\bibinfo {title}
  {M06-sx screened-exchange density functional for chemistry and solid-state
  physics},}\ }\href {\doibase 10.1073/pnas.1913699117} {\bibfield  {journal}
  {\bibinfo  {journal} {Proc. Natl. Acad. Sci. U. S. A.}\ }\textbf {\bibinfo
  {volume} {117}},\ \bibinfo {pages} {2294--2301} (\bibinfo {year}
  {2020})}\BibitemShut {NoStop}%
\bibitem [{\citenamefont {Zhao}\ and\ \citenamefont
  {Truhlar}(2007)}]{zhao2008m06}%
  \BibitemOpen
  \bibfield  {author} {\bibinfo {author} {\bibfnamefont {Y.}~\bibnamefont
  {Zhao}}\ and\ \bibinfo {author} {\bibfnamefont {D.~G.}\ \bibnamefont
  {Truhlar}},\ }\bibfield  {title} {\enquote {\bibinfo {title} {The m06 suite
  of density functionals for main group thermochemistry, thermochemical
  kinetics, noncovalent interactions, excited states, and transition elements:
  Two new functionals and systematic testing of four m06-class functionals and
  12 other functionals},}\ }\href {\doibase 10.1007/s00214-007-0310-x}
  {\bibfield  {journal} {\bibinfo  {journal} {Theor. Chem. Account}\ }\textbf
  {\bibinfo {volume} {120}},\ \bibinfo {pages} {215--241} (\bibinfo {year}
  {2007})}\BibitemShut {NoStop}%
\bibitem [{\citenamefont {Hui}\ and\ \citenamefont {Chai}(2016)}]{hui2016scan}%
  \BibitemOpen
  \bibfield  {author} {\bibinfo {author} {\bibfnamefont {K.}~\bibnamefont
  {Hui}}\ and\ \bibinfo {author} {\bibfnamefont {J.-D.}\ \bibnamefont {Chai}},\
  }\bibfield  {title} {\enquote {\bibinfo {title} {Scan-based hybrid and
  double-hybrid density functionals from models without fitted parameters},}\
  }\href {\doibase 10.1063/1.4940734} {\bibfield  {journal} {\bibinfo
  {journal} {J. Chem. Phys.}\ }\textbf {\bibinfo {volume} {144}},\ \bibinfo
  {pages} {044114} (\bibinfo {year} {2016})}\BibitemShut {NoStop}%
\bibitem [{\citenamefont {Martin}\ and\ \citenamefont
  {Kesharwani}(2014)}]{martin2014assessment}%
  \BibitemOpen
  \bibfield  {author} {\bibinfo {author} {\bibfnamefont {J.~M.~L.}\
  \bibnamefont {Martin}}\ and\ \bibinfo {author} {\bibfnamefont {M.~K.}\
  \bibnamefont {Kesharwani}},\ }\bibfield  {title} {\enquote {\bibinfo {title}
  {Assessment of ccsd(t)-f12 approximations and basis sets for harmonic
  vibrational frequencies},}\ }\href {\doibase 10.1021/ct500174q} {\bibfield
  {journal} {\bibinfo  {journal} {J. Chem. Theory Comput.}\ }\textbf {\bibinfo
  {volume} {10}},\ \bibinfo {pages} {2085--2090} (\bibinfo {year}
  {2014})}\BibitemShut {NoStop}%
\bibitem [{\citenamefont {Bertels}, \citenamefont {Lee},\ and\ \citenamefont
  {Head-Gordon}(2021)}]{bertels2021polishing}%
  \BibitemOpen
  \bibfield  {author} {\bibinfo {author} {\bibfnamefont {L.~W.}\ \bibnamefont
  {Bertels}}, \bibinfo {author} {\bibfnamefont {J.}~\bibnamefont {Lee}}, \ and\
  \bibinfo {author} {\bibfnamefont {M.}~\bibnamefont {Head-Gordon}},\
  }\bibfield  {title} {\enquote {\bibinfo {title} {Polishing the gold standard:
  {The} role of orbital choice in {CCSD(T)} vibrational frequency
  prediction},}\ }\href {\doibase 10.1021/acs.jctc.0c00746} {\bibfield
  {journal} {\bibinfo  {journal} {J. Chem. Theory Comput.}\ }\textbf {\bibinfo
  {volume} {17}},\ \bibinfo {pages} {742--755} (\bibinfo {year}
  {2021})}\BibitemShut {NoStop}%
\bibitem [{\citenamefont {Lee}\ and\ \citenamefont
  {Head-Gordon}(2018)}]{lee2018regularized}%
  \BibitemOpen
  \bibfield  {author} {\bibinfo {author} {\bibfnamefont {J.}~\bibnamefont
  {Lee}}\ and\ \bibinfo {author} {\bibfnamefont {M.}~\bibnamefont
  {Head-Gordon}},\ }\bibfield  {title} {\enquote {\bibinfo {title} {Regularized
  orbital-optimized second-order m{\o}ller--plesset perturbation theory: A
  reliable fifth-order-scaling electron correlation model with orbital energy
  dependent regularizers},}\ }\href@noop {} {\bibfield  {journal} {\bibinfo
  {journal} {J. Chem. Theory Comput.}\ }\textbf {\bibinfo {volume} {14}},\
  \bibinfo {pages} {5203--5219} (\bibinfo {year} {2018})}\BibitemShut {NoStop}%
\bibitem [{\citenamefont {Peterson}\ and\ \citenamefont
  {Dunning}(2002)}]{peterson2002accurate}%
  \BibitemOpen
  \bibfield  {author} {\bibinfo {author} {\bibfnamefont {K.~A.}\ \bibnamefont
  {Peterson}}\ and\ \bibinfo {author} {\bibfnamefont {T.~H.}\ \bibnamefont
  {Dunning}},\ }\bibfield  {title} {\enquote {\bibinfo {title} {Accurate
  correlation consistent basis sets for molecular core{\textendash}valence
  correlation effects: {The} second row atoms {Al{\textendash}Ar,} and the
  first row atoms {B{\textendash}Ne} revisited},}\ }\href {\doibase
  10.1063/1.1520138} {\bibfield  {journal} {\bibinfo  {journal} {J. Chem.
  Phys.}\ }\textbf {\bibinfo {volume} {117}},\ \bibinfo {pages} {10548--10560}
  (\bibinfo {year} {2002})}\BibitemShut {NoStop}%
\bibitem [{\citenamefont {Prascher}\ \emph {et~al.}(2010)\citenamefont
  {Prascher}, \citenamefont {Woon}, \citenamefont {Peterson}, \citenamefont
  {Dunning},\ and\ \citenamefont {Wilson}}]{prascher2011gaussian}%
  \BibitemOpen
  \bibfield  {author} {\bibinfo {author} {\bibfnamefont {B.~P.}\ \bibnamefont
  {Prascher}}, \bibinfo {author} {\bibfnamefont {D.~E.}\ \bibnamefont {Woon}},
  \bibinfo {author} {\bibfnamefont {K.~A.}\ \bibnamefont {Peterson}}, \bibinfo
  {author} {\bibfnamefont {T.~H.}\ \bibnamefont {Dunning}}, \ and\ \bibinfo
  {author} {\bibfnamefont {A.~K.}\ \bibnamefont {Wilson}},\ }\bibfield  {title}
  {\enquote {\bibinfo {title} {{Gaussian} basis sets for use in correlated
  molecular calculations. {VII.} valence, core-valence, and scalar relativistic
  basis sets for li, be, na, and mg},}\ }\href {\doibase
  10.1007/s00214-010-0764-0} {\bibfield  {journal} {\bibinfo  {journal} {Theor.
  Chem. Acc.}\ }\textbf {\bibinfo {volume} {128}},\ \bibinfo {pages} {69--82}
  (\bibinfo {year} {2010})}\BibitemShut {NoStop}%
\bibitem [{\citenamefont {Howard}\ and\ \citenamefont
  {Tschumper}(2015)}]{howard2015benchmark}%
  \BibitemOpen
  \bibfield  {author} {\bibinfo {author} {\bibfnamefont {J.~C.}\ \bibnamefont
  {Howard}}\ and\ \bibinfo {author} {\bibfnamefont {G.~S.}\ \bibnamefont
  {Tschumper}},\ }\bibfield  {title} {\enquote {\bibinfo {title} {Benchmark
  structures and harmonic vibrational frequencies near the ccsd (t) complete
  basis set limit for small water clusters:(h2o) n= 2, 3, 4, 5, 6},}\ }\href
  {\doibase 10.1021/acs.jctc.5b00225} {\bibfield  {journal} {\bibinfo
  {journal} {J. Chem. Theory Comput.}\ }\textbf {\bibinfo {volume} {11}},\
  \bibinfo {pages} {2126--2136} (\bibinfo {year} {2015})}\BibitemShut {NoStop}%
\bibitem [{\citenamefont {Howard}\ and\ \citenamefont
  {Tschumper}(2013)}]{howard2013n}%
  \BibitemOpen
  \bibfield  {author} {\bibinfo {author} {\bibfnamefont {J.~C.}\ \bibnamefont
  {Howard}}\ and\ \bibinfo {author} {\bibfnamefont {G.~S.}\ \bibnamefont
  {Tschumper}},\ }\bibfield  {title} {\enquote {\bibinfo {title} {N-body:
  Many-body qm: Qm vibrational frequencies: Application to small
  hydrogen-bonded clusters},}\ }\href {\doibase 10.1063/1.4829463} {\bibfield
  {journal} {\bibinfo  {journal} {J. Chem. Phys.}\ }\textbf {\bibinfo {volume}
  {139}},\ \bibinfo {pages} {184113} (\bibinfo {year} {2013})}\BibitemShut
  {NoStop}%
\bibitem [{\citenamefont {Hoja}\ and\ \citenamefont
  {Boese}(2022)}]{hoja2022v30}%
  \BibitemOpen
  \bibfield  {author} {\bibinfo {author} {\bibfnamefont {J.}~\bibnamefont
  {Hoja}}\ and\ \bibinfo {author} {\bibfnamefont {A.~D.}\ \bibnamefont
  {Boese}},\ }\bibfield  {title} {\enquote {\bibinfo {title} {The v30 benchmark
  set for anharmonic vibrational frequencies of molecular dimers},}\
  }\href@noop {} {\bibfield  {journal} {\bibinfo  {journal} {arXiv preprint,
  arXiv:2209.04392}\ } (\bibinfo {year} {2022})}\BibitemShut {NoStop}%
\bibitem [{\citenamefont {Lemke}(2017)}]{lemke2017structure}%
  \BibitemOpen
  \bibfield  {author} {\bibinfo {author} {\bibfnamefont {K.~H.}\ \bibnamefont
  {Lemke}},\ }\bibfield  {title} {\enquote {\bibinfo {title} {Structure and
  binding energy of the h2s dimer at the ccsd (t) complete basis set limit},}\
  }\href@noop {} {\bibfield  {journal} {\bibinfo  {journal} {J. Chem. Phys.}\
  }\textbf {\bibinfo {volume} {146}},\ \bibinfo {pages} {234301} (\bibinfo
  {year} {2017})}\BibitemShut {NoStop}%
\bibitem [{\citenamefont {Salmon}, \citenamefont {de~Lange},\ and\
  \citenamefont {Lane}(2016)}]{salmon2016structure}%
  \BibitemOpen
  \bibfield  {author} {\bibinfo {author} {\bibfnamefont {S.~R.}\ \bibnamefont
  {Salmon}}, \bibinfo {author} {\bibfnamefont {K.~M.}\ \bibnamefont
  {de~Lange}}, \ and\ \bibinfo {author} {\bibfnamefont {J.~R.}\ \bibnamefont
  {Lane}},\ }\bibfield  {title} {\enquote {\bibinfo {title} {Structure and
  abundance of nitrous oxide complexes in earth’s atmosphere},}\ }\href@noop
  {} {\bibfield  {journal} {\bibinfo  {journal} {J. Phys. Chem. A}\ }\textbf
  {\bibinfo {volume} {120}},\ \bibinfo {pages} {2096--2105} (\bibinfo {year}
  {2016})}\BibitemShut {NoStop}%
\bibitem [{\citenamefont {de~Lange}\ and\ \citenamefont
  {Lane}(2011)}]{de2011explicit}%
  \BibitemOpen
  \bibfield  {author} {\bibinfo {author} {\bibfnamefont {K.~M.}\ \bibnamefont
  {de~Lange}}\ and\ \bibinfo {author} {\bibfnamefont {J.~R.}\ \bibnamefont
  {Lane}},\ }\bibfield  {title} {\enquote {\bibinfo {title} {Explicit
  correlation and intermolecular interactions: Investigating carbon dioxide
  complexes with the ccsd (t)-f12 method},}\ }\href@noop {} {\bibfield
  {journal} {\bibinfo  {journal} {J. Chem. Phys.}\ }\textbf {\bibinfo {volume}
  {134}},\ \bibinfo {pages} {034301} (\bibinfo {year} {2011})}\BibitemShut
  {NoStop}%
\bibitem [{\citenamefont {Van~Dornshuld}\ and\ \citenamefont
  {Tschumper}(2016)}]{van2016big}%
  \BibitemOpen
  \bibfield  {author} {\bibinfo {author} {\bibfnamefont {E.}~\bibnamefont
  {Van~Dornshuld}}\ and\ \bibinfo {author} {\bibfnamefont {G.~S.}\ \bibnamefont
  {Tschumper}},\ }\bibfield  {title} {\enquote {\bibinfo {title} {Big changes
  for small noncovalent dimers: Revisiting the potential energy surfaces of
  (p2) 2 and (pccp) 2 with ccsd (t) optimizations and vibrational
  frequencies},}\ }\href@noop {} {\bibfield  {journal} {\bibinfo  {journal} {J.
  Chem. Theory Comput.}\ }\textbf {\bibinfo {volume} {12}},\ \bibinfo {pages}
  {1534--1541} (\bibinfo {year} {2016})}\BibitemShut {NoStop}%
\bibitem [{\citenamefont {Epifanovsky}\ \emph {et~al.}(2021)\citenamefont
  {Epifanovsky}, \citenamefont {Gilbert}, \citenamefont {Feng}, \citenamefont
  {Lee}, \citenamefont {Mao} \emph {et~al.}}]{epifanovsky2021software}%
  \BibitemOpen
  \bibfield  {author} {\bibinfo {author} {\bibfnamefont {E.}~\bibnamefont
  {Epifanovsky}}, \bibinfo {author} {\bibfnamefont {A.~T.}\ \bibnamefont
  {Gilbert}}, \bibinfo {author} {\bibfnamefont {X.}~\bibnamefont {Feng}},
  \bibinfo {author} {\bibfnamefont {J.}~\bibnamefont {Lee}}, \bibinfo {author}
  {\bibfnamefont {Y.}~\bibnamefont {Mao}},  \emph {et~al.},\ }\bibfield
  {title} {\enquote {\bibinfo {title} {Software for the frontiers of quantum
  chemistry: {An} overview of developments in the q-chem 5 package},}\
  }\href@noop {} {\bibfield  {journal} {\bibinfo  {journal} {J. Chem. Phys.}\
  }\textbf {\bibinfo {volume} {155}},\ \bibinfo {pages} {084801} (\bibinfo
  {year} {2021})}\BibitemShut {NoStop}%
\bibitem [{\citenamefont {Weigend}\ and\ \citenamefont
  {Ahlrichs}(2005)}]{weigend2005balanced}%
  \BibitemOpen
  \bibfield  {author} {\bibinfo {author} {\bibfnamefont {F.}~\bibnamefont
  {Weigend}}\ and\ \bibinfo {author} {\bibfnamefont {R.}~\bibnamefont
  {Ahlrichs}},\ }\bibfield  {title} {\enquote {\bibinfo {title} {Balanced basis
  sets of split valence, triple zeta valence and quadruple zeta valence quality
  for h to rn: {Design} and assessment of accuracy},}\ }\href {\doibase
  10.1039/b508541a} {\bibfield  {journal} {\bibinfo  {journal} {Phys. Chem.
  Chem. Phys.}\ }\textbf {\bibinfo {volume} {7}},\ \bibinfo {pages} {3297}
  (\bibinfo {year} {2005})}\BibitemShut {NoStop}%
\bibitem [{\citenamefont {Rappoport}\ and\ \citenamefont
  {Furche}(2010)}]{rappoport2010property}%
  \BibitemOpen
  \bibfield  {author} {\bibinfo {author} {\bibfnamefont {D.}~\bibnamefont
  {Rappoport}}\ and\ \bibinfo {author} {\bibfnamefont {F.}~\bibnamefont
  {Furche}},\ }\bibfield  {title} {\enquote {\bibinfo {title}
  {Property-optimized {Gaussian} basis sets for molecular response
  calculations},}\ }\href {\doibase 10.1063/1.3484283} {\bibfield  {journal}
  {\bibinfo  {journal} {J. Chem. Phys.}\ }\textbf {\bibinfo {volume} {133}},\
  \bibinfo {pages} {134105} (\bibinfo {year} {2010})}\BibitemShut {NoStop}%
\bibitem [{\citenamefont {Gill}, \citenamefont {Johnson},\ and\ \citenamefont
  {Pople}(1993)}]{gill1993}%
  \BibitemOpen
  \bibfield  {author} {\bibinfo {author} {\bibfnamefont {P.~M.}\ \bibnamefont
  {Gill}}, \bibinfo {author} {\bibfnamefont {B.~G.}\ \bibnamefont {Johnson}}, \
  and\ \bibinfo {author} {\bibfnamefont {J.~A.}\ \bibnamefont {Pople}},\
  }\bibfield  {title} {\enquote {\bibinfo {title} {A standard grid for density
  functional calculations},}\ }\href {\doibase 10.1016/0009-2614(93)80125-9}
  {\bibfield  {journal} {\bibinfo  {journal} {Chem. Phys. Lett.}\ }\textbf
  {\bibinfo {volume} {209}},\ \bibinfo {pages} {506--512} (\bibinfo {year}
  {1993})}\BibitemShut {NoStop}%
\bibitem [{\citenamefont {Dasgupta}\ and\ \citenamefont
  {Herbert}(2017)}]{dasgupta2017standard}%
  \BibitemOpen
  \bibfield  {author} {\bibinfo {author} {\bibfnamefont {S.}~\bibnamefont
  {Dasgupta}}\ and\ \bibinfo {author} {\bibfnamefont {J.~M.}\ \bibnamefont
  {Herbert}},\ }\bibfield  {title} {\enquote {\bibinfo {title} {Standard grids
  for high-precision integration of modern density functionals: {Sg-2} and
  {SG-3}},}\ }\href {\doibase 10.1002/jcc.24761} {\bibfield  {journal}
  {\bibinfo  {journal} {J. Comput. Chem.}\ }\textbf {\bibinfo {volume} {38}},\
  \bibinfo {pages} {869--882} (\bibinfo {year} {2017})}\BibitemShut {NoStop}%
\bibitem [{\citenamefont {Sitkiewicz}\ \emph {et~al.}(2022)\citenamefont
  {Sitkiewicz}, \citenamefont {Zale\'{s}ny}, \citenamefont {Ramos-Cordoba},
  \citenamefont {Luis},\ and\ \citenamefont {Matito}}]{sitkiewicz2022reliable}%
  \BibitemOpen
  \bibfield  {author} {\bibinfo {author} {\bibfnamefont {S.~P.}\ \bibnamefont
  {Sitkiewicz}}, \bibinfo {author} {\bibfnamefont {R.}~\bibnamefont
  {Zale\'{s}ny}}, \bibinfo {author} {\bibfnamefont {E.}~\bibnamefont
  {Ramos-Cordoba}}, \bibinfo {author} {\bibfnamefont {J.~M.}\ \bibnamefont
  {Luis}}, \ and\ \bibinfo {author} {\bibfnamefont {E.}~\bibnamefont
  {Matito}},\ }\bibfield  {title} {\enquote {\bibinfo {title} {How reliable are
  modern density functional approximations to simulate vibrational
  spectroscopies?}}\ }\href@noop {} {\bibfield  {journal} {\bibinfo  {journal}
  {J. Phys. Chem. Lett.}\ }\textbf {\bibinfo {volume} {13}},\ \bibinfo {pages}
  {5963--5968} (\bibinfo {year} {2022})}\BibitemShut {NoStop}%
\bibitem [{\citenamefont {Perdew}\ \emph {et~al.}(2005)\citenamefont {Perdew},
  \citenamefont {Ruzsinszky}, \citenamefont {Tao}, \citenamefont {Staroverov},
  \citenamefont {Scuseria},\ and\ \citenamefont
  {Csonka}}]{perdew2005prescription}%
  \BibitemOpen
  \bibfield  {author} {\bibinfo {author} {\bibfnamefont {J.~P.}\ \bibnamefont
  {Perdew}}, \bibinfo {author} {\bibfnamefont {A.}~\bibnamefont {Ruzsinszky}},
  \bibinfo {author} {\bibfnamefont {J.}~\bibnamefont {Tao}}, \bibinfo {author}
  {\bibfnamefont {V.~N.}\ \bibnamefont {Staroverov}}, \bibinfo {author}
  {\bibfnamefont {G.~E.}\ \bibnamefont {Scuseria}}, \ and\ \bibinfo {author}
  {\bibfnamefont {G.~I.}\ \bibnamefont {Csonka}},\ }\bibfield  {title}
  {\enquote {\bibinfo {title} {Prescription for the design and selection of
  density functional approximations: More constraint satisfaction with fewer
  fits},}\ }\href {\doibase 10.1063/1.1904565} {\bibfield  {journal} {\bibinfo
  {journal} {J. Chem. Phys.}\ }\textbf {\bibinfo {volume} {123}},\ \bibinfo
  {pages} {062201} (\bibinfo {year} {2005})}\BibitemShut {NoStop}%
\end{thebibliography}%
\end{document}